\documentclass[biblography,aps,rmp,preprint,groupedaddress,11pt,tightenlines,twocolumn]{revtex4-1}
\usepackage{graphics,bm}
\usepackage{epsfig}
\usepackage{graphicx}
\usepackage{feynmp}
\newcommand{\beq}{\begin{equation}}
\newcommand{\eeq}{\end{equation}}
\newcommand{\bqa}{\begin{eqnarray}}
\newcommand{\eqa}{\end{eqnarray}}

\def\sumint{\hbox{$\sum$}\!\!\!\!\!\!\int}
\def\square{\vcenter{\vbox{\hrule height.4pt
          \hbox{\vrule width.4pt height4pt
          \kern4pt\vrule width.3pt}\hrule height.4pt}}}

\begin{document}


\title{Phase diagram of QCD in a magnetic field: A review}


\author{Jens O. Andersen}
\author{William R. Naylor}
\affiliation{Department of Physics, 
Norwegian University of Science and Technology, 
H{\o}gskoleringen 5,
N-7491 Trondheim, Norway}
\author{Anders Tranberg}
\affiliation{Faculty of Science and Technology, University of Stavanger,
N-4036 Stavanger, Norway}

\date{\today}

\begin{abstract}
We review in detail recent advances in our understanding of
the phase structure and the phase transitions of 
hadronic matter in strong magnetic fields $B$ and zero quark chemical
potentials $\mu_f$.
Many aspects of QCD are described using low-energy effective
theories and models such as the MIT bag model,
the hadron resonance gas model, chiral perturbation theory, 
the Nambu-Jona-Lasinio (NJL) model, the quark-meson (QM) model and 
Polyakov-loop extended versions of the NJL and QM models. 
We critically examine their properties and applications.
This includes mean-field calculations as well as 
approaches beyond the mean-field approximation
such as the functional renormalization
group (FRG). Renormalization issues are discussed and
the influence of the vacuum fluctuations on the chiral
phase transition is pointed out.
Magnetic catalysis at $T=0$ is covered as well.
We discuss recent lattice results for the thermodynamics of
nonabelian gauge theories with emphasis on $SU(2)_c$ and
$SU(3)_c$. In particular, we focus on inverse magnetic catalysis
around the transition temperature $T_c$ as a competition between
contributions from valence quarks and sea quarks resulting 
in a decrease of $T_c$ as a function of $B$.
Finally, we discuss recent efforts
to modify models in order to reproduce the behavior observed on the
lattice.

\end{abstract}

\pacs{21}

\maketitle

\tableofcontents
\section{Introduction}
The phase structure of QCD is usually drawn in a phase diagram
spanned by the temperature $T$ and the baryon chemical potential $\mu_B$.
The first phase diagram was conjectured already in the 1970s suggesting
a confined low-temperature phase of hadrons and a deconfined 
high-temperature phase of quarks and gluons. 
Since the appearence of this phase diagram, huge efforts have been made
to map it out in detail. 
It turns out that the phase diagram of QCD
is surprisingsly rich, for example
there may be several color superconducting phases at low temperature 
depending on the baryon chemical potential
$\mu_B$~\cite{hatsudafuku,alfordcolor,will,hsub,alfordb}. 
Furthermore, the phase diagram can be generalized
in a variety of ways. For example, 
instead of using a baryon chemical potential $\mu_B$, i.e.~the same
chemical potential for each quark flavor, one can introduce an independent
chemical potential $\mu_f$ for each flavor. 
Equivalently (for two quark flavors), one can use a baryon chemical potential
$\mu_B={1\over2}(\mu_u+\mu_d$) and an isospin chemical potential 
$\mu_I={1\over2}(\mu_u-\mu_d)$. 
A nonzero isospin
chemical potential allows for new phases with pion condensation
once it exceeds the pion mass, $\mu_I\geq m_{\pi}$~\cite{sonmisha}.
One can also add one new axis for each quark mass $m_f$
in the system. It turns out that the nature of the chiral transition 
depends on the number of flavors and on their 
masses~\cite{stefanov,pisarskichiral}, 
and this information
has been conveniently displayed in the so called Columbia plot.
Finally, there are external parameters, such as an external magnetic field $B$,
that can be varied and are of phenomenological interest.

\begin{figure}
\begin{center}
\setlength{\unitlength}{1mm}
\includegraphics[width=7.0cm]{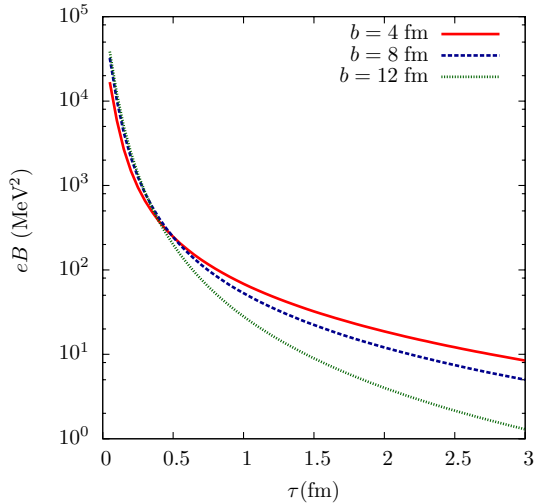}
\caption{Magnetic field 
as a function of proper time $\tau$ for three
different values of the impact parameter $b$. Figure
taken from \textcite{harmen}.}
\label{karsa}
\end{center}
\end{figure}

There are at least three 
areas of high-energy physics where strong magnetic fields
play an important role:
\begin{quote}
(1) Noncentral heavy-ion collisions
\\ \\
(2) Compact stars
\\ \\
(3) The early universe
\end{quote}
In noncentral heavy-ion collisions, very strong and time-dependent
magnetic fields are created. The basic mechanism is simple.
In the center-of-mass frame, the two nuclei represent electric currents
in opposite directions and according to Maxwell's equations, 
they produce a magnetic field $B$. The magnetic field depends on 
the energy of the ions, the impact parameter $b$, 
position as well as time.
Detailed calculations of these magnetic fields 
depend on a number of assumptions.
For example, it is common to ignore the contribution to the magnetic field
from the particles produced in the collision as the expansion of these
is almost spherical.
It is then sufficient to take into account only the colliding
particles.~\cite{harmen}.
The strength of these short-lived fields have been estimated to be
up to the order of $B\sim 10^{19}$ 
Gauss or $|qB|\sim 6m_{\pi}^2$, where $q$ is the electric charge of the pion.
Detailed calculations have been carried out 
by~\textcite{harmen,toneev,bzdak}. The result of such a calculation
is displayed in Fig.~\ref{karsa}, where the curves show the magnetic
field as a function of proper time $\tau$ 
for three different impact parameters.

There is a certain class of neutron stars, called magnetars, 
that is characterized by very high magnetic fields and 
relatively low rotation frequencies as compared to a typical 
neutron star~\cite{duncan}.
The strength of the 
magnetic fields on the surface of such stars
are believed to be on the order of $10^{14}-10^{15}$ Gauss.
The magnetic field strength depends on the density and is highest in the core of
the star. In the interior one expects 
magnetic fields on the order of $10^{16}-10^{19}$ Gauss.
This implies that in order to calculate the mass-radius relation for 
magnetars, a detailed knowledge of the equation of state of strongly
interacting matter in a large range of magnetic field strengths is required.
If the density in the core of the star is sufficiently large to
allow for quark matter, one must match the equation of state for
hadronic matter to that of deconfined quark matter.
The latter may again be color superconducting and perhaps even
inhomogeneous depending on the values of the relevant 
parameters~\cite{magfer,ferrob,magincera,warringab,shovstar,shovstar2}.

The situation is further complicated by the fact that the
star is (globally) electrically neutral as well as the fact that
the magnetic field breaks spherical symmetry.
The magnetic field then gives rise to an anisotropic pressure ${\cal P}$
whose components ${\cal P}_{ij}$ can be expressed in terms of the 
components of the energy-momentum tensor ${\cal T}_{ij}$, that enter
on the right-hand side of Einstein's field equations~\cite{mike+}.

In the absence of a magnetic field, the Minimal Standard Model
has a first-order transition for low Higgs masses $m_H$~\cite{peisa}.
With increasing $m_H$ the first-order transition becomes weaker~\cite{weakens} 
and
the first-order line eventually ends at a second-order point for
a critical value $m_H^c\approx72$ GeV~\cite{crithiggs}.
The universality class of the critical end point is that of the 
three-dimensional Ising model.
For larger Higgs masses, there is only a crossover.
In the presence of a (hyper)magnetic field, the transition becomes somewhat 
stronger~\cite{giov}.   
Allowing for a primordial hypermagnetic field of arbitrary magnitude, it is 
possible that even at the physical Higgs mass, the electroweak phase transition 
may be first order. If the magnetic fields are generated from bubble collisions
during the electroweak transition, they will typically be of order  
$B/T^2 \lesssim 0.5$~\cite{baym}, in which case 
non-perturbative numerical simulations suggest
that the transition is still not first order at the physical Higgs 
mass~\cite{peisa}.
Moreover, 
such magnetic fields could have other implications relevant for baryogenesis, 
for instance through its effect on sphaleron processes~\cite{simon}. 
In the Minimal Supersymmetric Standard Model, the transition is 
stronger than in the Standard Model, and even moderate magnetic fields may allow
for a first order transition
even at the physical Higgs mass.

There is now a large body of literature on QCD in a magnetic background
and we think that a review on the subject is timely.
In order to restrict the topics covered, we will focus on zero quark 
chemical potentials, $\mu_f=0$. Even with this restriction, we have to make
a selection of topics and papers that we consider in detail.
Such a selection is debatable, but hopefully we have covered the
field in a balanced way. Finally, a word of caution. Writing a review
is a challenge in terms of notation. We have tried to consistently 
use the same notation for a given quantity, but once in a while
we have changed the notation so as not to be in conflict with the
notation for other quantitites. Hopefully, it is clear from the
context which is which.

The review is organized as follows.
In Sec.~\ref{dspectrum}, we briefly discuss the solutions to the
Dirac equation in a constant magnetic field and explain 
that there is no sign problem in lattice QCD in a finite magnetic field.
In Sec.~\ref{freeenergies}, we calculate the one-loop free energy
densities
for fermions and bosons in a constant magnetic field
using dimensional regularization and $\zeta$-function regularization.
In Sec.~\ref{svinger}, we discuss Schwinger's classic results for
the vacuum energy of bosons and fermions 
in a constant magnetic background $B$.
In Sec.~\ref{efttm}, we discuss various low-energy models and theories 
that are being used to study the behavior
of hadronic systems at finite $T$ and $B$. These include 
the MIT bag model, chiral
perturbation theory (Chpt), the Nambu-Jona-Lasinio (NJL) model, 
and the quark-meson (QM) model.
The Polyakov extended versions of the NJL and QM models 
(PNJL and PQM models) are discussed in Sec.~\ref{extended}. 
In Sec.~\ref{funxio}, we review the functional renormalization
group and its application to hadronic matter at finite $B$.
In Sec.~\ref{magcal}, we discuss magnetic catalysis at zero temperature
and compare model - Dyson-Schwinger (DS), 
and lattice calculations.
In Sec.~\ref{latsec}, lattice results for $SU(2)_c$ and $SU(3)_c$
at finite temperature are reviewed, focussing on the mechanism behind
magnetic catalysis.
In Sec.~\ref{revisit}, we analyze recent efforts to
incorporate inverse magnetic catalysis in model calculations.
Finally, we discuss anisotropic pressure and magnetization
in Sec.~\ref{anis}.
The appendices provide the reader with our conventions and notation,
list of sum-integrals needed in the calculations, expansions of
some special functions, and some explicit calculations.

\section{Energy spectra for charged particles
in a constant magnetic field and their propagators}\label{dspectrum}
In this section, we briefly discuss the spectra of fermions and bosons 
in a constant magnetic background $B$.
We first consider fermions.
The Dirac equation for a single fermion of mass $m_f$ 
in a background gauge field $A_{\mu}$ is given by
\bqa
\left(iD\!\!\!\!/-m_f\right)\psi&=&0\;,
\label{dirac}
\eqa
where $D\!\!\!\!/=\gamma^{\mu}D_{\mu}$, 
$\gamma^{\mu}$ are the $\gamma$-matrices in Minkowski space,
$D_{\mu}=\partial_{\mu}-iq_fA_{\mu}$ is the covariant derivative, and $q_f$
is the electric charge.
In the case where the zeroth component of the gauge field vanishes,
$A_0=0$, the stationary solutions can be written as
\bqa
\psi&=&e^{-iEt}
\left(
\begin{array}{c}
\phi\\
\chi\\
\end{array}\right)\;,
\label{station}
\eqa
where $\phi$ and $\chi$ are two-component spinors.
Inserting Eq.~(\ref{station}) into Eq.~(\ref{dirac}) and using the 
Dirac representation of the $\gamma$-matrices, we obtain
the coupled equations
\bqa
\label{set11}
\left(E-m_f\right)\phi
&=&-i\left({\boldsymbol \sigma}\cdot{\boldsymbol D}\right)\chi\;, 
\\
\left(E+m_f\right)\chi
&=&-i\left({\boldsymbol \sigma}\cdot{\boldsymbol D}\right)\phi\;.
\label{set22}
\vspace{1cm}
\eqa
Eliminating $\chi$ from the Eqs.~(\ref{set11})--(\ref{set22}), we find
an equation for $\phi$
\bqa
\left(E^2-m_f^2\right)\phi&=&
-\left({\boldsymbol \sigma}\cdot{\boldsymbol D}\right)^2\phi\;.
\label{eli}
\eqa
Specializing to a constant magnetic field, we choose 
the Landau gauge, $A_{\mu}=(0,0,-Bx,0)$.\footnote{Another common choice
is the symmetric gauge,  $A_{\mu}=\mbox{$1\over2$}(0,By,-Bx,0)$.} 
Eq.~(\ref{eli}) then reads
\begin{widetext}
\bqa
\left[E^2-m_f^2
+{\partial^2\over\partial x^2}
+\left({\partial\over\partial y}+iq_fBx\right)^2
+{\partial^2\over\partial z^2}
+\sigma_zq_fB
\right]\phi&=&0\;.
\eqa
The solution is now written as $\phi=e^{is_{\perp}(q_fB)p_yy+ip_zz}f(x)$, 
where $s_{\perp}(q_fB)={\rm sign}(q_fB)$. 
The equation for $f(x)$ then becomes
\bqa
\left[
-{d^2\over dx^2}+\left(s_{\perp}p_y+q_fBx\right)^2-\sigma_zq_fB
\right]f(x)&=&\left[E^2-m_f^2-p_z^2\right]f(x)\;.
\eqa
This is a $2\times2$ matrix equation. However, the two equations
decouple and 
the solutions can then be written as 
\bqa
f(x)=\mbox{$\left(
\begin{array}{c}
f_+(x)\\
0
\end{array}\right)$}
\hspace{1cm}
{\rm and}
\hspace{1cm}
f(x)=
\left(
\begin{array}{c}
0\\
f_{-}(x)
\end{array}\right)\;, 
\eqa
where the subscript $\pm$ indicates that the solutions are
eigenvectors of $\sigma_z$ with eigenvalues $\pm1$, respectively.
The equation for $f_{\pm}(x)$ finally becomes
\bqa
\left[
-{d^2\over dx^2}+\left(s_{\perp}p_y+q_fBx\right)^2
\right]f_{\pm}(x)&=&
\left[E^2-m_f^2-p_z^2\pm q_fB\right]f_{\pm}(x)\;.
\eqa
\end{widetext}
This is the equation for a harmonic oscillator with known solutions
involving the Hermite polynomials $H_k(x)$.
The solutions are
\bqa\nonumber
\phi&=&
ce^{-{1\over2}({x\over l}+p_yl)^2}
H_k\left(\mbox{${x\over l}+p_yl$}\right)
\\&&\times 
e^{i(s_{\perp}p_yy+p_zz)}\;,
\label{solutions}
\eqa
where $c$ is a normalization constant and $l=1/\sqrt{|q_fB|}$.
The spectrum is 
\bqa
E_k^2&=&m_f^2+p_z^2+|q_fB|(2k+1-s)\;,
\label{fermspec}
\eqa
where $k=0,1,2,...$ and $s=\pm1$.
We note that
there is a two-fold degeneracy due to the spin variable $s$
for all values of $k$ except for $k=0$.
Moreover, the energy is independent of $p_y$ and 
the 
energy levels are therefore degenerate in this variable. 
The degeneracy is associated with the position of the center
of the Landau levels. Assume that we use a quantization volume
$V=L^3$, where $L$ is the length of the side of the box.
Since the characteristic size of a Landau level
is $1/\sqrt{|q_fB|}$, the degeneracy $N$ associated with the quantum number
$p_y$ is
$N={|q_fB|\over2\pi}L^2\;.$
The sum over states in the quantization volume $V$ is then given by
a sum over spin $s$, Landau levels $k$, and the $z$-component of the 
momentum $p_z$ multiplied by $N$:
\bqa
{1\over V}{|q_fB|\over2\pi}L^2\sum_{s=\pm1}\sum_{k=0}^{\infty}\sum_{p_z}\;.
\label{diskret}
\eqa
In the thermodynamic limit, the sum over $p_z$ is replaced by an 
integral such that the expression in Eq.~(\ref{diskret}) is replaced by 
\bqa
{|q_fB|\over2\pi}\sum_{s=\pm1}\sum_{k=0}^{\infty}\int_{-\infty}^{\infty}
{dp_z\over2\pi}\;.
\label{subst1}
\eqa
In the thermodynamic limit and for $B=0$, the sum over 
three-momenta ${\boldsymbol p}$ is replaced by an integral in the usual way
\bqa
{1\over V}\sum_{\boldsymbol p}\rightarrow\int{d^3p\over(2\pi)^3}\;.
\label{subst2}
\eqa
Once we have found a complete set of eigenstates, we can calculate the
fermion propagator. We then need the expression for 
the two-component spinor $\chi$ as well. 
The fermion propagator at $T=0$ for a fermion with electric charge $q_f$
in Minkowski space is given by the expression~\cite{shovcat2}
\begin{widetext}
\bqa
S(x,x^{\prime})&=&
e^{i\Phi({\bf x}_{\perp},{\bf x}_{\perp}^{\prime})}
\int{d^4p\over(2\pi)^4}
e^{-ip(x-x^{\prime})}\tilde{S}({\bf p}_{\perp},{\bf p}_{\parallel})
\;,
\eqa
where $x=(t,{\bf x})$, ${\bf x}_{\perp}=(x^1,x^2)$, 
$p=(p_0,{\bf p})$,
${\bf p}_{\perp}=(p_1,p_2)$, ${\bf p}_{\parallel}=(p_0,p_3)$
and with 
$\Phi({\bf x}_{\perp},{\bf x}_{\perp}^{\prime})$
and
$\tilde{S}({\bf p}_{\perp},{\bf p}_{\parallel})$ given by
\bqa
\Phi({\bf x}_{\perp},{\bf x}_{\perp}^{\prime})
&=&s_{\perp}{(x^1+x^{\prime1})(x^2-x^{\prime2})\over 2l^2}\;,
\\ 
\nonumber
\tilde{S}({\bf p}_{\perp},{\bf p}_{\parallel})
&=&
\int_0^{\infty}{ds}\,\exp\left\{is\left[p^2_{\parallel}
-m_f^2\right]
-i{p^2_{\perp}\over|q_fB|}\tan\left(|q_fB|s\right)
\right\}
\\
&&
\times\left[
\left(\gamma^0p_0-\gamma^3p_3+m)(1+\gamma^1\gamma^2\tan(q_fBs)\right)
-{\boldsymbol\gamma}_{\perp}\cdot{\bf p}_{\perp}\left(1+\tan^2(q_fBs)\right)
\right]
\eqa
The prefactor $\Phi({\bf x}_{\perp},{\bf x}_{\perp}^{\prime})$
is the so-called Schwinger phase and the 
term $\tilde{S}({p}_{\perp},{p}_{\parallel})$ is translationally
invariant.  The translationally invariant part can be
decomposed into contributions from the different Landau levels
\bqa
\tilde{S}({\bf p}_{\perp},{\bf p}_{\parallel})
&=&ie^{-{p_{\perp}^2\over|q_fB|}}\sum_{k=0}^{\infty}
{(-1)^kD_k({\bf p}_{\perp},{\bf p}_{\parallel})
\over p^2_{\parallel}-m_f^2-2|q_fB|k}\;,
\label{decomp}
\eqa
where
\bqa\nonumber
D_k({\bf p}_{\perp},{\bf p}_{\parallel})
&=&
(\gamma^0p_0-\gamma^3p_3+m)\left[
(1-i\gamma^1\gamma^2s_{\perp})L_k\left(2\mbox{$p_{\perp}^2\over|q_fB|$}\right)
-(1+i\gamma^1\gamma^2s_{\perp})L_{k-1}\left(2\mbox{$p_{\perp}^2\over|q_fB|$}\right)
\right]
\\ &&
+4({\bf\gamma}_{\perp}\cdot{\bf p}_{\perp})
L_{k-1}^1\left(2\mbox{$p_{\perp}^2\over|q_fB|$}\right)
\;,
\eqa
\end{widetext}
and $L_k^a(x)$ are the generalized Laguerre polynomials.
Note that $L_{-1}^a(x)=0$.

The spectrum for bosons with mass $m$ and charge $q$
can be found using the same techniques.
In this case, the differential operator does not involve the 
term $|q_fB|\sigma_z$ and so the resulting eigenvalue
equation is easier to solve. The spectrum is obtained immediately
by setting $s=0$:
\bqa
E_k^2&=&m^2+p_z^2+|qB|(2k+1)\;,
\label{ebos}
\eqa
where $k=0,1,2,...$ . The eigenfunction are again given by 
Eq.~(\ref{solutions}).
Once we have a complete set of eigenfunctions, we can 
derive the propagator. We derive the bosonic
propagator in Appendix~\ref{propapp} and at $T=0$
it reads
\begin{widetext}
\bqa
\label{bosprop}
\Delta(x,x^{\prime})&=&
e^{i\Phi({\bf x}_{\perp},{\bf x}_{\perp}^{\prime})}
\int{d^4p\over(2\pi)^4}
e^{-ip(x-y)}\Delta_B({p}_{\perp},{p}_{\parallel})\;,
\eqa
where the translationally invariant part is  
\bqa
\Delta_B({\bf p}_{\perp},{\bf p}_{\parallel})
&=&\int_0^{\infty}{ds\over\cos(|qB|s)}\exp\left\{is\left[
p_{\parallel}^2-m^2\right]-ip_{\perp}^2{\tan(|qB|s)\over|qB|}
\right\}\;.
\eqa
\end{widetext}

We close this section with briefly commenting on the sign problem
of QCD. It is straightforward to show that there is no sign problem
in QCD in an external Abelian gauge field $A_{\mu}$.
In order to show this, we go to Euclidean space.
The partition function of QCD can be written as
\bqa\nonumber
{\cal Z}&=&
\int{\cal D}\bar{\psi}{\cal D}\psi{\cal D}A_{\mu}
e^{-\int d^3x\int_0^{\beta}d\tau\,\bar{\psi}[(D\!\!\!\!/+m_f)]\psi}e^{-S_{g}}
\\ 
&=&\int{\cal D}A_{\mu}
e^{-S_{g}}\det(D\!\!\!\!/+m_f)
\;,
\eqa
where $\beta=1/T$ and $S_g$ is the Euclidean action for the gluons,
\bqa
S_g&=&{1\over4}\int_0^{\beta}d\tau\int d^3x\,
{\rm Tr}\left[G_{\mu\nu}G_{\mu\nu}\right]\;.
\eqa
$S_g>0$ and the exponent can be regarded as a positive probability weight. 
We also have to check the sign of the fermion determinant.
It is convenient to use the 
chiral representation of the $\gamma$-matrices.
The matrix $D\!\!\!\!/$ can then be written as
\bqa
D\!\!\!\!/
&=&
\left(
\begin{array}{cc}
0&iX\\
iX^{\dagger}&0
\end{array}\right)\;,
\eqa
where 
$iX=D_0+i{\boldsymbol\sigma}\cdot{\boldsymbol D}$.
The fermion determinant then takes the form
\bqa
\det(D\!\!\!\!/+m_f)
&=&\det\left[XX^{\dagger}+m_f^2\right]\;,
\label{deteq}
\eqa
which shows that it is manifestly positive. 
QCD in a magnetic field is therefore
free of the sign problem and one can use standard lattice techniques
based on importance sampling.

\section{One-loop free energy densities}\label{freeenergies}
In this review, we are often concerned with Euclidean Lagrangian densities
of the form
\bqa\nonumber
{\cal L}&=&\bar{\psi}_f\gamma_{\mu}D_{\mu}\psi_f+m_f\bar{\psi}_f\psi_f
+(D_{\mu}\Phi)^{\dagger}(D_{\mu}\Phi)
\\ &&
+m^2\Phi^{\dagger}\Phi
+{\cal L}_{\rm int}\;,
\eqa
where $\psi_f$ is a fermion field of flavor $f$ and $\Phi$ is a complex
scalar field. Unless otherwise stated, we consider two flavors, $N_f=2$
and $f=u,d$. Moreover,
$D_{\mu}=\partial_{\mu}-iqA_{\mu}$ is the covariant derivative
for bosons and 
$D_{\mu}=\partial_{\mu}-iq_fA_{\mu}$ is the covariant derivative
for fermions. Here
$q=\pm e$ is the electric charge
for the charged scalars, 
$q_u=2/3e$ and $q_d=-1/3e$ are
the electric charges for $u$-quarks and $d$-quarks, respectively.
$m$ and $m_f$ are the tree-level masses of 
the bosons and fermions. 
${\cal L}_{\rm int}$ is the interacting part of the Lagrangian.
It may contain bosonic and fermionic four-point interactions as
well as Yukawa-type couplings between the bosons and fermions.

In the functional approach to the imaginary-time formalism, the
partition function ${\cal Z}$ is given by a path integral
\bqa
{\cal Z}&=&\int{\cal D}\Phi^{\dagger}{\cal D}\Phi
{\cal D}\bar{\psi}{\cal D}\psi
e^{-S[\Phi^{\dagger},\Phi,\bar{\psi},\psi]}\;,
\eqa
where the action is given by
\bqa
S[\Phi^{\dagger},\Phi,\bar{\psi},\psi]
&=&\int_0^{\beta}d\tau\int d^dx\,{\cal L }\;,
\eqa
and $d$ is the number of spatial dimensions.
In many cases, we approximate the free energy density 
${\cal F}=-{T\over V}\log{\cal Z}$ (where $V$ is the spatial volume)
of the system by a 
one-loop calculation. We therefore need to perform Gaussian integrals
over bosonic or fermionic fields. These are given by the standard
expressions.
The one-loop free energy density ${\cal F}_1$
for a boson is
\bqa
{\cal F}_1&=&
{1\over\beta V}{1\over2}{\rm Tr}\ln D_0^{-1}\;,
\eqa
and for a fermion
\bqa
{\cal F}_1&=&-{1\over\beta V}
{\rm Tr}\ln D_0^{-1}\;,
\eqa
where 
$D_0^{-1}$ is the free inverse propagator.
Here the trace is over spacetime, field indices, and 
Dirac indices in the case of fermions.
These expressions are general as they 
apply whether or not the particle couples to an external
magnetic field. Of course, the 
explicit expressions after evaluating the traces 
and making the substitutions,~(\ref{subst1})
or~(\ref{subst2}), are different.
For example, the one-loop free energy density
for a neutral boson with mass $m$
reads
\bqa
{\cal F}_1&=&{1\over2}\sumint_P\ln\Big[P_0^2+p^2+m^2\Big]\;,
\eqa 
where the sum-integral is defined in Eq.~(\ref{defsumint1}) and 
involves a sum over Matsubara frequencies $P_0$
and an integral over three-momenta ${\bf p}$.
The explicit expression for this sum-integral 
as well as others needed are listed in Appendix~\ref{sumint}.
The one-loop free energy density
for a boson
with electric charge $q$ and for a fermion with electric charge $q_f$
as a function of $B$ are given by the sum-integrals
\begin{widetext}
\bqa
{\cal F}_1&=&{1\over2}\sumint_{P}^B
\ln\Big[
P_0^2+p_z^2+m^2 
+|qB|(2k+1)
\Big]\;,
\label{sumintbb}
\\
{\cal F}_1&=&-\sumint_{\{P\}}^B
\ln\Big[
P_0^2+p_z^2+m_f^2
+|q_fB|(2k+1-s)
\Big]\;,
\label{sumintf}
\eqa
where the sum-integrals are defined in Eqs.~(\ref{defsumint3}) 
and~(\ref{defsumint4}), and
involves a sum over spin $s$, Landau levels $k$ and 
Matsubara frequencies $P_0$, as well as an integral
over $p_z$.
We next evaluate the sum-integral (\ref{sumintf}) in some detail.
We first sum over the Matsubara frequencies using Eq.~(\ref{matsusum}).
This yields
\bqa
{\cal F}_1&=&-{|q_fB|\over2\pi}\sum_{s=\pm1}\sum_{k=0}^{\infty}
\int_{p_z}\Bigg\{\sqrt{p_z^2+M_B^2}
+2T\ln\left[1+e^{-\beta\sqrt{p_z^2+M_B^2}}\right]\Bigg\}
\;,
\label{free}
\eqa
where $M_B^2=m_f^2+|q_fB|(2k+1-s)$.
Let us first consider the temperature-independent
term. Using dimensional regularization in $d-2=1-2\epsilon$ 
dimensions
to regulate
the ultraviolet divergences, we obtain
\bqa
\int_{p_z}\sqrt{p_z^2+M_B^2}
&=&-{M_B^2\over4\pi}\left({e^{\gamma_E}\Lambda^2\over M_B^2}\right)^{\epsilon}
\Gamma(-1+\epsilon)\;,
\label{vacuumres}
\eqa
where $\Lambda$ is the renormalization scale associated with
the modified minimal subtraction scheme $\overline{\rm MS}$.
The sum over Landau levels $k$ involves the term $M_B^{2-2\epsilon}$
and is divergent for $\epsilon=0$. We will regularize the sum
using $\zeta$-function regularization. 
The sum over spin $s$ and Landau levels $k$ can then be written as
\bqa\nonumber
\sum_{s=\pm1}\sum_{k=0}^{\infty}M_B^{2-2\epsilon}
&=&2(2|q_fB|)^{1-\epsilon}\sum_{k=0}^{\infty}
\left[k+{m_f^2\over2|q_fB|}\right]^{1-\epsilon} 
-m_f^{2-2\epsilon}
\\ 
&=&2(2|q_fB|)^{1-\epsilon}\zeta(-1+\epsilon,x_f)
-m_f^{2-2\epsilon}\;, 
\label{sum}
\eqa
where we have defined $x_f={m_f^2\over2|q_fB|}$ and the Hurwitz 
$\zeta$-function $\zeta(s,q)$ is defined by
\bqa
\zeta(s,q)&=&
\sum_{k=0}^{\infty}(q+k)^{-s}\;.
\eqa
Inserting Eq.~(\ref{sum}) into Eq.~(\ref{free}), the
temperature-independent part of the free energy density, ${\cal F}_1^{T=0}$,
becomes
\bqa
{\cal F}_1^{T=0}&=&
{(q_fB)^2\over2\pi^2}\left({e^{\gamma_E}\Lambda^2\over2|q_fB|}\right)^{\epsilon}
\Gamma(-1+\epsilon)
\left[\zeta(-1+\epsilon,x_f)-{1\over2}x_f^{1-\epsilon}\right]\;.
\label{f0div}
\eqa
Expanding Eq.~(\ref{f0div}) in powers of $\epsilon$ through order
$\epsilon^0$ gives
\bqa\nonumber
{\cal F}_1^{T=0}&=&
{1\over(4\pi)^2}\left({\Lambda^2\over2|q_fB|}\right)^{\epsilon}
\Bigg[
\left({2(q_fB)^2\over3}+m_f^4\right)\left({1\over\epsilon}+1\right)
-8(q_fB)^2\zeta^{(1,0)}(-1,x_f)
\\ &&
-2|q_fB|m_f^2\ln x_f+{\cal O}(\epsilon)\Bigg]\;,
\label{div}
\eqa
\end{widetext}
where we have defined
\bqa
\zeta^{(1,0)}(-a,x)={\partial\zeta(-a+\epsilon,x)\over \partial\epsilon}
\bigg|_{\epsilon=0}\;.
\eqa
The expression~(\ref{div}) has simple poles in $\epsilon$.
One of the divergences is proportional to $(q_fB)^2$ while the other
is proportional to $m_f^4$. Later we will show how one can eliminate
these divergences by renormalization.

The temperature-dependent part of the free energy density
in 
Eq.~(\ref{free}), ${\cal F}_1^T$, can be integrated
by parts and this gives
\begin{widetext}
\bqa\nonumber
{\cal F}_1^T&=&
-{|q_fB|\over\pi}T
\sum_{s=\pm1}\sum_{k=0}^{\infty}\int_{p_z}
\ln\left[1+e^{-\beta\sqrt{p_z^2+M_B^2}}\right]
\\ \nonumber
&=&-
{|q_fB|\over2\pi^2}\left({e^{\gamma_E}\Lambda^2}\right)^{\epsilon}
{\Gamma(\mbox{${1\over2}$})\over\Gamma(\mbox{${3\over2}$}-\epsilon)}
\sum_{s=\pm1}\sum_{k=0}^{\infty}
\int_0^{\infty}
{p_z^{2-2\epsilon}dp_z\over\sqrt{p_z^2+M_B^2}}{1\over e^{\beta\sqrt{p_z^2+M_B^2}}+1}\;.
\\ 
&=&-{2\over(4\pi)^2}
\left({e^{\gamma_E}\Lambda^2\over2|q_fB|}\right)^{\epsilon}
K_0^B(\beta m_f)|q_fB|T^2\;,
\label{secondterm}
\eqa
\end{widetext}
where $K_0^B(\beta m_f)$ is defined in Eq.~(\ref{knb}).
The sum of Eqs.~(\ref{div}) and~(\ref{secondterm}) is then 
Eq.~(\ref{sumfermi}).
The other sum-integrals needed can be calculated using the same 
techniques. They are listed in Appendix~\ref{sumint}.

We close this section with some comments on regulators.
By changing variables $p_{\perp}^2=2k|q_fB|$, summing over $s$ and 
taking the limit $B\rightarrow0$ in Eq.~(\ref{free}), the first
term reduces to
\bqa
{\cal F}_1^{T=0}&=&-2
\int_p\sqrt{p^2+m_f^2}\;,
\eqa
where $p^2=p_{\perp}^2+p_z^2$. Using dimensional regularization, this becomes
\bqa\nonumber
{\cal F}_1^{T=0}&=&
{1\over(4\pi)^2}\left({\Lambda^2\over m_f^2}\right)^{\epsilon}
\left[
\left(
{1\over\epsilon}+{3\over2}
\right)m_f^4+{\cal O}(\epsilon)
\right]\;.
\\ &&
\eqa
This is the same result one finds if one takes the limit $B\rightarrow0$
in Eq.~(\ref{div}) and uses the large-$x_f$ expansion of
$\zeta^{(1,0)}(-1,x_f)$ given by Eq.~(\ref{largefermi}).
Using a sharp three-dimensional cutoff $\Lambda$, one obtains
\begin{widetext}
\bqa
{\cal F}_{T=0}&=&
{1\over(4\pi)^2}
\left\{
-2\Lambda\sqrt{\Lambda^2+m_f^2}\,(2\Lambda^2+m_f^2)
+2m_f^4
\ln\left[{{\Lambda\over m_f}+\sqrt{1+{\Lambda^2\over m_f^2}}}\,\,\right]
\right\}\;.
\eqa
If the starting point is the expression for the free energy density as
as a four-dimensional Euclidean integral, one finds
by imposing a four-dimensional cutoff $\Lambda$,
\bqa\nonumber
{\cal F}_{T=0}&=&
-2\int{d^4p\over(2\pi)^4}
\ln\left[
p^2+m_f^2\right]
\\
&=&{1\over(4\pi)^2}
\left\{
{1\over2}\Lambda^4-\Lambda^4\ln\left[1+{m^2_f\over\Lambda^2}\right]
-\Lambda^2m_f^2
+m_f^4\ln\left[1+{\Lambda^2\over m^2_f}\right]
\right\}\;.
\label{4dcut}
\eqa
\end{widetext}
We notice that the coefficient of the logarithimic term is independent of
the regulator, while the power divergences (for $\Lambda\rightarrow\infty$)
depend on the regulator. In particular, they are all set to zero
in dimensional regularization while the logarithimic divergence in the
cutoff scheme corresponds to a pole in $\epsilon$
in dimensional regularization.

\section{Schwinger's results}
\label{svinger}
In this section, we rederive Schwinger's classic results~\cite{schwinger} 
for the
vacuum energy of a boson and a fermion in a constant magnetic field $B$.
In the original derivation, the result was given for an arbitrary
constant electromagnetic field.
Not only is the calculation useful to see the connection with the derivation
in the previous section, but the one-loop expression 
for the vacuum energy also takes form such that it is straightforward to
use a simple ultraviolet cutoff $\Lambda$ instead of
dimensional regularization and zeta-function regularization.
This will be useful when we consider Nambu-Jona-Lasinio models 
in which the ultraviolet divergences often are regulated by
a simple UV cutoff.

The starting point is the zero-temperature expression for the
one-loop free energy density for a charged boson with mass $m$ and charge $q$,
and its antiparticle with mass $m$ and charge $-q$.
In the limit $T\rightarrow0$, the sum over Matsubara frequencies
approaches an integral over the continuous variable $p_0$.
The free energy density reads
\bqa\nonumber
{\cal F}_1&=&{|qB|\over2\pi}\sum_{k=0}^{\infty}\int_{-\infty}^{\infty}{dp_0\over2\pi}
\int_{p_z}
\ln\left[p_0^2+p_z^2+M_B^2\right]\;.
\\ &&
\eqa
where $M^2_B=m^2+|qB|(2k+1)$.
The derivative of ${\cal F}_1$ with respect to $M^2$
is 
\bqa
\nonumber
{\partial{\cal F}_1\over\partial m^2}
&=&
{|qB|\over2\pi}\sum_{k=0}^{\infty}\int_{-\infty}^{\infty}{dp_0\over2\pi}
\int_{p_z}
{1\over p_0^2+p_z^2+M_B^2}\;.
\\ &&
\label{dfdm}
\eqa
The effective propagator in momentum space $1/(p_0^2+p_z^2+M_B^2)$
has the integral representation
\bqa
{1\over p_0^2+p_z^2+M_B^2}
&=&\int_0^{\infty}{e^{-s(p_0^2+p_z^2+M_B^2)}}ds\;.
\label{repre}
\eqa
Inserting Eq.~(\ref{repre}) into Eq.~(\ref{dfdm}), we obtain
\bqa\nonumber
{\partial{\cal F}_1\over\partial m^2}
&=&
{|qB|\over2\pi}\sum_{k=0}^{\infty}\int_{-\infty}^{\infty}{dp_0\over2\pi}
\\ &&
\times 
\int_{p_z}\int_0^{\infty}e^{-s(p_0^2+p_z^2+M_B^2)}\,ds\;.
\label{dfdm1}
\eqa
The integral over $p_z$ is finite for $\epsilon=0$
and 
after integration over $p_z$ and $p_0$, Eq.~(\ref{dfdm1}) reduces to
\bqa
{\partial{\cal F}_1\over\partial m^2}
&=&
{|qB|\over8\pi^2}\sum_{k=0}^{\infty}
\int_0^{\infty}{e^{-sM_B^2}\over s}ds\;.
\label{dfdm2}
\eqa
Likewise, the sum over Landau levels is convergent and
after summation over $k$,
Eq.~(\ref{dfdm2}) reduces to
\bqa
{\partial{\cal F}_1\over\partial m^2}
&=&{1\over(4\pi)^2}\int_0^{\infty}{ds\over s^2}e^{-sm^2}
{|qB|s\over\sinh(|qB|s)}\;.
\eqa
Finally integrating over $m^2$, we obtain
the one-loop free energy density
\bqa
{\cal F}_1
&=&-{1\over(4\pi)^2}
\int_0^{\infty}
{ds\over s^3}e^{-sm^2}
{|qB|s\over\sinh(|qB|s)}
\;,
\label{fm1}
\eqa
where the constant of integration has been set to zero.
The result, Eq.~(\ref{fm1}), is divergent at $s=0$.
Since $s$ has mass dimension $-2$, this corresponds to
an ultraviolet divergence in momentum space.
It is therefore
convenient to organize the result by adding and subtracting
divergent terms to Eq.~(\ref{fm1}), writing it as
\begin{widetext}
\bqa\nonumber
{\cal F}_{0+1}&=&
{1\over2}B^2+{(qB)^2\over6(4\pi)^2}\int_0^{\infty}
{ds\over s}e^{-sm^2}
-{1\over(4\pi)^2}\int_0^{\infty}
{ds\over s^3}e^{-sm^2}
\\ &&
-{1\over(4\pi)^2}
\int_0^{\infty}
{ds\over s^3}e^{-sm^2}
\left[{|qB|s\over\sinh(|qB|s)}
-1+{(qBs)^2\over6}
\right]\;,
\label{divint}
\eqa
where we have added the tree-level term ${1\over2}B^2$.
The first and second integrals are divergent at $s=0$, while the 
third integral is finite. 
The divergent integrals are regulated by
introducing an ultraviolet cutoff $\Lambda$ via $s=1/\Lambda^2$
and evaluating Eq.~(\ref{divint}), we obtain
\bqa\nonumber
{\cal F}_{0+1}&=&
{1\over2}B^2\left[1+{q^2\over3(4\pi)^2}
\left(\ln{\Lambda^2\over m^2}-\gamma_E\right)\right]
-{1\over2(4\pi)^2}\left[
\Lambda^4-2\Lambda^2m^2
+m^4\left(\ln{\Lambda^2\over m^2}
-\gamma_E+{3\over2}\right)
\right]
\\ &&
+{4(qB)^2\over(4\pi)^2}
\left[
\zeta^{(1,0)}(-1,\mbox{${1\over2}$}+x)
+{1\over4}x^2-{1\over2}x^2\ln x+{1\over24}\ln x+{1\over24}
\right]\;,
\eqa
where $x={m^2\over2|qB|}$.
In most applications, one omits the $\Lambda^4$-term 
as it is independent of $m$ and $B$.

For fermions with mass $m_f$ and electric
charge $q_f$, one obtains in a similar manner the result
\bqa\nonumber
{\cal F}_{0+1}&=&
{1\over2}B^2+{4(q_fB)^2\over3(4\pi)^2}\int_0^{\infty}{ds\over s}
e^{-sm_f^2}
+{2\over(4\pi)^2}\int_0^{\infty}{ds\over s^3}e^{-sm_f^2}
\\ \nonumber
&&+{2\over(4\pi)^2}\int_0^{\infty}{ds\over s^3}e^{-sm_f^2}
\left[
|q_fB|s\coth(|q_fB|s)-1-{1\over3}(q_fBs)^2
\right]\\ \nonumber
&=&
{1\over2}B^2\left[1+{4q_f^2\over3(4\pi)^2}
\left(\ln{\Lambda^2\over m_f^2}-\gamma_E\right)\right]
+{1\over(4\pi)^2}\left[
\Lambda^4-2\Lambda^2m_f^2
+m_f^4\left(\ln{\Lambda^2\over m_f^2}-\gamma_E+
{3\over2}
\right)
\right]
\\ &&
-{8(q_fB)^2\over(4\pi)^2}\left[
\zeta^{(1,0)}(-1,x_f)
+{1\over4}x_f^2-{1\over2}x_f^2\ln x_f
+{1\over2}x_f\ln x_f-{1\over12}\ln x_f-{1\over12}
\right]
\;.
\eqa
We end this section by noting that there
is alternative way of regularing the divergent integrals over $s$.
Instead of performing these in integrals in one dimension, we use 
dimensional regularization. For example, the first integral 
in Eq.~(\ref{divint})
is replaced by
\bqa\nonumber
{\left(e^{\gamma_E}\Lambda^2\right)^{\epsilon}
\over(4\pi)^2}
\int_0^{\infty}{ds\over s^{3-\epsilon}}{e^{-sm^2}}
={m^4\over(4\pi)^2}\left({e^{\gamma_E}\Lambda^2\over m^2}\right)^{\epsilon}
\Gamma(-2+\epsilon)
={m^4\over2(4\pi)^2}\left[{1\over\epsilon}+{3\over2}
+\ln\left({\Lambda^2\over m^2}\right)
+{\cal O}(\epsilon)\right]\;.
\\ &&
\label{rightie}
\eqa
With the extra factor of $e^{\gamma_E\epsilon}$, the result~(\ref{rightie})
is identical to that obtained in the $\overline{\rm MS}$ scheme,
cf. Eq.~(\ref{sumb}).

\end{widetext}

\section{Effective theories and models}
\label{efttm}

\subsection{MIT bag model}
\label{baggie}
The MIT bag model was introduced in the 1970s as a simple phenomenological
model for the confinement of quarks inside hadrons~\cite{jaffe}.
The quarks are confined to a spherical cavity by requiring that the
quark vector current vanishes on the boundary. The quarks inside the
bag are considered non-interacting which is justified by appealing
to asymptotic freedom of QCD. 
The idea is that the vacuum energy density of the 
perturbative vacuum (inside the hadron) is larger than 
than that of the nonperturbative vacuum outside the bag.
Equivalently, the vacuum pressure inside the bag is smaller than that
outside the bag
and the radius $R$ of a hadron is (heurestically) given by 
the balance between this difference and the pressure generated by the 
quarks inside the bag. 
The bag constant $B$ can be estimated as follows~\cite{johnson}. 
The pressure generated by the quarks inside a spherical cavity
is, by the uncertainty relation, on the order of $1/R$.
Balancing this contribution and that from the bag, which 
is on the order of $R^3$, one finds a relation between the mass and the 
radius of hadron.
Minimizing the total mass with respect to $R$ gives
$R\sim B^{-{1\over4}}$ and using the mass of a proton gives a 
bag constant on the order of $100$ MeV.

~\textcite{chakra} was the first to discuss the
MIT bag model in a magnetic field in the context of 
compact stars and the stability of strange quark matter.
More recently, the deconfinement transition has been investigated
using the bag model~\cite{fragamit}.

One can investigate the phase structure of QCD by
calculating the pressure in the hadronic phase as well as in the
deconfined phase as a function of temperature, particle masses
and magnetic field $B$. The phase with the larger pressure
wins. The transition takes place when the pressure in the two
phases is equal and the deconfinement temperature 
therefore satisfies $P_{\rm HHG}=P_{\rm QGP}$.
Thus there is no order parameter for the deconfinement transition in 
the bag model.
This is different from 
other models we will be discussing later, such as the
Polyakov-loop extended Nambu-Jona-Lasinio model and the
Polyakov-loop extended quark-meson model.
In these models, we analyze the behavior of the Polyakov loop variable
which is an order parameter for confinement in pure-glue QCD
and an approximate order parameter in QCD with dynamical 
quarks~\cite{benjamin1,benjamin2}.
Finally, the MIT bag model tells us nothing about the chiral transition
in QCD since this is governed by the behavior of the quark condensate
as a function of $T$ (and $B$).

The free energy density of the hadronic phase is 
approximated by that of an ideal gas 
of massive pions and reads
\begin{widetext}
\bqa
{\cal F}_{\rm HHG}&=&{1\over2}B^2+
{1\over2}\sumint_P\ln\left[P^2+m_{\pi}^2\right]
+\sumint_P^B\ln\left[P_0^2+p_z^2+M_B^2\right]\;,
\label{mitbos}
\eqa
where the first term is the tree-level 
contribution from the constant magnetic field and 
the second term is from the
neutral pion which does not couple
to the magnetic field. The third term is the contribution from the
charged pions.
This expression is the same as one obtains from a one-loop
approximation to the free energy density
in chiral perturbation theory.
We will discuss Chpt later.

Using the expressions for the bosonic sum-integrals given by
Eqs.~(\ref{sumb}) and~(\ref{sumbose}),
Eq.~(\ref{mitbos}) can be written as 
\bqa\nonumber
{\cal F}_{\rm HHG}
&=&
{1\over2}B^2+{1\over2(4\pi)^2}
\left({\Lambda^2\over|2qB|}\right)^{\epsilon}\bigg[
\left({(qB)^2\over3}-m_{\pi}^4\right)\left({1\over\epsilon}+1\right)
+8(qB)^2\zeta^{(1,0)}(-1,\mbox{${1\over2}$}+x)
\\ &&
-2J_0^B(\beta m_{\pi})|qB|T^2\bigg]
-{1\over4(4\pi)^2}
\left({\Lambda^2\over m^2_{\pi}}\right)^{\epsilon}
\left[m_{\pi}^4\left({1\over\epsilon}+{3\over2}\right)
+2J_0(\beta m_{\pi})T^4\right]\;,
\label{hhg}
\eqa
where the thermal integrals $J_n(\beta m)$ and $J_n^B(\beta m)$
are defined in Appendix~\ref{sumint}.
The first divergence is proportional to $(qB)^2$
and is removed by wavefunction renormalization of
the term ${1\over2}B^2$ in Eq.~(\ref{hhg}).
This is done by making the replacement $B^2\rightarrow Z^2B^2$,
where $Z$ is the wavefunction renormalization term
The second divergence, which is proportional to $m_{\pi}^4$, can be removed
by adding an appropriately chosen vacuum counterterm 
$\Delta{\cal E}$ to the free energy density. 
The counterterms are given by 
\bqa
Z^2=1-{q^2\over3(4\pi)^2\epsilon}\;,
\hspace{1cm}
\Delta{\cal E}={3m_{\pi}^4\over4(4\pi)^2\epsilon}\;.
\label{minsub}
\eqa
After renormalization, the free energy density of the hot hadronic gas is
\bqa\nonumber
{\cal F}_{\rm HHG}&=&{1\over2}B^2\left[
1+{q^2\over3(4\pi)^2}\ln{\Lambda^2\over2|qB|}
\right]
-{3m_{\pi}^4\over4(4\pi)^2}\left[\ln{\Lambda^2\over m_{\pi}^2}+{3\over2}\right]
+{4(qB)^2\over(4\pi)^2}
\bigg[
\zeta^{(1,0)}(-1,\mbox{${1\over2}$}+x)
\\ &&
+{1\over4}x^2-{1\over2}x^2\ln x+{1\over24}
\bigg]
-{1\over2(4\pi)^2}\left[
J_0(\beta m_{\pi})T^4+2J_0^B(\beta m_{\pi})|qB|T^2
\right]
\;.
\label{HHG}
\eqa
\end{widetext}
Note that here and in the following, the thermal integrals
$J_n(\beta m)$, $J_n^B(\beta m)$,
$K_n(\beta m)$, and $K_n^B(\beta m)$
are always evaluated at $\epsilon=0$ whenever they appear in 
renormalized expressions for the free energy density
and other physical quantities.

The free energy density in the quark-gluon plasma phase is
\begin{widetext}
\bqa
{\cal F}_{\rm QGP}&=&{1\over2}B^2+(N_c^2-1)\sumint_P\ln(P^2)
-N_c\sum_f\sumint_{\{P\}}^B\ln\left[
P_0^2+p_z^2+M_B^2\right]+{\cal B}\;,
\label{qgp}
\eqa
where the first term is from the constant magnetic field,
the second term is from the gluons, the third term
is from the quarks, and the last term ${\cal B}$ is the bag constant.
This term represents the difference in the vacuum energy between the
two phases.
Using the expressions for the bosonic and fermionic 
sum-integrals 
Eqs.~(\ref{sumb}) and ~(\ref{sumfermi}),
we find
\bqa\nonumber
{\cal F}_{\rm QGP}&=&
-(N_c^2-1){\pi^2T^4\over45}+{1\over2}B^2
+{N_c\over(4\pi)^2}\left({\Lambda^2\over|2q_fB|}\right)^{\epsilon}
\sum_f
\left[
\left({2(q_fB)^2\over3}+m_f^4\right)\left({1\over\epsilon}+1\right)
\right.\\&&\left.
-8(q_fB)^2\zeta^{(1,0)}(-1,x_f)-2|q_fB|m_f^2\ln x_f
-2K_0^B(\beta m_f)|q_fB|T^2
\right]
+{\cal B}\;.
\eqa
Again, the ultraviolet divergences are removed by
wavefunction renormalization and by adding a vacuum counterterm.
This amounts to the substitutions
$B^2\rightarrow Z^2B^2$ and ${\cal B}\rightarrow{\cal B}+\Delta{\cal B}$,
where
\bqa
Z^2=1-N_c\sum_f{4q_f^2\over3(4\pi)^2\epsilon}\;,\hspace{1cm}
\Delta{\cal B}=-{N_c}\sum_f{m_f^4\over(4\pi)^2\epsilon}\;.
\label{repl1}
\eqa
The renormalized free energy density
in the quark-gluon plasma phase then
reduces to
\bqa\nonumber
{\cal F}_{\rm QGP}&=&
-(N_c^2-1){\pi^2T^4\over45}
+{1\over2}B^2\left(
1+N_c\sum_f{4q_f^2\over3(4\pi)^2}
\ln{\Lambda^2\over|2q_fB|}
\right)
+{N_c\over(4\pi)^2}\sum_fm_f^4\left[\ln{\Lambda^2\over m_f^2}+{3\over2}\right]
\\&&\nonumber
-{8N_c\over(4\pi)^2}\sum_f(q_fB)^2\bigg[\zeta^{(1,0)}(-1,x_f)
+{1\over4}x_f^2
-{1\over2}x_f^2\ln x_f
+{1\over2}x_f\ln x_f
-{1\over12}
\bigg]\\ &&
-{2N_c\over(4\pi)^2}\sum_fK_0^B(\beta m_f)|q_fB|T^2
+{\cal B}\;.
\eqa

\begin{figure}
\begin{center}
\setlength{\unitlength}{1mm}
\includegraphics[width=9.0cm,angle=0]{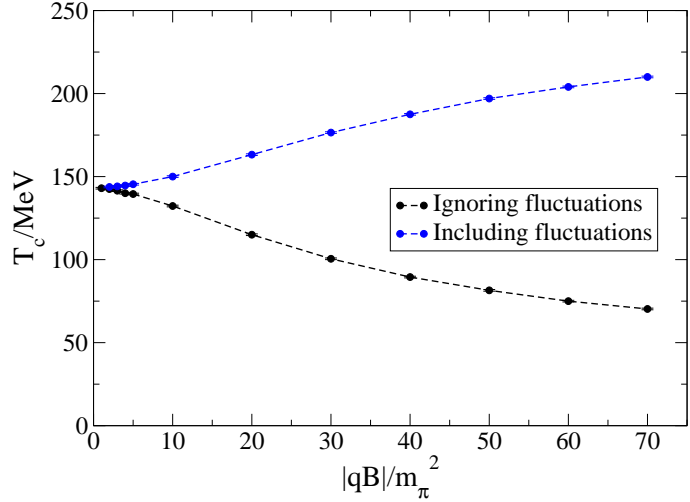}
\caption{Critical temperature as a function of $|qB|/m_{\pi}^2$
in the MIT bag model. See main text for details.}
\label{bag1}
\end{center}
\end{figure}

~\textcite{fragamit}
take a slightly different approach to the renormalization of
the MIT bag model than the one presented so far.
The divergent terms 
$(qB)^2/\epsilon$ and $(q_fB)^2/\epsilon$ in the two phases
remain after the subtraction of the vacuum energy 
at $T=B=0$.
These divergences can be removed as done above, but leaves
us with some finite terms.
They argue that the finite terms proportional to $(qB)^2$
and $(q_fB)^2$
must be
subtracted in an ad hoc fashion since the charges that generate
the magnetic field are not included in the description.
They therefore subtract all mass-independent terms
that are proportional to $(qB)^2$ or $(q_fB)^2$, which 
leads to free energies densities in the two phases
\bqa\nonumber
{\cal F}_{\rm HHG}&=&
{4(qB)^2\over(4\pi)^2}
\left[\zeta^{(1,0)}(-1,\mbox{${1\over2}$}+x)
-\zeta^{(1,0)}(-1,\mbox{${1\over2}$})
+{1\over4}x^2
-{1\over2}x^2\ln x
\right]
\\ 
&&
-{1\over2(4\pi)^2}\left[
J_0(\beta m_{\pi})T^4+2J_0^B(\beta m_{\pi})|qB|T^2
\right]\;,
\;\\ \nonumber
{\cal F}_{\rm QGP}&=&
-{N_c\over2\pi^2}\sum_f(q_fB)^2\left[\zeta^{(1,0)}(-1,x_f)
-\zeta^{(1,0)}(-1,0)
+{1\over4}x_f^2-{1\over2}x_f^2\ln x_f+{1\over2}x_f\ln x_f\right]
\\ && 
-(N_c^2-1){\pi^2T^4\over45}
-{2N_c\over(4\pi)^2}\sum_fK_0^B(\beta m_f)|q_fB|T^2
+{\cal B}
\;.
\eqa
In Fig.~\ref{bag1} we show the critical temperature $T_c$ 
for the phase transition as a function of $|qB|/m_{\pi}^2$
for $N_c=3$ and $N_f=2$.
We have used $m_{\pi}=140\,{\rm MeV}$, $m_u=m_d=5\,{\rm MeV}$
as well as $\Lambda=600\,{\rm MeV}$, and ${\cal B}=(200\,{\rm MeV})^4$.  
The black curve is with the $B=0$ vacuum
fluctuations and the red curve is
where the $B=0$ vacuum fluctuations have been subtracted.
Clearly, the figure demonstrates the importance of how one treats the
vacuum fluctuations in the model. In both cases, we have an effective
$B$-dependent bag constant, which can be easily found by absorbing
all the $T=0$ terms into ${\cal B}$.
An obvious quantity to calculate is a $B$-dependent effective 
bag constant ${\cal B}(B)$ that reproduces the critical temperature
determined by lattice simulations.

\end{widetext}

We close this section by mentioning two related 
calculations~\cite{fedorov,orla1}.
Instead of a bag constant, the pressure contains another constant term
arising from the gluonic condensate.
The energy density term in the hadrodnic phase is of the form
\bqa
{\cal E}_{\rm vac}&=&-{b\over8(4\pi)^2}\langle G^2\rangle\;,
\label{energyb}
\eqa
where $b=(33N_c-2N_f)/3$ and $G^2=(g_sG_{\mu\nu}^a)^2$.
At temperatures around the transition temperature, 
the condensate is approximately half the value at $T=0$.
Lattice calculations at zero magnetic field give the value
$\langle G^2\rangle=0.87$ GeV$^4$ and a critical temperature of $177$  MeV.
~\textcite{fedorov} showed that the critical temperarature as well as
the latent heat decrease as functions of the magnetic field $B$.
The deconfinement transition is first order as defined
by a nonzero latent heat between the two phases for 
magnetic fields smaller than$\sqrt{|qB|}\sim 600$ MeV.
The transition is a crossover for magnetic fields larger than this value.

As pointed out by~\textcite{orla1}, the masses of the pions are 
strongly dependent on the magnetic field and should be 
taken into account. Similarly, the vacuum energy density~(\ref{energyb}) 
also depends on $B$~\cite{ozaki}. 
This calls for a more systematic 
study at finite magnetic field.

\subsection{Chiral perturbation theory}
\label{subchi}
Chiral perturbation theory is an effective low-energy theory for 
QCD in the hadronic phase~\cite{weinberg,leutwyler1,gasser}.
It is a model-independent framework in the sense that it only
depends on the symmetries of QCD, the symmetry breaking pattern of
QCD in the vacuum, and the relevant degrees of freedom.
At sufficiently low energy or temperature, 
only the pseudo-Goldstone bosons are relevant
degrees of freedom, although other degrees of freedom 
can be systematically added.
In massless QCD with $N_f$ flavors,
the chiral Lagrangian has a global $SU(N_f)_L\times SU(N_f)_R$
symmetry
describing $N_f^2-1$ massless excitations. If the quarks have equal masses,
this symmetry is explicitly broken to $SU(N_f)_V$.
Explicit symmetry breaking in the chiral Lagrangian can be
systematically included by adding terms to the Lagrangian 
that respect the $SU(N_f)_V$ symmetry. 

In QCD, when one couples the quarks to the electromagnetic field,
the flavor symmetry is broken. One can no longer freely transform 
a $u$-quark into a $d$-quark  or an $s$-quark.
For massless QCD with $N_f=2$, the $SU(2)_L\times SU(2)_R$
symmetry is broken down to $U(1)_V\times U(1)_A$ by electromagnetic
interactions. The $U(1)_V$-symmetry 
corresponds to the invariance under 
a rotation of a $u$-quark by an angle $\alpha$, 
$u\rightarrow e^{i\alpha}u$ and a rotation of a $d$-quark by
the opposite angle, $d\rightarrow e^{-i\alpha}d$.
The $U_A(1)$ symmetry corresponds to the invariance under a chiral
rotation $u\rightarrow e^{i\gamma_5\alpha}u$
and $d\rightarrow e^{-i\gamma_5\alpha}d$.

Chiral perturbation theory is not an expansion in some small coupling
constant, but is an expansion in powers of momenta $p$, where a
derivative in the Lagrangian counts as one power and a quark
mass counts as two powers. Chpt
is a nonrenormalizable quantum field theory, implying 
that a calculation at a given order $n$ in momentum
$p$, requires that one adds higher-order
operators in order to cancel the divergences that arise in the calculations 
at order $n$. One needs more and more couplings
as one goes to higher loop orders, and therefore more
measurements to determine them.
However, this poses no problem; as long as one 
is content with
finite precision, a nonrenormalizable field theory has predictive power
and is as good as any other field theory.
This is is the essence of effective field theory.

In this section, we restrict ourselves to two-flavor QCD.
Chpt is then an effective theory for the three
pions and 
the effective Lagrangian can be written as a power series
\bqa
{\cal L}_{\rm eff}&=&{\cal L }^{(2)}+{\cal L }^{(4)}+...
\eqa
where the superscript indicates the order in momentum.
In Euclidean space, the leading term is given by
\begin{widetext}
\bqa
{\cal L}^{(2)}&=&
{1\over4}F^2{\rm Tr}\left[
\left(D_{\mu}U\right)^{\dagger}\left(D_{\mu}U\right)
-M^2(U+U^{\dagger})
\right]
\;,
\label{chirall}
\eqa
where $M$ and $F$ are the tree-level values of the
pion mass and pion decay constant,
respectively. Moreover
$U=e^{i\tau_i\pi_i/F}$ is a unitary $SU(2)$ matrix,
$\tau_i$ are the Pauli matrices, $\pi_i$ are the pion fields,
and $D_{\mu}$ is the covariant derivative defined by
\bqa
D_{\mu}U&=&\partial_{\mu}U
-i[Q,U]A_{\mu}
\;,
\eqa
where $Q$ is the charge matrix of the quarks, 
$Q={\rm diag}(\mbox{$2\over3$},-\mbox{$1\over3$})e$.
As explained earlier, a constant magnetic field $B$ 
explicitly breaks the global chiral symmetry
that transforms $u$ and $d$ quarks into each other, but leaves
a residual $U(1)_A$ symmetry.
Due to this reduced symmetry, there is only
one true Goldstone boson namely the neutral pion $\pi^0\equiv\pi_3$. If
the magnetic field is sufficiently strong, the charged
pions are very heavy and expected to decouple from the low-energy dynamics.
In this regime, the low-energy field theory involves a single 
massless particle.
The space-time symmetry is 
$SO(1,1)\times SO(2)$, which are Lorentz boosts in the $x_0x_3$-plane as well
as rotations in the $x_1x_2$-plane perpendicular to $B$.
We therefore need to consider separately the 
derivative operators $\partial_{\perp}=(\partial_1,\partial_2)$
and $\partial_{\parallel}=(\partial_0,\partial_3)$ and build our invariants 
from these. The effective Lagrangian for $\pi^0$
then reads
\bqa
{\cal L}_{\rm eff}
&=&{1\over4}F^{(1)}_{\perp}(\partial_{\perp}U_{\perp})
(\partial_{\perp}U_{\perp})^{\dagger}
+{1\over4}F^{(1)}_{\parallel}
(\partial_{\parallel}U_{\parallel})
(\partial_{\parallel}U_{\parallel})^{\dagger}
+...,
\label{spezial}
\eqa
where $U_{\perp}=e^{i\pi^0/F_{\perp}}$and $U_{\parallel}=e^{i\pi^0/F_{\parallel}}$. 
Note that we must 
allow for two different decay constants 
$F_{\perp}^{(1)}$ and $F_{\parallel}^{(1)}$~\cite{sado,kami}.
The Lagrangian~(\ref{spezial}) is a special case
(albeit with different notation)
of the general case with $N_u$ up quark flavors and 
$N_d$ down quark flavors considered by~\textcite{shovmir}.
In this case the symmetry is 
$SU(N_u)_L\times SU(N_u)_R\times SU(N_d)_L\times SU(N_d)_R\times U(1)_A$
which is broken down to the diagonal subgroup
$SU(N_u)_V\times SU(N_d)_V$. This gives rise to 
$N_u^2+N_d^2-1$ massless Goldstone particles.

The Lagrangian ${\cal L}^{(2)}$, Eq.~(\ref{chirall})
can be expanded in powers
of the pion fields. Through fourth order, we obtain
\bqa\nonumber
{\cal L}^{(2)}&=&-F^2M^2+
{1\over2}\left(\partial_{\mu}\pi^0\right)^2
+{1\over2}M^2\left(\pi^0\right)^2
+(\partial_{\mu}+iqA_{\mu})\pi^+(\partial_{\mu}-iqA_{\mu})\pi^-
+M^2\pi^+\pi^-
\\ &&\nonumber
-{M^2\over24F^2}\left[
(\pi^0)^2+2\pi^+\pi^-
\right]^2
+{1\over6F^2}\bigg\{
2\pi^0[\partial_{\mu}\pi^0][\partial_{\mu}(\pi^+\pi^-)]
-2\pi^+\pi^-(\partial_{\mu}\pi^0)^2
 \\&& 
-2\left[\left(\pi^0\right)^2+2\pi^+\pi^-\right]
(\partial_{\mu}\pi^+)(\partial_{\mu}\pi^-)
+[\partial_{\mu}(\pi^+\pi^-)]^2
\bigg\}
\;,
\label{trunc}
\eqa
where we have defined the complex pion fields
as $\pi_{\pm}={1\over\sqrt{2}}(\pi_1\pm i\pi_2)$
In the same manner, we can expand the Lagrangian 
${\cal L}^{(4)}$ to second order in the pion fields
\bqa\nonumber
{\cal L}^{(4)}
&=&
{1\over4}F_{\mu\nu}^2+
{2l_5\over F^2}(qF_{\mu\nu})^2\pi^+\pi^-
+{2il_6\over F^2}qF_{\mu\nu}\left[
(\partial_{\mu}\pi^-)(\partial_{\nu}\pi^+)
+iqA_{\mu}\partial_{\nu}(\pi^+\pi^-)
\right]
+(l_3+l_4){M^4\over F^2}\left(\pi^0\right)^2
\\ && 
+2(l_3+l_4){M^4\over F^2}\pi^+\pi^-
+l_4{M^2\over F^2}(\partial_{\mu}\pi^0)^2+
2l_4{M^2\over F^2}
(\partial_{\mu}+iqA_{\mu})\pi^+
(\partial_{\mu}-iqA_{\mu})\pi^-\;.
\label{l44}
\eqa
\end{widetext}
The Lagrangian ${\cal L}^{(6)}$ contains more than fifty terms for
two flavors. However, in a two-loop calculation of the pressure at finite $B$
only one term contributes, namely 
$M^2(qF_{\mu\nu})^2$~\cite{agassian00,werbos,werbos2}.

As mentioned above, the parameters $M$ and $F$ in the Lagrangian can be
interpreted as the tree-level values of the 
pion mass $m_{\pi}$
and pion decay constant $F_{\pi}$, respectively. However, these quantities
receive loop corrections and they can no longer be identified 
with the bare parameters of the Lagrangian ${\cal L}$.
The loop integrals are ultraviolet divergent and the divergences
are cancelled by the renormalization of the low-energy constants
$l_i$ ($l_i=1,2,3...$) that appear in the Lagrangian. The relation between the
bare low-energy constants $l_i$ and their renormalized counterparts
$\bar{l}_i$ is
\bqa
l_i&=&-{\gamma_i\over2(4\pi)^2}\left[{1\over\epsilon}
+\bar{l}_i
\right]\;,
\label{renbare}
\eqa
evaluated at the scale $\Lambda=M$.
The coefficients $\gamma_i$ are tabulated in the paper by~\textcite{leutwyler1}.
In the actual calculations, we present below, we need
$\gamma_3=\mbox{$1\over2$}$ and $\gamma_4=2$.
Note that our relation Eq.~(\ref{renbare})
does not involve the renormalization scale $\Lambda$ as 
in~\textcite{leutwyler1} since it is a part of the definition of 
the sum-integrals. Moreover, our expression~(\ref{trunc}) for the
truncated Lagrangian ${\cal L}^{(2)}$ differs from the expression found
in for example the papers 
by~\textcite{smilga,werbos} since they use a different 
parametrization
for the unitary matrix $U$, namely the Weinberg parametrization.
However, we obtain the same expressions for physical quantities
independent of parametrization~\cite{kapustelsen}. This simply reflects
that physical quantities are independent of the coordinate system used.

To leading order in chiral perturbation theory and second order
in the pion fields, the Lagrangian 
describes free bosons in a magnetic field. Thus a one-loop
calculation of the free energy density is the same as the one 
we did in the hadronic phase for the MIT bag model and the
renormalized result is given by Eq.~(\ref{HHG}), except that one must
add the tree-level term $-M^2F^2$ from Eq.~(\ref{trunc}), i.e.
${\cal F}_1^{\rm Chpt}=-M^2F^2+{\cal F}_{\rm HHG}$.
The vacuum energy to one-loop order at finite $B$ was first
calculated by ~\textcite{schwinger} and generalized to two loops 
by~\textcite{agassian00}.
The two-loop result for the free energy density
at finite temperature
first appeared recently~\cite{jensa,jensb}.

\subsubsection{Quark condensate, pion mass, and pion decay constant}
The zero-temperature
quark condensate at one-loop in the chiral limit was
first derived by~\textcite{smilga}
and later generalized to finite quark mass, i.e. finite $m_{\pi}$
by~\textcite{werbos}. They also generalized their result
to constant electromagnetic fields. 
The two-loop result for the chiral condensate in the chiral limit was
calculated by ~\textcite{agassian00}
and generalized to finite pion mass by~\textcite{werbos2}.
Agasian and Shushpanov also calculated the finite-temperature
quark condensate at one-loop~\cite{agarenner}, which was 
extended to two loops by~\textcite{jensa}.

Let us write the free energy density
through $n$ loops 
as ${\cal F}_n={\cal F}_n^{\rm vac}+{\cal F}_n^{T}+{\cal F}_n^{B}$,
where ${\cal F}_n^{\rm vac}$ is the contribution in the vacuum, i.e.
for $B=T=0$, ${\cal F}_n^B$ is the zero-temperature
contribution due to a finite magnetic field, and ${\cal F}^T_n$
is the finite-temperature contribution. 
The chiral condensate is given by~\cite{leutwyler3}
\bqa
\langle\bar{q}q\rangle_B&=&\langle\bar{q}q\rangle_0
\left[1-{c\over F^2}{\partial({\cal F}_n^B+{\cal F}_n^T)\over\partial m_{\pi}^2}
\right]\;,
\eqa
where $\langle\bar{q}q\rangle_0$ denotes the quark condensate at 
$T\!=B\!=\!0$, 
$c=-F^2{\partial m_{\pi}^2\over\partial m}\langle\bar{q}q\rangle_0^{-1}$.
Using 
${\cal F}^B_1+{\cal F}^T_1=-M^2F^2+{\cal F}_{\rm HHG}-{\cal F}_{\rm HHG}(B=0)$
and the fact that $c=1$ in the chiral limit~\cite{leutwyler3}, one obtains
the one-loop result for the quark condensate
\begin{widetext}
\bqa
\langle\bar{q}q\rangle_B&=&\langle\bar{q}q\rangle_0
\left[1+{1\over(4\pi)^2F^2}\left(I_B(B)
-J_1(\beta M)T^2-2J_1^B(\beta M)|qB|
\right)
\right]\;,
\label{condchpt}
\eqa
\end{widetext}
where the function $I_B(M)$ is defined by 
\bqa \nonumber
I_B(M)&=&
M^2\ln{M^2\over2|qB|}
-M^2
\\ &&
-2|qB|\zeta^{(1,0)}(0,\mbox{$1\over2$}+x)\;.
\eqa
Using the first term of the small-$x$ expansion~(\ref{small4}), one finds
that at $T=0$,
the condensate grows linearly with the field in the chiral 
limit~\cite{smilga},
$\langle\bar{q}q\rangle_B=\langle\bar{q}q\rangle_0
\left[1+{|qB|\ln2\over(4\pi)^2F^2}\right]$.

\begin{figure}
\begin{center}
\setlength{\unitlength}{1mm}
\includegraphics[width=7.0cm]{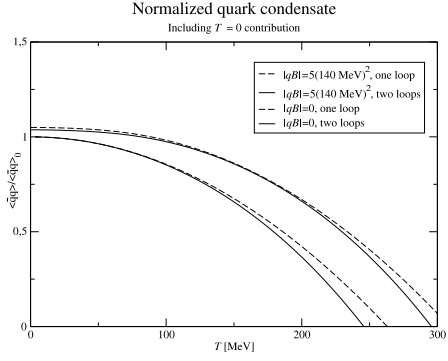}
\caption{Normalized quark condensate at one and two loops
in the chiral limit as a function of $T$ for 
$B=0$ and $|qB|=5(140$ MeV)$^2$. Figure taken from~\textcite{jensa}.}
\label{quark}
\end{center}
\end{figure}

In Fig.~\ref{quark}, we show the one- and two-loop results for the
normalized quark condensate $\langle\bar{q}q\rangle_B/\langle\bar{q}q\rangle_0$
in the chiral limit
as a function of $T$ for $|qB|=5(140$ MeV)$^2$
and for $B=0$ for comparison.
The vacuum contribution has been included and amounts to an increase
of the chiral condensate of about 5\% in this case.
The effects are large due to the very strong magnetic field, and for 
weaker fields, the difference between the two sets of curves is smaller.
The curves suggest that the critical temperature for the chiral transition
is increasing as a function of $B$, but since Chpt breaks down at
perhaps $T\sim$150 MeV, one should be careful making quantitative statements.
For temperatures where Chpt can be trusted the curves suggest
that perturbation theory in a magnetic field converges at least
as well as for $B=0$.

\begin{figure}
\begin{center}
\setlength{\unitlength}{1mm}
\includegraphics[width=3cm]{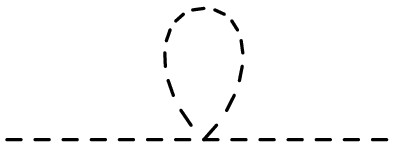}
\hspace{2mm}
\includegraphics[width=3cm]{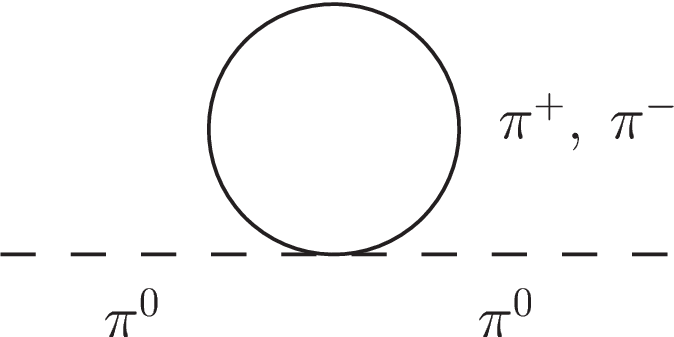}
\caption{One-loop graphs contributing to the self-energy of the neutral pion.}
\label{selfenergy}
\end{center}
\end{figure}

We next consider the correction to the neutral pion mass $m_{\pi^0}$
due to a magnetic field. The Feynman diagrams contributing to 
the one-loop self-energy $\Pi(P_0,{\bf p})$ are 
shown in Fig.~\ref{selfenergy}.
The inverse propagator can be written as
\bqa
\Gamma^{(2)}(P_0,{\bf p})&=&P^2+M^2+\Pi(P_0,{\bf p})\;,
\label{n2def}
\eqa
where the the one-loop self-energy is given by 
the expression
\begin{widetext}
\bqa\nonumber
\Pi(P_0,{\bf p})&=&
-{2\over3F^2}
P^2\sumint_K^B{1\over K_0^2+k_z^2+M_B^2}
+{1\over6F^2}M^2\left[
2\sumint_K^B{1\over K_0^2+k_z^2+M_B^2}
-3\sumint_P{1\over K^2+M^2}
\right]
\\ 
&&+2l_4P^2{M^2\over F^2}
+2(l_3+l_4){M^4\over F^2}
\;,
\eqa
where the terms in the second line are counterterms coming from 
${\cal L}^{(4)}$.
Collecting all terms proportional to $P^2$, we redefine the field
$\pi^0$
such that the coefficient of $P^2$ in Eq.~(\ref{n2def}) equals 
unity~\cite{villa}.
This yields
\bqa
\Gamma^{(2)}(P_0,{\bf p})=P^2+m_{\pi^0}^2\;,
\eqa
where the physical pion mass squared $m_{\pi^0}^2$ is
\bqa\nonumber
m^2_{\pi^0}&=&
M^2+{M^2\over F^2}\sumint_P^B{1\over P_0^2+p_z^2+M_B^2}
-{1\over2}{M^2\over F^2}\sumint_P{1\over P^2+M^2}+2l_3{M^4\over F^2}
\\
&=&M^2
\left[
1-{1\over(4\pi)^2F^2}\left(I_B(M)
+{1\over2}J_1(\beta M)T^2-J_1^B(\beta M)|qB|\right)\right]\;.
\label{chmpi0}
\eqa

This result was first obtained 
at zero temperature by~\textcite{smilga} and later generalized
to finite temperature by~\textcite{agassian00}.
We note that $m_{\pi^0}$ vanishes in the chiral limit $M\rightarrow0$,
as it must since the neutral pion is a Goldstone boson.

We next consider the pion decay constant for the neutral pion, $F_{\pi^0}$.
The components of the axial current ${\cal A}_{\mu}^0$ are given by 
\bqa
{\cal A}_{\mu}^0
&=&-F\partial_{\mu}\pi^0
+{2\over3F}\left[
2\pi^+\pi^-\partial_{\mu}\pi^0
-\pi^0\partial_{\mu}(\pi^+\pi^-)
-2{M^2}l_4\partial_{\mu}\pi^0
\right]
\;.
\eqa
\end{widetext}
In order to calculate the pion decay constant, we need to
evaluate the matrix element $F_{\pi^0}=\langle0|{\cal A}_{\mu}^0|\pi^0\rangle$.
In a consistent one-loop calculation, one needs to take into account
wavefunction renormalization of the tree-level term
$-F\partial_{\mu}\pi^0$.\footnote{This wavefunction renormalization 
counterterm is the same we used to obtain Eq.~(\ref{chmpi0}).}
Calculating the matrix element, one finds
\bqa\nonumber
F_{\pi^0}&=&
F\left[1-{1\over F^2}\sumint_P^B{1\over P_0^2+p_z^2+M_B^2}
+{M^2\over F^2}l_4\right]\;.
\\ &&
\eqa
After renormalization, we find
\bqa\nonumber
F_{\pi^0}&=&F\left[1
+{1\over(4\pi)^2F^2}
\left(
I_B(M)-\right.\right.\\ &&\left.\left.
J_1^B(\beta M)|qB|\right)\right]\;, 
\label{fpipm0}
\eqa
Using Eqs.~(\ref{chmpi0}) and~(\ref{fpipm0}), we see that 
\bqa
m^2_{\pi^0}F_{\pi^0}^2&=&m\langle\bar{q}q\rangle_B\;,
\eqa
which is the Oakes-Gell-Mann-Renner relation in a magnetic field.
This relation was first shown by~\textcite{agarenner}.

\subsection{Nambu-Jona-Lasinio model}
The Nambu-Jona-Lasinio (NJL) model was originally proposed as a theory for
interacting nucleons and pions in the 1960s before the discovery of 
quarks~\cite{njl1,njl2}.
After the discovery of quarks and the formulation of QCD as the theory
of the strong interactions, the fermion fields in the Lagrangian
were reinterpreted as quark fields and the NJL model as an effective
low-energy model for QCD.
We will make a few remarks on the NJL model below, but for
a detailed discussion of its properties, 
we refer to the reviews by~\textcite{klevansky1} and~\textcite{buballa1}.
In the NJL model, one-gluon exchange between the quarks is replaced
by local four-point quark interactions. Thus there are no gauge fields
in the model and the local $SU(N_c)$ gauge 
symmetry of QCD is replaced by a global $SU(N_c)$ symmetry. 
As a result, two of the most prominent features
of QCD - asymptotic freedom and confinement - are lost.
The latter can be seen by the fact that the polarization function 
for pions, $\Pi_{\rm M}(p^2)$,
develops an imaginary part for $p^2>4M^2$
where $M$ is the quark mass and the pions become
unstable against decay to their constitutent parts~\cite{buballa1}.

Another important aspect of QCD, namely that of chiral symmetry breaking
in the vacuum, is taken into account by the NJL model.
The spontaneous breaking of chiral symmetry guarantees via the
Goldstone theorem the appearance of massless, or light if
chiral symmetry is explicitly broken, bosonic excitations in the spectrum.
For $N_f=2$, these
particles are the (light) pions and the explanation of the 
low pion mass was a success of the NJL model.
We note in passing that in Lorentz invariant theories, the number
of Goldstone bosons equals the number of broken generators.
When Lorentz invariance is broken, for example at finite density,
the number of of massless excitations maybe strictly smaller than the 
number of broken generators~\cite{chadha,tomasgb} and some of them
have a quadratic dispersion relation $E\sim p^2$ instead of a linear one.

However, chiral symmetry breaking is not seen at any finite order
in perturbation theory and one needs to sum an infinite number
of a certain class of diagrams to obtain a nonzero chiral condensate. 
This is done by introducing a set of collective bosonic or 
auxiliary fields such that the Lagrangian becomes bilinear in the 
quark fields. One can then integrate out exactly the fermions in the
path integral and afterwards expand the resulting functional determinant
in powers of the collective fields and their derivatives.
This expansion is an expansion in $1/N_c$. 
To leading order in $1/N_c$, i.e. in the large-$N_c$ limit,
this gives rise to a gap equation for the chiral condensate.
This is also referred to as the mean-field approximation, since
the collective fields are replaced by their expectation values.
At next-to-leading order, this expansion generates kinetic terms
for the bosonic fields and so they become propagating
quantum fields.
At next-to-next-to-leading order, the expansion 
generates interaction terms among the 
bosons~\cite{eguchi,klevansky1,boom1}.

The NJL model is nonrenormalizable in the sense that 
loop diagrams generate divergences that cannot be cancelled by local
counterterms of the same type as those appearing in the 
original Lagrangian. One therefore needs to add new operators
to cancel the these divergences.
The operators that are induced this way are
suppressed by some power of some (large) mass scale $\Lambda$.
This mass scale signals new physics that is not captured by the model,
but as long as we stay well below this scale, this is not a problem.
At finite precision, only a finite number of operators contribute
to a given physical quantity.
One way of dealing with the ultraviolet divergences in the momentum integrals
is by cutting them off using a sharp three-dimensional cutoff $\Lambda$
or a smooth ultraviolet cutoff. 
In the case of a sharp cutoff,
the momentum scale $\Lambda$ can be interpreted as an upper scale
below which the model or theory is valid.
A soft cutoff is often referred to as a 
form factor and denoted by $F(p)$, where $p$ is the three-momentum.
A form factor that mimics asymptotic freedom is
\bqa
F(p)&=&{\Lambda^2\over\Lambda^2+p^2}\;,
\eqa 
where $\Lambda$ is a mass scale. The function $F(p)$
guarantees that loop integrals converge for large momentum $p$.
A three-dimensional 
cutoff breaks Lorentz invariance, but this may be less severe 
at finite temperature, where it is broken anyway.
There are other form factors that are tailored to the problem of
a magnetic background and we will briefly discuss them below.

The Minkowski space Lagrangian of the NJL model with $N_f=2$ can be written as
\bqa
{\cal L}&=&{\cal L}_0+{\cal L}_{\bar{q}q}+{\cal L}_{\rm det}\;,
\label{njll}
\eqa
where the various terms are
\bqa
{\cal L}_0&=&i\bar{\psi}/\!\!\!\!D\psi-m_0\bar{\psi}\psi\;,
\\ \nonumber
{\cal L}_{\bar{q}q}&=&G_1\left[
(\bar{\psi}\psi)^2+(\bar{\psi}{\boldsymbol\tau}\psi)^2
+(\bar{\psi}i\gamma_5\psi)^2
\right. \\ &&\left.
+(\bar{\psi}i\gamma_5{\boldsymbol\tau}\psi)^2\right]\;,
\\ \nonumber
{\cal L}_{\rm det}&=&G_2
\left[(\bar{\psi}\psi)^2-(\bar{\psi}{\boldsymbol\tau}\psi)^2
-(\bar{\psi}i\gamma_5\psi)^2
\right. \\ &&\left.
+(\bar{\psi}i\gamma_5{\boldsymbol\tau}\psi)^2\right]
\;,
\eqa
where ${\boldsymbol\tau}$ are the Pauli spin matrices,
and $/\!\!\!\!D=\gamma^{\mu}D_{\mu}$. 
$D_{\mu}=\partial_{\mu}-iQA_{\mu}$ 
is the covariant derivative where $Q={\rm diag}({2\over3},-{1\over3})e$
is the charge matrix.
$G_1$ and $G_2$ are coupling constants and
$m_0$ is the mass matrix, $m_0={\rm diag}(m_u,m_d)$.
As is normally done in the literature, we use $m_u=m_d$.
For $N_f=2$, $\psi$ is an isospin doublet,
\bqa
\psi&=&
\left(\begin{array}{c}
u\\
d\\
\end{array}\right)\;.
\label{d0}
\eqa
The terms ${\cal L}_0+{\cal L}_{\bar{q}q}$ are
invariant under the global symmetries
$SU(N_c)\times SU(2)_L\times SU(2)_R\times U(1)_B\times U(1)_A$
in the chiral limit and 
$SU(N_c)\times SU(2)_{L+R}\times U(1)_B$ at the physical
point.\footnote{This is for the case $m_u=m_d$. 
If $m_u\neq m_d$, the symmetry $SU(2)_{L+R}$
reduces to $U(1)_{I_3}$.}
These are the symmetries of QCD, except that the
color symmetry is global and not local.
The term ${\cal L}_{\rm det}$ breaks the $U(1)_A$ symmetry while
preserving the others.
This term is 't Hooft's instanton-induced interaction and mimics
the breaking of the axial $U(1)_A$ symmetry in the QCD 
vacuum~\cite{tuft}. 
It is necessary to explain the relatively large mass
of the $\eta$ particle.
We note that such a term is a six-quark interaction in
three-flavor QCD.

We next consider the two nonzero quark condensates
$\langle\bar{u}u\rangle$ and $\langle\bar{d}d\rangle$. These
can be expressed in terms of 
$\langle\bar{\psi}\psi\rangle$ and $\langle\bar{\psi}\tau_3\psi\rangle$
as
$\langle\bar{u}u\rangle\pm\langle\bar{d}d\rangle$.
and a nonzero $\langle\bar{\psi}\tau_3\psi\rangle$ implies that
$\langle\bar{u}u\rangle\neq\langle\bar{d}d\rangle$.
Hence we can write
$(\bar{\psi}\psi)^2=
(\bar{\psi}\psi-\langle\bar{\psi}\psi\rangle)^2
+2\langle\bar{\psi}\psi\rangle\bar{\psi}\psi
-\langle\bar{\psi}\psi\rangle^2$
and
$(\bar{\psi}\tau_3\psi)^2=
(\bar{\psi}\tau_3\psi-\langle\bar{\psi}\tau_3\psi\rangle)^2
+2\langle\bar{\psi}\tau_3\psi\rangle\bar{\psi}\tau_3\psi
-\langle\bar{\psi}\tau_3\psi\rangle^2$.
In the mean-field approximation, we linearize the 
interaction terms in presence of the two condensates
$\langle\bar{\psi}\psi\rangle$ and $\langle\bar{\psi}\tau_3\psi\rangle$, 
i.e. we neglect the fluctuations around the mean field.
Hence, we approximate the quartic terms by
\bqa
\label{appp1}
(\bar{\psi}\psi)^2
&\approx&
2\langle\bar{\psi}\psi\rangle\bar{\psi}\psi
-\langle\bar{\psi}\psi\rangle^2\;,
\\ 
(\bar{\psi}\tau_3\psi)^2
&\approx&
2\langle\bar{\psi}\tau_3\psi\rangle\bar{\psi}\tau_3\psi
-\langle\bar{\psi}\tau_3\psi\rangle^2\;.
\label{appp2}
\eqa
Substituting Eqs.~(\ref{appp1}) and~(\ref{appp2})
into Eq.~(\ref{njll}), we obtain 
the Lagrangian which is bilinear in the fermion fields:
\bqa\nonumber
{\cal L}_{\rm bilinear}&=&
-{(M_0-m_0)^2\over4G_0}-{M_3^2\over4(1-2c)G_0}
\\ &&
+\bar{\psi}\left[/\!\!\!\!D_{\mu}-{M}\right]\psi\;,
\label{bilin}
\eqa
where ${M}=M_0+\tau_3M_3$ and we have introduced
\bqa
M_0&=&m_0-2G_0\langle\bar{\psi}\psi\rangle\;,
\\
M_3&=&-2(1-2c)G_0\langle\bar{\psi}\psi\rangle\;,
\\
G_1&=&(1-c)G_0\;,\hspace{1cm}
\\
G_2&=&cG_0\;.
\eqa
The parameter $c$ controls the instanton interaction or 
the amount of explicit breaking of the $U(1)_A$ symmetry.
Beyond the mean-field approximation, 
the number of scalars and pseudo-scalars depends on $c$;
For $c={1\over2}$, only the sigma and the pions are present, while for
all other values, also the $\eta$ and the ${\bf a}$ are in the spectrum.
The constituent quark masses for the $u$ and $d$ quarks can be 
expressed as $M_u=M_0+M_3$ and $M_d=M_0-M_3$. Generally these constituent
quark masses are different, only for 
$G_1=G_2$ are they identical. This corresponds to $c={1\over2}$.
The fact that $\langle\bar{u}u\rangle$ generally is
different from $\langle\bar{d}d\rangle$ should come as no surprise as
the electric charge of the $u$-quark is different from that of the
$d$-quark.

The Lagrangian~(\ref{bilin})
is bilinear in the quark fields and we can integrate
over them exactly. 
The vacuum energy is evaluated using dimensional
regularization with zeta-function regularization in the usual way
and whose $M$-independent (divergent and finite) terms are 
omitted~\footnote{This corresponds to ignoring 
wavefunction renormalization of the the tree-level term 
${1\over2}B^2$ in the free energy density.}.
The remaining divergences can be isolated by subtracting and adding
the vacuum energy for $B=0$. The difference between the two
vacuum energies is finite, while the subtracted vacuum energy is
evaluated using a hard three-dimensional cutoff $\Lambda$.
This yields the free energy 
density~\cite{mened,boom2}
\begin{widetext}
\bqa\nonumber
{\cal F}_{0+1}&=&
{(M_0-m)^2\over4G_0}+{M_3^2\over4(1-2c)G_0}
+{N_c\over8\pi^2}\sum_f\left[
M_f^4\ln\left({\Lambda\over M_f}+\sqrt{1+{\Lambda^2\over M_f^2}}
\,\,\,\right)
\right.
\\ &&\nonumber\left.
-M_f\Lambda\left(M_f^2+2\Lambda^2\right)\sqrt{1+{\Lambda^2\over M_f^2}}
\,\,\,\right]
-{8N_c\over(4\pi)^2}\sum_f(q_fB)^2\left[\zeta^{(1,0)}(-1,x_f)
\right.\\ &&\left.
+{1\over4}x_f^2-{1\over2}x_f^2\ln x_f+{1\over2}x_f\ln x_f\right]
-{2N_cT^2\over(4\pi)^2}\sum_fK_0^B(\beta M_f)|q_fB|\;,
\label{njleffpot1}
\eqa
\end{widetext}
where 
$x_f=M_f/|2q_fB|$. A similar expression can be found in the paper
by~\textcite{ebert20}, where a four-dimensional cutoff is used, 
cf. Eq.~(\ref{4dcut}).
\subsubsection{Quark condensates}
The calculations discussed in this subsection were 
presented in the paper by~\textcite{boom2}. 
In these calculations, they used 
$m_0=6$ MeV, $\Lambda=590$ MeV, and $G_0\Lambda^2=2.435$.
These values lead to a pion mass of $140.2$ MeV, a pion decay constant
of $92.6$ MeV, and a quark condensate 
$\langle\bar{u}u\rangle=\langle\bar{d}d\rangle=(-241.3{\rm\,\,MeV})^3$,
all in the vacuum.

\begin{figure}
\begin{center}
\setlength{\unitlength}{1mm}
\includegraphics[width=7.0cm]{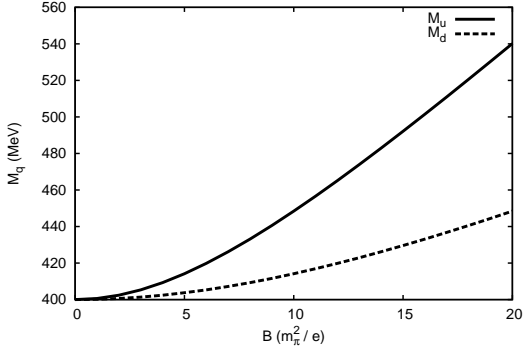}
\caption{Constituent quark masses $M_u$
and $M_d$ as functions of $B$ measured in units of $m_{\pi}^2/e$ for
$c=0$. Figure taken from~\textcite{boom2}.}
\label{boomud}
\end{center}
\end{figure}

The values of $M_u$ and $M_d$ are obtained by solving the gap equations
\bqa
{\partial {\cal F}_{0+1}\over\partial M_f}&=&0\;,
\hspace{1cm}f=u,d
\;.
\eqa
In Fig.~\ref{boomud}, the constituent quark masses $M_u$
and $M_d$ are shown 
as a function of $B$ measured in units of $m_{\pi}^2/e$ for
$c=0$, i.e. with the $U(1)_A$ symmetry intact.
Notice that $M_u=M_d$ for $B=0$, while they split for finite magnetic field
and the splitting increases as $B$ grows.
A nonzero $c$ will bring the masses closer together and 
at $c=\mbox{$1\over2$}$ they are equal.

\begin{figure}
\begin{center}
\setlength{\unitlength}{1mm}
\includegraphics[width=7.0cm]{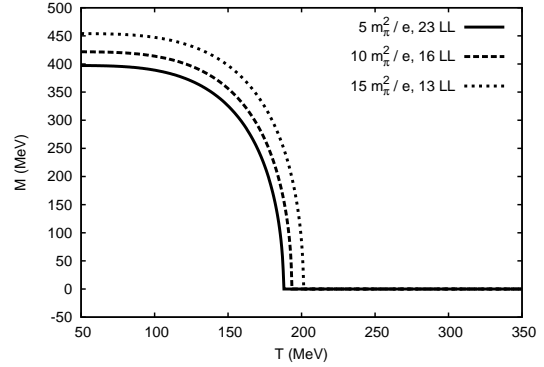}
\caption{Constitutent quark mass $M$ as a function of $T$ 
for three different values of the magnetic field and in the chiral limit.
Figure taken from~\textcite{boom2}.}
\label{ffas2}
\end{center}
\end{figure}

In Fig.~\ref{ffas2}, the constitutent quark mass 
$M_u=M_d=M$ is shown as a function of $T$
for three different values of the magnetic field.
The results are in the chiral limit and 
for $c={1\over2}$.
The transition is second order with mean-field critical exponents
for all values of the magnetic field~\cite{inagaki,boom2}.
The order of the phase transition in various approximations
will be discussed further in Secs.~\ref{qmsec} and~\ref{funxio}.
The number $x$ in $x$ LL is the number of Landau levels one must include
such that the error is less than 1\%. The stronger the magnetic field,
the fewer Landau levels are needed to be included in the sum in order
to obtain a certain accuracy. The reason is that the effective
mass of the fermions increases with the magnetic field and that 
more Landau levels are effective Boltzmann suppressed.

\begin{figure}
\begin{center}
\setlength{\unitlength}{1mm}
\includegraphics[width=7.0cm]{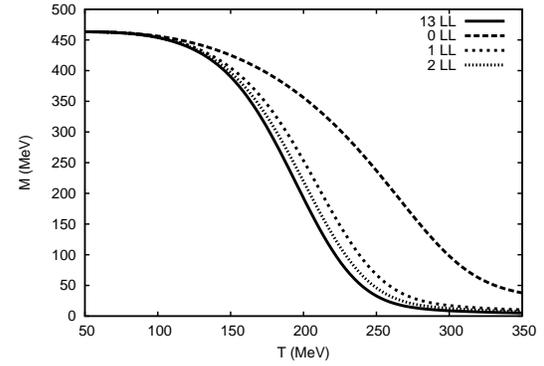}
\caption{
Constitutent quark mass $M$ as a function of $T$ 
for four different values of the magnetic field and at the physical point.
Figure taken from~\textcite{boom2}.}
\label{ffas1}
\end{center}
\end{figure}

In Fig.~\ref{ffas1}, the constitutent quark mass 
$M_u=M_d=M$ for $c={1\over2}$ and at the physical point
is shown as a function of $T$ for four different values of the magnetic field.
The constituent quark mass is a strictly positive continuous function of $T$
and hence the transition is a crossover.

\subsubsection{Other condensates}
So far we have discussed the quark condensates $\langle\bar{u}u\rangle$
and $\langle\bar{d}d\rangle$ as functions of the magnetic field.
However, due to the external magnetic field, the symmetry of the
system is reduced and other nonzero condensates are possible.
\textcite{ferrer14} considered a one-flavor NJL model
in a constant external magnetic field. A constant magnetic field
breaks Lorentz invariance down to $SO(1,1)\times SO(2)$, where the latter
latter corresponds to rotations around the 
axis in the direction of the magnetic field $B$.
The standard interaction term they consider is
\bqa
{\cal L}_{\rm int}^{(1)}&=&{G\over2}
\left[
(\bar{\psi}\psi)^2+(\bar{\psi}i\gamma^5\psi)^2
\right]\;,
\label{int1b}
\eqa
where $G$ is the usual coupling in the NJL model
The new interaction 
term that respects chiral symmetry and is invariant
under $SO(1,1)\times SO(2)$ is
\bqa
{\cal L}_{\rm int}^{(2)}&=&{G^{\prime}\over2}
\left[
(\bar{\psi}\Sigma^3\psi)^2+(\bar{\psi}i\gamma^5\Sigma^3\psi)^2
\right]\;,
\label{int2b}
\eqa
where $\Sigma^3={i\over2}[\gamma^1,\gamma^2]$
is the spin operator along the field direction
and $G^{\prime}$ is a new coupling constant.
The interaction terms in Eqs.~(\ref{int1b})--(\ref{int2b})
can be derived from one-gluon exchange in QCD using Fierz identities.
The value of the coupling constant $G^{\prime}$ 
is unknown, but vanishes in the limit $B\rightarrow0$.
The reduced symmetry gives rise to a new nonzero condensate
\bqa
\xi&=&\langle\bar{\psi}\gamma^1\gamma^2\psi\rangle\;,
\eqa
in addition to the chiral condensate $\sigma=\langle\bar{\psi}\psi\rangle$.
Calculating the thermodynamic potential
$\Omega$ in the mean-field approximation and
using the equations of motion ${\partial\Omega\over\partial\sigma}=0$
and ${\partial\Omega\over\partial\xi}=0$, they show that
\bqa
\xi&=&{G^{\prime}\over G}\sigma\;.
\label{propcond}
\eqa
Thus the spin condensate vanishes in zero magnetic field as expected and
for nonzero magnetic field it is proportional to the quark condensate.
Due to Eq.~(\ref{propcond}), the two condensates evaporate at the same
critical temperature. The same behavior has been found on the 
lattice~\cite{xibali}, where $\xi$ and $\sigma$
drop to zero at around $T=160$ MeV.

The spectrum of fermionic excitations was calculated 
by~\textcite{ferrer14} and reads
\bqa
E_0^2&=&p_z^3+(\sigma+\xi)^2\;,n=0\;,
\\ 
\nonumber
E_n^2&=&p_z^3+\sigma^2+(\sqrt{2|q_fB|n}\pm\xi)^2
\;,n\geq1\;,
\\ &&
\eqa
where $\pm$ correspond to the positive and negative spin projections, 
respectively. 
Here we notice that there is a Zeeman splitting in the spectrum for $n\geq1$
but not so for $n=0$. 
The interpretion of the term involving
$\xi$ is that it arises from an anomalous magnetic moment of the 
quarks and antiquarks.

The critical temperature $T_c$ for the system changes due to the
existence of the condensate $\xi$. If the magnetic field is sufficiently
strong that all the particles are in the lowest Landau level (LLL), one can 
calculate the critical temperature analytically.
In this case, the dynamical mass of the quarks in the LLL is
given by $M_{\xi}=\sigma+\xi$ and the critical temperature is proportional
to $M_{\xi}$. A calculation by~\textcite{ferrer14} shows that
\bqa
T_c&\approx&0.8M_{\xi}\;.
\eqa
Thus the temperature increases linearly with $M_{\xi}$ and is governed
by the coupling $G^{\prime}$.

We close this section with a brief discussion of a possible new phase
in the QCD vacuum at very strong magnetic fields.
In thise phase, charged $\rho^{\pm}$ mesons condense and as a result
the vacuum behaves as a superconductor. The idea goes back to 
~\textcite{nielsen,olesen1,olesen2} 
who showed that the $W^{\pm}$ condense in a sufficiently strong
magnetic field: the energy of a $W^{\pm}$ boson becomes purely imagniary
signalling an instabilty of the electroweak vacuum and the 
formation of a condensate.
The dispersion relation for a charged $\rho$ meson in 
a magnetic field is
\bqa
E_k^2&=&m_{\rho}^2+p_z^2+|qB|(2k+1-2s)\;,
\eqa
where $s=\pm1$. For a particle in the lowest Landau level 
with zero longitudinal momentum $p_z$, the energy becomes
purely imaginary when the magnetic field exceeds 
$B_{c}=m_{\rho}^2/q$, with $m_{\rho}=775$ MeV.
This suggests that the QCD vacuum is unstable
against condensation of charged $\rho$ 
mesons~\cite{cherno1,cherno2,callebaut}.
The condensate breaks a $U(1)$ symmetry, 
however this is not in conflict with the Vafa-Witten theorem as no massless 
Nambu-Goldstone bosons appear in the spectrum~\cite{cherno3,yama2}.

\subsection{Quark-meson model}
\label{qmsec}
Introducing the collective fields and integrating over the quark
fields in the NJL model, leads to a fermion determinant in the expression for
the effective action. This functional determinant is a function
of the background fields $\sigma$ and $\boldsymbol\pi$.
As explained earlier, the mean-field approximation consists of 
setting $\sigma$ equal to its expectation value and ignore
fluctuations of the fields $\sigma$ and $\boldsymbol \pi$.
Expanding the fluctuation determinant around the expectation
value of the $\sigma $ field, one generates kinetic terms for the
mesons as well as interaction terms. The terms that are generated
are in principle all those that are 
consistent with the symmetries of the NJL model.
These terms can be organized according to the powers of the fields
and their derivatives. If one truncates the series at
second order in derivatives and fourth order in the fields, 
we are effectively left with a quark-meson model whose coupling
constants depend on the parameters of the NJL model
and some one-loop fermionic integrals.

The Euclidean Lagrangian of the two-flavor quark-meson
model can be written as 
\begin{widetext}
\bqa\nonumber
{\cal L}&=&
\bar{\psi}\bigg[
\gamma_{\mu}D_{\mu}
+g(\sigma-i\gamma_5{\boldsymbol \tau}\cdot{\boldsymbol \pi})\bigg]\psi
\label{lagra}
+{1\over2}\bigg[
(\partial_{\mu}\sigma)^2
+({\partial}_{\mu}{\boldsymbol \pi})^2
\bigg]
+{1\over2}m^2\bigg[\sigma^2+{\boldsymbol \pi}^2\bigg]
+{\lambda\over24}\bigg[
\sigma^2+{\boldsymbol \pi}^2\bigg]^2
-h\sigma\;.
\\ &&
\label{ggglag}
\eqa
\end{widetext}
This Lagrangian has an $O(4)$ symmetry for $h=0$, which is explicitly
broken to $O(3)$ for nonzero $h$.\footnote{In addition to a
global $SU(N_c)$ symmetry.} This term gives rise to nonzero
pion masses after spontaneous symmetry breaking.
However, once we couple the quark-meson model to an Abelian gauge field,
the model is only $O(2)\times O(2)$ 
or $U(1)_V\times U(1)_A$ invariant.
At the quark level, this was explained in Sec.~\ref{subchi}.
At the mesonic level, the vector and axial
phase transformations are
\bqa
\Delta\rightarrow e^{2i\alpha}\Delta 
\;,
\hspace{0.5cm}
v\rightarrow v\;,
\\
\Delta\rightarrow\Delta\;,
\hspace{0.5cm}
v\rightarrow e^{2i\gamma_5\alpha}v\;,
\eqa
where we have introduced
$\Delta={1\over\sqrt{2}}(\pi_1+i\pi_2)=\pi^+$ and
$v={1\over\sqrt{2}}(\sigma+i\gamma_5\pi_0)$.
This implies that we have two invariants
$\sigma^2+\pi_0^2$ and $\pi^+\pi^-$ 
instead of $\sigma^2+{\boldsymbol \pi}^2$. We therefore have two mass
parameters $m^2_1$ and $m_2^2$ instead of a single mass parameter $m^2$,
and three coupling constants $\lambda_1$, $\lambda_2$, and $\lambda_3$
instead of a single coupling $\lambda$.
Finally, the Yukawa interaction term splits into
the two terms $g_1\bar{\psi}(\sigma -i\gamma_5\tau_3\pi_3)\psi$
and $-g_2\bar{\psi}i\gamma_5(\tau_1\pi_1+\tau_2\pi_2)\psi$.
These couplings are in principle functions of the magnetic field $B$
but we do not know their $B$-dependence, only that their values
are identical for $B=0$.
As is commonly done in the literature, we therefore set
all the couplings and masses equal and equal
to their values in the vacuum. 

At one loop, the effective potential receives contributions from 
the bosonic as well as the fermionic fields via the functional determinants.
A common approximation in the QM model is to neglect the
quantum and thermal fluctuations of the mesons~\cite{scavenius,fragi}.
We make this approximation in this section and discuss the
inclusion of mesonic fluctuations in the section on the 
functional renormalization group.

We first shift the sigma field and write it as a sum
of a classical background field
$\phi$ and a quantum field $\tilde{\sigma}$
\bqa
\sigma&\rightarrow&\phi+\tilde{\sigma}\;.
\eqa
The tree-level potential then becomes
\bqa
\label{tree}
{\cal F}_0&=&{1\over2}B^2+{1\over2}m^2\phi^2+{\lambda\over24}\phi^4-h\phi\;.
\eqa
The tree-level masses of the sigma, the 
pions, and the quark are (before coupling to $B$)
\bqa
\label{msigmi}
m_{\sigma}^2&=&m^2+{\lambda\over2}\phi^2\;,
\\
m_{\pi}^2&=&m^2+{\lambda\over6}\phi^2\;,
\\
m_q&=&g\phi\;.
\eqa
The pion mass satisfies $h=\phi m_{\pi}^2$ at the 
minimum of the tree-level potential 
and therefore vanishes for $h=0$,
in agreement with the Goldstone theorem.
The one-loop potential then becomes
\begin{widetext}
\bqa\nonumber
{\cal F}_{0+1}&=&{1\over2}B^2
+{1\over2}m^2\phi^2+{\lambda\over24}\phi^4-h\phi
-{1\over4(4\pi)^2}\left({\Lambda^2\over m_{\sigma}^2}\right)^{\epsilon}
\left[\left({1\over\epsilon}+{3\over2}\right)m^4_{\sigma}
+2J_0(\beta m_{\sigma})
\right]
\\ && \nonumber
-{1\over4(4\pi)^2}\left({\Lambda^2\over m_{\pi^0}^2}\right)^{\epsilon}
\left[\left({1\over\epsilon}+{3\over2}\right)m^4_{\pi^0}
+2J_0(\beta m_{\pi^0})\right]
+{1\over2(4\pi)^2}\left({\Lambda^2\over|2qB|}\right)^{\epsilon}
\left[\left({(qB)^2\over3}-m_{\pi^0}^4\right)
\left({1\over\epsilon}+1\right)
\right.\\ &&\left. \nonumber
+8(qB)^2\zeta^{(1,0)}(-1+\mbox{$1\over2$}+x)
-2J_0^B(\beta m_{\pi^0})\right]
+{N_c\over(4\pi)^2}
\sum_f
\left({\Lambda^2\over2|q_fB|}\right)^{\epsilon}
\bigg[
\left({2(q_fB)^2\over3}+m^4_{q}\right)\left({1\over\epsilon}+1\right)
\\&&
-8(q_fB)^2\zeta^{(1,0)}(-1,x_f)
-2|q_fB|m_q^2\ln x_f
-2K_0^B(\beta m_q)|q_fB|T^2
+{\cal O}(\epsilon)
\bigg]\;.
\eqa
The $B$-dependent divergence is removed in the usual way by
making the replacement $B^2\rightarrow Z^2B^2$, where 
\bqa
Z^2&=&\left[1-{q^2\over3(4\pi)^2\epsilon}
-N_c\sum_f{4q_f^2\over3(4\pi)^2\epsilon}\right]
\eqa
The other divergences are removed by making the replacement
$m^2\rightarrow m^2+\Delta m^2$,
$\lambda\rightarrow\lambda+\Delta\lambda$,
and adding a vacuum energy counterterm $\Delta{\cal E}_0$,
where 
\bqa
\Delta m^2={\lambda m^2\over(4\pi)^2\epsilon}\;,\hspace{1cm}
\Delta\lambda={\lambda^2\over8\pi^2}-
{3N_cN_fg^4\over2\pi^2\epsilon}\;.\hspace{1cm}
\Delta{\cal E}={m^4\over(4\pi)^2\epsilon}\;.
\eqa
The renormalized one-loop effective potential becomes
\bqa\nonumber
{\cal F}_{0+1}&=&
{1\over2}B^2\left[
1+{q^2\over3(4\pi)^2}\ln{\Lambda^2\over|2qB|}
+N_c\sum_f{4q_f^2\over3(4\pi)^2}\ln{\Lambda^2\over|2q_fB|}
\right]
+{1\over2}m^2\phi^2
+{\lambda\over24}\phi^4-h\phi
\\&& \nonumber
-{m_{\sigma}^4\over4(4\pi)^2}\left[\ln{\Lambda^2\over m_{\sigma}^2}
+{3\over2}\right]
-{3m_{\pi^0}^4\over4(4\pi)^2}\left[\ln{\Lambda^2\over m_{\pi^0}^2}+{3\over2}\right]
+{N_cm_q^4\over(4\pi)^2}\sum_f
\left[\ln{\Lambda^2\over m_q^2}+{3\over2}\right]
\\ && \nonumber
+{4(qB)^2\over(4\pi)^2}\left[
\zeta^{(1,0)}(-1,\mbox{$1\over2$}+x)
+{1\over4}x^2-{1\over2}x^2\ln x+{1\over24}\right]
-{8N_c\over(4\pi)^2}\sum_f(q_fB)^2\bigg[
\zeta^{(1,0)}(-1,x_f)
\\ && \nonumber
+{1\over4}x_f^2-{1\over2}x_f^2\ln x_f+{1\over2}x_f\ln x_f-{1\over12}
\bigg]
-{1\over2(4\pi)^2}\left[J_0(\beta m_{\sigma})T^4
+J_0(\beta m_{\pi^0})T^4
\right.\\ &&\left.
+2J_0^B(\beta m_{\pi^0})|qB|T^2
\right]
-{2N_cT^2\over(4\pi)^2}\sum_fK_0^B(\beta m_q)|q_fB|
\;.
\label{finalle}
\eqa
As mentioned above, it a common approximation to neglect the bosonic
contribution to the one-loop effective potential and as a result
our expression reduces to
\bqa\nonumber
{\cal F}_{0+1}&=&
{1\over2}B^2\left[
1
+N_c\sum_f{4q_f^2\over3(4\pi)^2}\ln{\Lambda^2\over|2q_fB|}
\right]
+{1\over2}m^2\phi^2+{\lambda\over24}\phi^4-h\phi
+{N_cm_q^4\over(4\pi)^2}\sum_f
\left[\ln{\Lambda^2\over m_q^2}+{3\over2}\right]
\\&& \nonumber
-{8N_c\over(4\pi)^2}
\sum_f(q_fB)^2\bigg[
\zeta^{(1,0)}(-1,x_f)
+{1\over4}x_f^2
-{1\over2}x_f^2\ln x_f
+{1\over2}x_f\ln x_f
-{1\over12}
\bigg]
\\ && 
-{2N_cT^2\over(4\pi)^2}\sum_fK_0^B(\beta m_q)|q_fB|
\;.
\label{finalle2}
\eqa
\end{widetext}
In the literature, the parameters are often fixed at tree level.
The parameters in the Lagrangian~(\ref{ggglag}) can then be expressed
in terms of the sigma mass, the pion mass, and the pion decay
constant as
\bqa
\label{fix1}
m^2&=&-{1\over2}(m_{\sigma}^2-3m_{\pi}^2)\;,\\
\lambda&=&{3(m_{\sigma}^2-m_{\pi}^2)\over f_{\pi}^2}\;,
\\
h&=&f_{\pi}m_{\pi}^2\;.
\label{fix3}
\eqa
Having determined the parameters as described above, 
the tree-level potential has its minimum at the
correct value $\phi=f_{\pi}= 93$ MeV, while the minimum of 
the one-loop potential depends on the renormalization scale.
We can choose $\Lambda$ such that 
the minimum of the one-loop effective potential~(\ref{finalle2})
in the vacuum, i.e. $T=B=0$,
still is at $\phi=f_{\pi}=93$ MeV. This is done by requiring
\bqa
{d{\cal F}_{0+1}\over d\phi}\bigg|_{\phi=f_{\pi}}&=&0\;.
\label{minnie}
\eqa
A straigthforward calculation yields 
\bqa
{N_cg^4f_{\pi}^2\over(2\pi)^2}
\left[
\ln{\Lambda^2\over g^2f_{\pi}^2}+1\right]&=&0\;.
\eqa
Using $f_{\pi}=93$ MeV and $g=3.2258$ (see below), this yields
$\Lambda=181.96$ MeV. 
Even if we use this value for the renormalization scale such that 
${\cal F}_{0+1}$ has its minimum at $\phi=f_{\pi}$,
it is strictly speaking not correct to use the 
parameters~Eqs.~(\ref{fix1})--(\ref{fix3}) in the one-loop
effective potential. The reason is that the sigma and pion masses
receive radiative corrections, which must be taken into account
in the equations that relate the physical masses and the
parameters of the theory. In other words, Eqs.~(\ref{fix1})--(\ref{fix3})
receive corrections. The sigma mass is often defined by 
the curvature or the second derivative of the effective potential.
At tree-level, this is given by Eq.~(\ref{msigmi}), but 
the expression for $m_{\sigma}$ changes if we take the one-loop correction
to the effective potential into account. To illustrate the dramatic difference,
we have calculated the two.
Using parameters such that the sigma mass at tree level
is $750$ MeV and the pion mass is $140$ MeV, the curvature of the one-loop 
effective potential with $\Lambda=182$ MeV,
corresponds to a sigma mass of $530$ MeV.
Even if one takes radiative corrections into account and determine
the parameters at the one-loop level, this procedure is not
entirely correct. Determining the sigma mass by the curvature of
the effective potential corresponds to including its self-energy 
evaluated at zero external momentum, 
$p^2=0$. However, the physical mass of the sigma 
is given by the pole of the propagator, which involves the
self-energy evaluated self-consistently at $p^2=-m_{\sigma}^2$.
The difference between the self-energy evaluated at these two points
gives rise to a finite shift of the sigma mass that is normally
not taken into account. A similar remark applies to the
pion mass at the physical point.

\begin{figure}
\includegraphics[width=7.0cm]{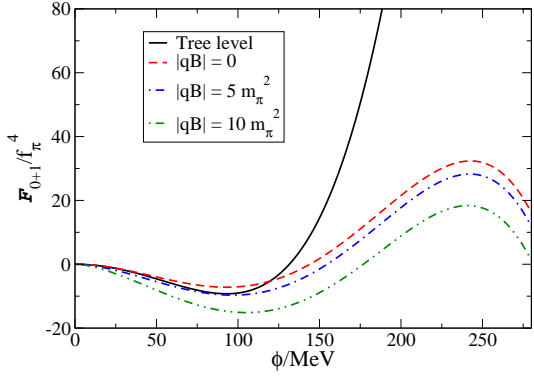}
\caption{Normalized effective potential ${\cal F}_{0+1}/f_{\pi}^4$
in the chiral limit for $T=0$.
Tree level (solid curve), 
one-loop with $|qB|=0$ (dashed curve),
one loop with $|qB|=5m_{\pi}^2$ (dotted curve), and
one loop with $|qB|=10m_{\pi}^2$ (dash-dotted curve).}
\label{vacpot}
\end{figure}

\begin{figure}
\setlength{\unitlength}{1mm}
\includegraphics[width=6.0cm]{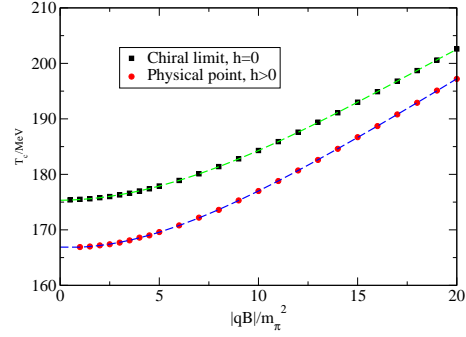}
\caption{The critical temperature for the 
chiral transition as a function of $|qB|/m_{\pi}^2$.
Black points are the chiral limit and red points are the physical point.}
\label{bdep}
\end{figure}

\begin{figure}
\setlength{\unitlength}{1mm}
\includegraphics[width=7.0cm]{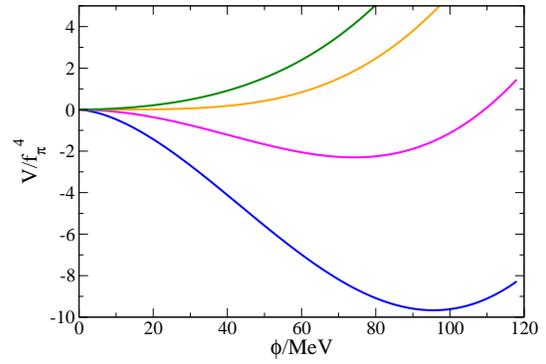}
\caption{The effective potential normalized by $f_{\pi}^4$ for four different
temperatures and $|qB|=5m_{\pi}^2$, where we have 
included the vacuum fluctuations.}
\label{potfluc}
\end{figure}

\begin{figure}
\setlength{\unitlength}{1mm}
\includegraphics[width=7.0cm]{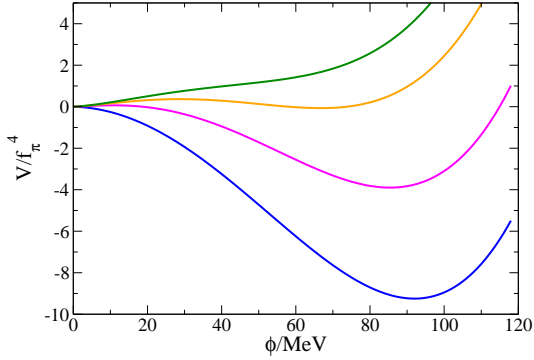}
\caption{The effective potential normalized  by $f_{\pi}^4$ for four different
temperatures and $|qB|=5m_{\pi}^2$, where we have 
omitted the vacuum fluctuations.}
\label{nopotfluc}
\end{figure}

We next present some numerical results based on the effective
potential~(\ref{finalle2}).
At the physical point, we 
use a pion mass of $m_{\pi}=140$ MeV, a sigma mass of
$m_{\sigma}=800$ MeV, 
a constituent quark mass of $m_q=300$ MeV, 
and a pion decay constant of $f_{\pi}=93$ MeV.
This yields the parameters
$m^2=-290600$ MeV$^2$ , $\lambda=215.29$, and $g=3.2258$. 
In the chiral limit, we instead use $m_{\pi}=0$, which yields
$m^2=-320000$ MeV$^2$, $\lambda=222$, and $g=3.2258$. 
We use these parameter values 
to generate the results presented in Figs.~\ref{vacpot}--\ref{nopotfluc}.

In Fig.~\ref{vacpot}, we show the renormalized effective potential at
$T=0$ normalized to $f_{\pi}^4$ as a function of $\phi$ for different
values of the magnetic field. 
The minimum is moving to the right as the 
magnetic field increases and so the model exhibit magnetic catalysis as
expected. Moreover, the effective potential is unstable for large values
of the field $\phi$. This instability is present already for $B=0$
and is due to a term 
$\sim m_q^4\ln{\Lambda^2\over m_q^2}$, 
which dominates the effective potential
at large $\phi$ and goes negative for 
$m_q=g\phi\geq\Lambda$.\footnote{This is the leading term
in the large-$x$ expansion~(\ref{largefermi}).}
The one-loop bosonic term is also of the form $m^4\ln{\Lambda^2\over m^2}$
with the opposite sign and may stabilize the effective potential
if the prefactor is suffiently large.
However, perturbative calculations typically break down for 
large value of the field. In fact,
a renormalization group improvement is necessary to make large values of
$\phi$ accessible by removing large logarithms. These issues
have discussed
in detail in e.g.~\textcite{rg1,rg2,rg3}.

In Fig.~\ref{bdep}, we show the transition temperature for the QM
model as a function of $|qB|/m_{\pi}^2$ in the chiral limit (black)
and at the physical point (red). Both are growing functions of the
magnetic field, which shows that the QM model exhibits 
magnetic catalysis.

In Fig.~\ref{potfluc}, we show the normalized effective potential
for four different temperatures and
$|qB|=5m_{\pi}^2$ and with vacuum fluctuations 
(blue line: $T=0$, pink line $T=140$ MeV, orange line: $T=T_c=177.9$ MeV, and
green line: $T=185$ MeV). The curves clearly show that the phase transition
is of second order.

In Fig.~\ref{nopotfluc}, we show the normalized effective potential
for four different temperatures, $|qB|=5m_{\pi}^2$ and no vacuum fluctuations 
(blue line: $T=0$, pink line $T=120$ MeV, orange line: $T=T_c=158.0$ MeV, and
green line: $T=176$ MeV). The curves clearly show that the phase transition
is of first order. Moreover, we notice that including the vacuum 
fluctuations gives a significantly higher critical temperature $T_c$.

We have now seen numerically
that the phase transition is first order
if the fermionic vacuum fluctuations are neglected and second order if they are
included. It turns out that the role of vacuum fluctuations in 
the quark-meson model is the same also in the absence of a magnetic field.
This case was carefully analyzed by \textcite{skokovfri}
in the case of the chiral transition in the QM model with $B=0$.
Recently, the analysis was generalized to finite $B$ field by
\textcite{greco}. 

One would like to get some analytic understanding of this
result. The basic idea is to construct a Ginzburg-Landau 
effective potential of the form
\bqa
V_{\rm GL}&=&{1\over2}\alpha_2m_q^2+{1\over4}{\alpha_4}m_q^4\;,
\eqa
where $m_q=g\phi$, and
$\alpha_2$ and $\alpha_4$ are parameters that depend on the
temperature $T$, the parameters of the Lagrangian, and the magnetic field $B$.
A temperature $T_c^*$
is defined by a vanishing coefficient $\alpha_2$,
i.e. $\alpha_2(T_c^*)=0$. If $\alpha_4(T_c^*)>0$, then the transition is
second order and the critical temperature is $T_c=T_c^*$.
If $\alpha_4(T_c^*)<0$, the effective potential has two minima for
$T$ slightly larger than $T_c^*$. The 
transition is first order and the critical temperature is $T_c>T_c^*$.

\begin{figure}
\begin{center}
\setlength{\unitlength}{1mm}
\includegraphics[width=7.0cm]{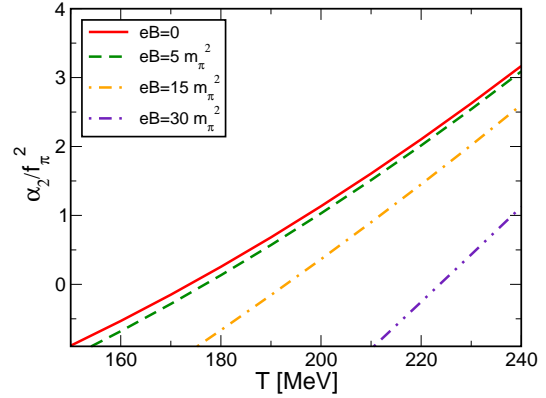}
\caption{Coefficient $\alpha_2$ normalized by $f_{\pi}^2$ as a function of
temperature $T$ for different values of the magnetic field.
Figure taken from~\textcite{greco}.}
\label{alpha2}
\end{center}
\end{figure}

In the plots shown below, the authors used parameters that 
correspond to $f_{\pi}=92.4$ MeV, $m_q=335$ MeV, $m_{\pi}=0$,
and $m_{\sigma}=700$ MeV.
We first consider the renormalized case, i.e. the case where 
one adds counterterms for the mass and coupling and imposes
some appropriate renormalization conditions.
In Fig.~\ref{alpha2}, the normalized coefficient $\alpha_2/f_{\pi}^2$
is shown as a function of $T$ for different values of 
the magnetic field $B$. We see that $\alpha_2$ is an increasing function of
$T$ and $T_c^*$ is an increasing function of $|qB|$.

\begin{figure}
\begin{center}
\setlength{\unitlength}{1mm}
\includegraphics[width=7.0cm]{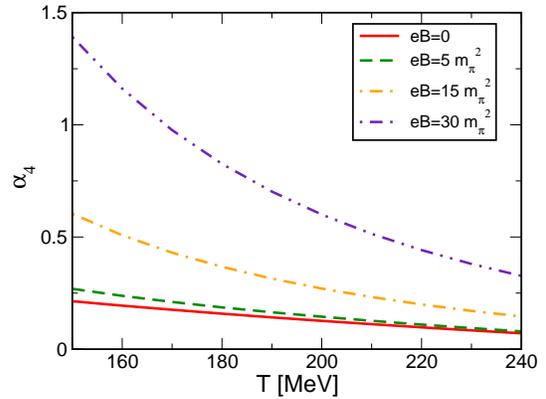}
\caption{Coefficient $\alpha_4$ as a function of
temperature $T$ for different values of the magnetic field.
Figure taken from~\textcite{greco}.}
\label{alpha4}
\end{center}
\end{figure}

In Fig.~\ref{alpha4}, the dimensionless coefficient $\alpha_4$
is shown as a function of $T$ for different values of 
the magnetic field $B$. The coefficient is positive for all values of $T$
implying that the transition is second order.

\begin{figure}
\begin{center}
\setlength{\unitlength}{1mm}
\includegraphics[width=7.0cm]{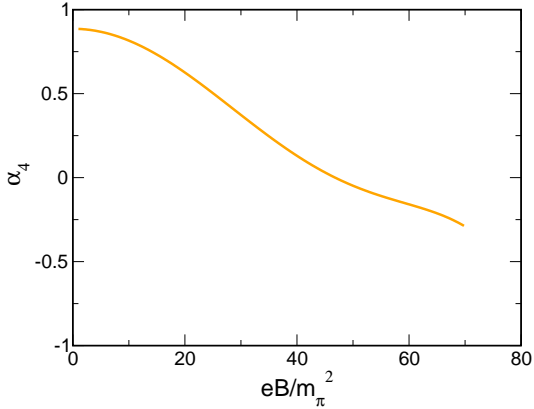}
\caption{Coefficient $\alpha_4$ as a function of
magnetic field $B$ at $T_c^*$.
Figure taken from~\textcite{greco}.}
\label{nralpha4}
\end{center}
\end{figure}

We next consider the unrenormalized case, i.e. the case where on
regularizes the divergences by a sharp ultraviolet cutoff $\Lambda=550$ MeV.
In Fig.~\ref{nralpha4}, the coefficient $\alpha_4$
is shown as a function of $B$ evaluated at $T_c^*$. 
The coefficient is negative for $|qB|>|qB_c|\approx47$ $m_{\pi}^2$ 
and so the transition is first order
for large magnetic fields. The position of $B_c$ obviously depends on
$\Lambda$ and for $\Lambda\rightarrow\infty$, one recovers 
the results in the renormalized case.
On the other hand, if the cutoff $\Lambda$ is below a critical value 
$\Lambda_c$, the transition is first order for all values of $B$.
The sensitivity to the value of the sharp cutoff suggests that one should
be careful and in particular not choose a cutoff below the 
scale set by the particles in the theory.

\subsection{Hadron resonance gas model}
In this subsection, we briefly discuss the hadron resonance gas (HRG) model
in a magnetic background. This model was studied in detail 
by~\textcite{endrodires}. It can be used to access the low-temperature
phase of QCD, even at nonzero chemical potential. The partition function
is given by a sum of partition functions of noninteracting hadrons and
resonances. This approach gives a surprisingly good description of the
thermodynamics of QCD in the confined phase, up to temperatures just below
the transition region~\cite{borsahrg,karschhrg,huovinenhrg}. 

We can schematically write
the free energi density as 
\bqa
{\cal F}&=&\sum_hd_h{\cal F}_h(B,T,m_h,q_h,s_h,g_h)\;,
\label{hadronsum}
\eqa
where $d_h$, 
$m_h$, $q_h$, $s_h$, and $g_h$ are the multiplicity,
mass, electric charge, spin, and
gyromagnetic ratio of hadron $h$, respectively. For simplicity,
the gyromagnetic ratio in the paper by~\textcite{endrodires}
was set to $g_h=2q_h/e$. The hadrons taken into
account in the sum are $\pi^{\pm}, \pi^0,...,\sum^0$.

\begin{figure}
\begin{center}
\setlength{\unitlength}{1mm}
\includegraphics[width=7.0cm]{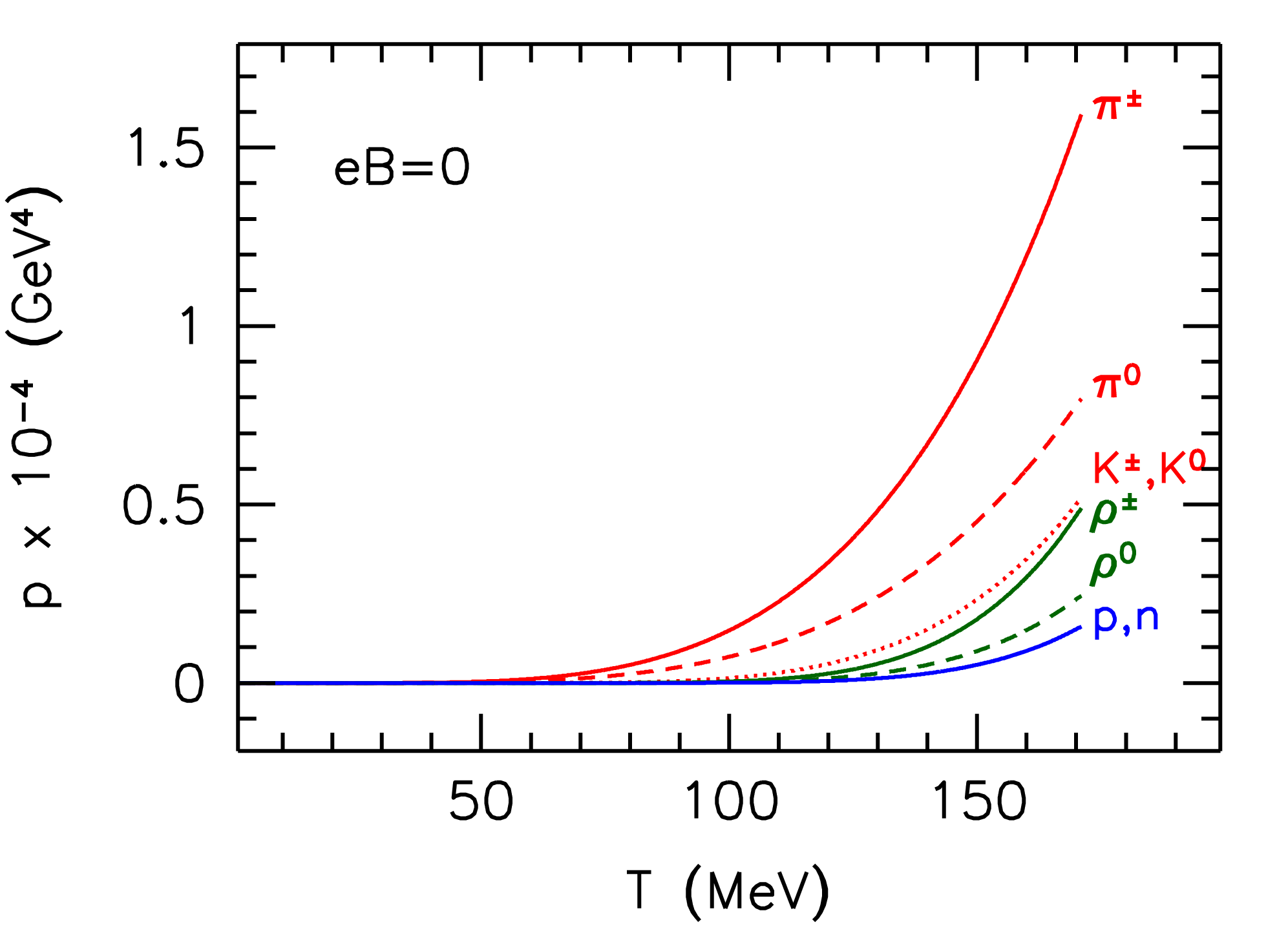}
\caption{Indivudual contributions to the HRG pressure as function of the
temperature $T$ for $B=0$.
Figure taken from~\textcite{endrodires}.}
\label{contribution}
\end{center}
\end{figure}

\begin{figure}
\begin{center}
\setlength{\unitlength}{1mm}
\includegraphics[width=7.0cm]{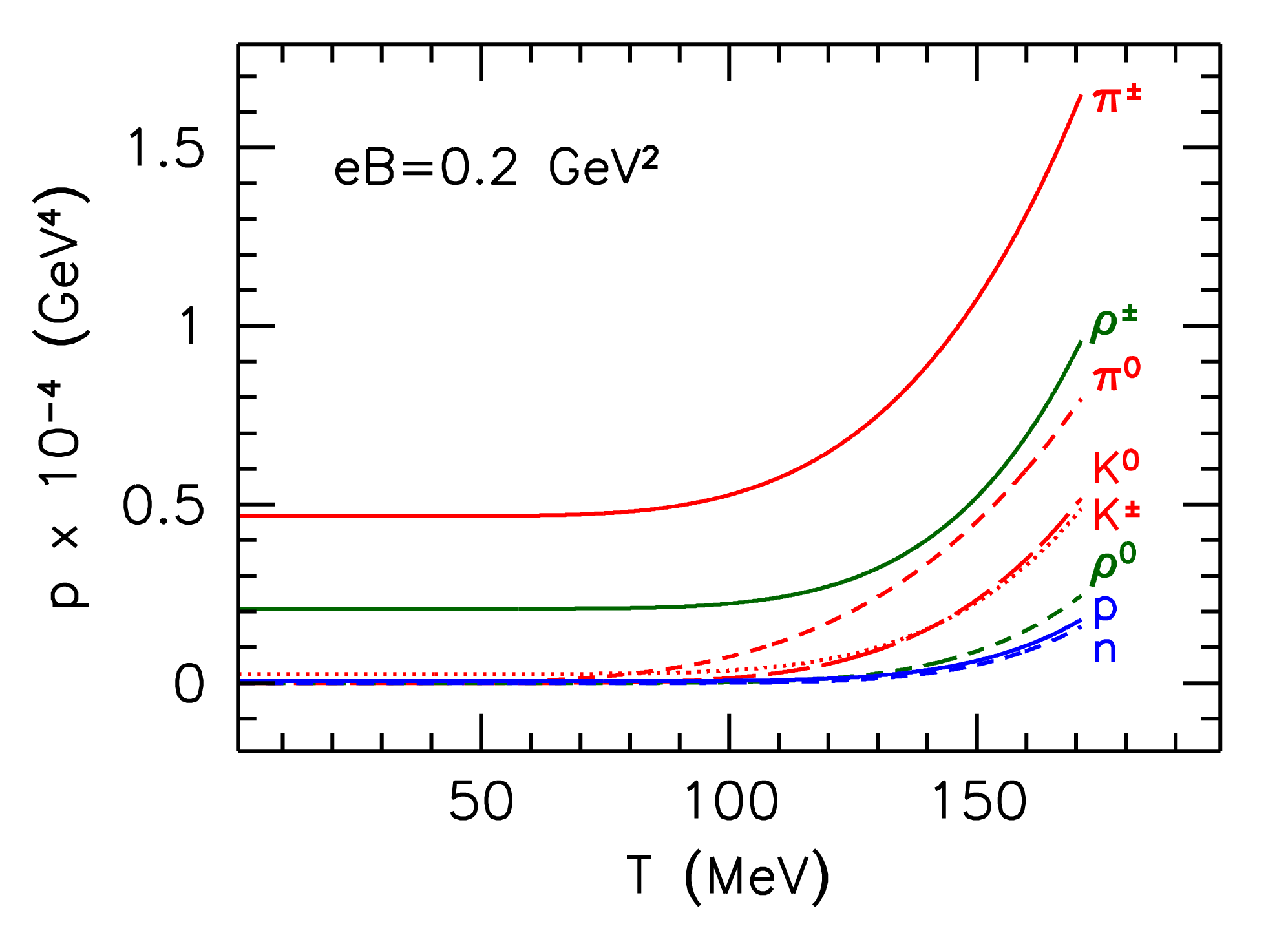}
\caption{Indivudual contributions to the HRG pressure as function of the
temperature $T$ for $|qB|=0.2$ GeV$^2$.
Figure taken from~\textcite{endrodires}.}
\label{contribution2}
\end{center}
\end{figure}

Since the free energy density is a sum of the free energy densities of 
noninteracting mesons and baryons,
Eq.~(\ref{hadronsum}) is given by a sum of the one-loop
terms Eqs.~(\ref{sumbose}) and~(\ref{sumfermi}).
Renormalization can be performed using minimal subtraction as
discussed previously. 
However, the author used Schwinger's renormalization scheme 
which involves an extra logarithmic term. For example, 
the wavefunction counterterm is~\footnote{There is an extra factor
of $\ln{4\pi\over e^{\gamma_E}}$ since the $\rm MS$-scheme is used.} 
\bqa
Z^2&=&\left[1-
{q^2\over3(4\pi)^2\epsilon}-{q^2\over3(4\pi)^2}\ln{m^2\over\Lambda^2}
\right]\;.
\label{endren}
\eqa
Defining the renormalized magnetic field $B_r$ via $B_r^2=B^2Z^2$
and subtracting the free energy density at $B=0$,
the one-loop free energy density from a boson is
\begin{widetext}
\bqa
{\cal F}_{0+1}
&=&{1\over2}B_r^2
+{4(qB)^2\over(4\pi)^2}\left[
\zeta^{(1,0)}(-1,\mbox{$1\over2$}+x)+{1\over4}x^2-{1\over2}x^2\ln x
+{1\over24}\left(\ln x+1\right)
\right]\;.
\label{endren1}
\eqa
\end{widetext}
This is turn implies that the renormalized 
free energy density approaches zero in the
limit $m_h\rightarrow\infty$ instead of growing logarithmically.
However, the prescription 
cannot be used in the massless limit due to infrared divergences
coming from this extra term. 
Furthermore, the order-$B^2$ term is given by the first term in 
Eq.~(\ref{endren}) as the leading term in the bracket start at
${\cal O}(B^4)$. 
Figs.~\ref{contribution} and~\ref{contribution2}
show the individual contributions to the HRG
pressure as a function of the temperature $T$ for 
zero magnetic field 
and for $|qB|=0.2$ GeV$^2$, respectively.
We first note that the contribution from the neutral particles
is independent of $B$ as the gyromagnetic ratio was set to zero.
The relative contribution of the charged particles is changing with $B$.
The reason is that the effective mass is essentially 
$m_{\rm eff}=\sqrt{m_h^2+|q_hB|(1-2s)}$, and this increases for e.g. $m_{\pi^{\pm}}$
($s=0$)
and decreases for e.g. $\rho^{\pm}$ ($s=1$)
and so the Boltzmann weight changes with $B$.
This is clearly seen in Figs.~\ref{contribution} and~\ref{contribution2}.

Moreover, the pressure is an increasing function of the magnetic field
for fixed temperature $T$. This implies that the magnetization is 
positive and hence that the hadronic phase is paramagnetic.
We will discuss this further in Sec.~\ref{anis}.
The speed of sound $c_s$ (not shown) 
displays a peak, which moves to lower temperatures
as the magnetic fields grows. The position of the peak on the $T$-axis
is a possible definition of the transition temperature and so the results
suggest that $T_c$ decreases with $B$. However, the HRG model
is valid only at low temperatures and not too large magnetic 
fields~\footnote{Clearly, $m_{\rm eff}^2<0$ for the $\rho$ meson
when $|qB|>m_{\rho}^2$.}~\cite{endrodires} so one must be cautious.

\section{Polyakov-loop extended models}
\label{extended}
As mentioned earlier, the NJL model is not confining.
Likewise, the QM model is an effective theory that consists of 
deconfined quarks as well as mesons as effective degrees of 
freedom~\cite{bowman}.
This is probably a good description at temperatures around the transition
temperature, but for very low temperatures it is certainly 
not. At low temperatures, the thermodynanmics is dominated by the light 
pions. While these models incorporate chiral symmetry breaking, they 
are not confining.
This is a serious shortcoming as an effective low-energy description of QCD,
Below, we shall see that we can mimic - in a statistical 
sense - the effects of confinement by coupling the chiral models to
a nontrivial $SU(3)$ background gauge field $A_{\mu}$~\cite{pnjlfuku}.
One can express this background gauge field in terms of 
the complex-valued Polyakov loop variable $\Phi$ and so the 
effective potential becomes a function of the expectation value of
the chiral condensate as well as the expecation value of the Polykov loop.
Finally, one adds the contribution to the free energy density 
from the 
gluons via a polyakov loop potential.

\subsection{Coupling to the Polyakov loop}
In pure gauge theory, the Polyakov loop $\Phi$
is an order parameter for 
deconfinement~\cite{benjamin1,benjamin2}.
For QCD with dynamical quarks, it is an approximate 
order parameter, just like the chiral condensate.
It is defined as the trace of the thermal Wilson line, where
the thermal Wilson $L$ line is given by 
\bqa
L({\bf x})&=&{\cal P}\exp\left[i\int_0^{\beta}d\,\tau\,A_4({\bf x},\tau)\right]\;,
\label{defwilson}
\eqa
where $A_4=iA_0$ and $A_0=t_aA^a_0$. Here $A_{\mu}^a$ are the
$SU(3)_c$ gauge felds and the generators
$t^a={1\over2}\lambda^a$, where $\lambda^a$
are the Gell-Mann matrices. ${\cal P}$ denotes path ordering.
The Polyakov $\Phi$ and its complex conjugate
are then given by 
\bqa
\Phi&=&{1\over N_c}
{\rm Tr}L
\;,
\\
\bar{\Phi}&=&{1\over N_c}
{\rm Tr}L^{\dagger}
\;.
\eqa
The Polykov loop transforms nontrivially under the center group
$Z_{N_c}$ of the gauge group $SU(N_c)$, $\Phi\rightarrow e^{2\pi i n/N_c}\Phi$,
where $n=0,1,2,...N_c-1$. Its behavior in the pure gauge theory is
\bqa
\langle\Phi\rangle&\sim&0\;,\;{\rm confinement\,\,at\,\,low\,\,}T\;,
\\\langle\Phi\rangle&\sim&1\;,\;{\rm deconfinement\,\,at\,\,high\,\,}T\;,
\eqa
and so the center symmetry $Z_{N_c}$ is broken in the high-temperature phase.

A constant nonabelian background is now introduced via the covariant derivative
which takes the form
\bqa
D_{\mu}&=&\partial_{\mu}-iq_fA_{\mu}^{\rm EM}-iA_{\mu}\;,
\label{covdev}
\eqa
where $A_{\mu}^{\rm EM}=(0,0,Bx,0)$ and $A_{\mu}=\delta_{\mu0}A_0$.
In the Polyakov gauge, we can write the background gauge field 
$A_4=iA_0$ as
\bqa
A_4&=&t_3A_4^{(3)}+t_8A_4^{(8)}\;,
\eqa
For constant gauge fields,
the thermal Wilson line can be written as
\bqa
L=\left(
\begin{array}{ccc}
e^{i(\phi_1+\phi_2)}&0&0\\
0&e^{i(-\phi_1+\phi_2)}&0 \\
0&0&e^{-2i\phi_2}
\end{array}\right)\,,
\label{lexp}
\eqa
where $\phi_1={1\over2}\beta A_4^{(3)}$
and $\phi_2={1\over2\sqrt{3}}\beta A_4^{(8)}$.
In the perturbative vacuum $\phi_1=\phi_2=0$ and in the confining vacuum
$\phi_1=\phi_2={\pi\over3}$.
We can use $\phi_1=\phi_2$ and the Polyakov loop variable itself reduces to
\bqa
\Phi&=&{1\over3}\left[1+2\cos(\phi_1)\right]\;.
\eqa
Note that the Polyakov loop is real for $\mu_B$
and therefore $\Phi=\bar{\Phi}$.
In the expressions below, we will however, keep
$\Phi$ and $\bar{\Phi}$ so the expressions below
agree with those in the literature.

The zeroth component of the gauge field acts as a chemical
potential in the covariant derivative~(\ref{covdev}).
With 
this observation and the definition Eq.~(\ref{defwilson}),
we can immediately make the following replacement
for a fermion in the background field:
\begin{widetext}
\bqa
\ln\left[1+e^{-\beta E_q}\right]
&\rightarrow&
{1\over2N_c}{\rm Tr}\ln\left[1+Le^{-\beta E_q}\right]
+{1\over2N_c}{\rm Tr}\ln\left[1+L^{\dagger}e^{-\beta E_q}\right]\;,
\label{replacelog}
\eqa
where the trace on the right-hand side is in color space
and $E_q$ is the energy of the fermionic excitations.
Performing the trace of the first term 
in Eq.~(\ref{replacelog}) using Eq.~(\ref{lexp}),
one obtains
\bqa\nonumber
{1\over N_c}{\rm Tr}\ln\left[1+Le^{-\beta E_q}\right]
&=&
{1\over3}\ln\left[1+3\left(\Phi+\bar{\Phi}e^{-\beta E_q}\right)
e^{-\beta E_q}+e^{-3\beta E_q}
\right]\;,
\eqa
and where the second term in Eq.~(\ref{replacelog}) can be obtained
by Hermitean conjugation.
The temperature-dependent part of the
one-loop fermionic contribution to the free energy density
can 
then be written as
\bqa\nonumber
{\cal F}_1^T&=&-T\sum_f{|q_fB|\over\pi}\sum_{s=\pm1}\sum_{k=0}^{\infty}
\int_0^{\infty}\Bigg\{
\ln\left[1+3\left(\Phi+\bar{\Phi}e^{-\beta E_q}\right)
e^{-\beta E_q}+e^{-3\beta E_q}\right]
\\ &&
+\ln\left[1+3\left(\bar{\Phi}+\Phi e^{-\beta E_q}\right)e^{-\beta E_q}+e^{-3\beta E_q}
\right]
\Bigg\}\;.
\eqa
It reduces to the second term in Eq.~(\ref{free}) in the limit 
$\Phi,\bar{\Phi}\rightarrow1$, with an extra factor of $N_c=3$, as it should.
We note in passing that the vacuum part of the one-loop fermionic 
free energy density is unchanged and therefore
the PNJL model reduces to the NJL model
at $T=0$.

Taking the 
trace in color space,
the Fermi-Dirac distribution is generalized to
\bqa
n_F(\beta E)&=&
\frac{1+2\bar{\Phi}e^{\beta E_q} 
+ \Phi e^{2\beta E_q}}{1+3\bar{\Phi} e^{\beta E_q}+3\Phi e^{2\beta E_q}
+e^{3\beta E_q}}\;.
\label{fermipol}
\eqa
\end{widetext}
It is instructive to look at the behavior of Eq.~(\ref{fermipol})
at very low and at very high temperatures.
At low temperatures, we have
$\Phi\approx0$ and therefore the Fermi-Dirac distribution reduces to
\bqa
n_F(\beta E_q)&\approx&{1\over e^{3\beta E_q}+1}\;,
\label{tclt}
\eqa
which is the distribution function of a fermion with energy $3E_q$.
Thus the contribution 
from the fermions to the
effective potential is suppressed at low temperature as compared to
the corresponding chiral model without the Polyakov loop. 
This is referred to as statistical confinement.
In the same manner, we see that 
Eq.~(\ref{fermipol}) at high temperature behaves as 
\bqa
n_F(\beta E_q)&\approx&{1\over e^{\beta E_q}+1}\;,
\eqa
where we have used that $\Phi\approx1$.
This is distribution function of a fermion with energy $E_q$, i.e. 
that of deconfined quarks. A word of caution here is appropriate.
The same behavior as Eq.~(\ref{tclt}) is found in two-color
QCD with quarks in the adjoint representation~\cite{riske} and so 
the number $x$ in $x\beta E_q$ does not necessarily give the correct number
of quarks to form a color singlet.

We have now coupled the Polyakov loop variable to the matter sector
of theory. However, we must also include the contribution to the 
free energy density from the gauge sector and this is done 
by adding a phenomenological 
Polyakov loop
potential $U(\Phi,\bar{\Phi})$. This potential is required
to reproduce the pressure for pure-glue QCD as 
calculated on the lattice for the temperatures
around the transition temperature.

A number of forms for the polyakov loop potentials have been proposed 
and investigated at the mean-field level for the PNJL model~\cite{lourenco} 
and the PQM model with $\mu_B=0$~\cite{schaefer2010}. 
In the following we will review three different Polyakov loop potentials.
Since the Polyakov loop variable is the order parameter for the $Z_3$
center symmetry of pure-glue QCD, a Ginzburg-Landau type potential
should incorporate this. A polynomial expansion then leads 
to~\cite{polyakovpot1}
\begin{widetext}
\bqa
\frac{U_{\rm poly}}{T^4} &= -\frac{1}{2}b_2(T) \Phi{\bar\Phi}  -\frac{1}{6}b_3
\big( \Phi^3+{\bar\Phi}^{3} \big) + \frac{1}{4}b_4 \big( \Phi{\bar\Phi} \big)^2 
\; , \label{poly}
\eqa
where the coefficients are
\bqa
b_2(T) &=& 6.75 - 1.95\left(\frac{T_0}{T}\right) + 2.624\left(\frac{T_0}
{T}\right)^2 - 7.44\left(\frac{T_0}{T}\right)^3 \; , \\
b_3 &=& 0.75\;,\\
b_4 &=& 7.5\;.
\eqa
The parameter $T_0$ is the transition temperature 
for pure-glue QCD lattice calculations, $T_0=270$ MeV~\cite{tc270}.
\textcite{polyakovpot2,polyakovpot3} proposed another form of the 
Polyakov loop potential based on the $SU(3)$ Haar measure:
\bqa
	\frac{U_{\rm log}}{T^4} &= -\frac{1}{2}a(T) \Phi{\bar\Phi}  +b(T) \ln 
\left[ 1- 6\,{\bar\Phi}\Phi + 4 \big( \Phi^3+{\bar\Phi}^{3} \big) - 
3\big( {\bar\Phi}\Phi \big)^2 \right] \; , \label{log}
\label{correct}
\eqa
where the coefficients are
\bqa
	a(T) &=& 3.51 - 2.47\left(\frac{T_0}{T}\right) + 15.2\left(\frac{T_0}{T}
\right)^2 \; , \\
	b(T) &= &-1.75\left(\frac{T_0}{T}\right)^3\;.
\eqa
We note that the logarithmic term ensures that the magnitude of 
$\Phi$ and $\bar{\Phi}$ are constrained to be in the region between 
$-1$ and $1$, i.e.\ the possible attainable 
values for the normalized trace of an element of $SU(3)$.
The coefficient $a(T)$ approaches $16{\pi^2\over90}\approx3.51$
as $T\rightarrow\infty$
such that the potential Eq.~(\ref{correct}) reproduces the
Stefan-Boltzmann limit.
Finally, ~\textcite{polyakovpot4} proposed the following
Polyakov loop potential 
\bqa
\frac{U_{\rm Fuku}}{T^4} &= -\frac{b}{T^3} \bigg(54e^{-a\,T_0/T}\Phi{\bar\Phi}
  +\ln \left[ 1- 6\,\Phi{\bar\Phi} + 4 \big( \Phi^3+{\bar\Phi}^{3} \big) 
- 3\big( \Phi{\bar\Phi} \big)^2 \right] \bigg)\; , \label{Fuku}
\eqa
\end{widetext}
where the constants are $a=664/270$ and $b=(196.2\textrm{ MeV})^3$.
This potential differs from the logarithimic potential~(\ref{correct})
only by the coefficient of the first term. 

A problem with all the Polyakov loop potentials proposed is that
they are independent of the number of flavors and of the baryon chemical
potential. However, we know that, for example, the transition temperature
for the deconfinement transition is a function of $N_f$.
In other words, one ought to incorporate the back-reaction from the 
fermions to the gluonic sector. \textcite{bj} 
use perturbative arguments to estimate the effects of the number
of flavors and the baryon chemical potential on the transition temperature
$T_0$. The functional form of $T_0$ is~\cite{herbst} for $\mu_B=0$ is
\begin{equation}
	T_0 = T_\tau e^{-1/(\alpha_0 \, b(N_f))}\;,
\label{nfmub}
\end{equation}
where
\begin{equation}
b(N_f) = \frac{1}{6\pi}(11N_c - 2N_f)\;,
\label{bpar}
\end{equation}
and the the parameters are
$T_\tau = 1.77 \textrm{ GeV}$ and $\alpha_0 = 0.304$. 
This yields a transition temperature of $240$ MeV and $208$ MeV for
$N_f=1$ and $N_f=2$, respectively.
Another way of including the back-reaction from the fermions has been 
implemented by~\textcite{rainer1,rainer2}.
They calculate the  glue potential as a function of a background gauge
field with and without dynamical fermions using the functional renormalization
group. They compare the two potentials and found a mapping between them
and this mapping is used to modify the Polyakov loop potential
discussed above.

The phase stucture of the Polyakov-loop extended model is then found by solving
simultaneously the gap equations
\bqa
{\partial{\cal F}\over\partial M_0}
=0\;,\hspace{0.5cm}
{\partial{\cal F}\over\partial\Phi}=0\;,
\hspace{0.5cm}({\rm PNJL})\;,
\\
{\partial{\cal F}\over\partial\phi}
=0\;,\hspace{0.8cm}
{\partial{\cal F}\over\partial\Phi}=0\;,
\hspace{0.5cm}({\rm PQM})\;.
\eqa
where ${\cal F}$ is the sum of the free energy density
from the fermions and
the Polyakov loop potential $U$.

\textcite{Gatto} considered the
PNJL model using the logarithmic potential Eq.~(\ref{log}).
They used $G_1=G_2$ and added an eight-quark interaction term
of the form
\bqa
\delta{\cal L}&=&
G_8
\left[
(\bar{\psi}\psi)^2
+(\bar{\psi}i\gamma_5{\boldsymbol\tau}\psi)^2\right]^2\;,
\eqa
where $G_8$ is a coupling constant. In this case, the 
constituent quark mass reads 
$M_0=m_0-2G_0\langle\bar{\psi}\psi\rangle-4G_8\langle\bar{\psi}\psi\rangle^3$.
They also used a form factor of the form
\bqa
F(p)&=&{\Lambda^{2N}\over\Lambda^{2N}+(p_z^2+2|q_fB|k)^N}\;,
\eqa
choosing the value $N=5$. 

\begin{figure}
\begin{center}
\setlength{\unitlength}{1mm}
\includegraphics[width=7.0cm]{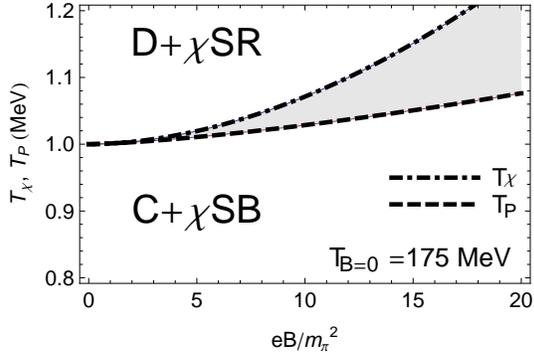}
\caption{Phase diagram in the $B$--$T$ plane.
Figure taken from~\textcite{Gatto}.}
\label{ruggi}
\end{center}
\end{figure}

Fig.~\ref{ruggi} shows the phase diagram in the $B$--$T$ plane
for the chiral (dot-dashed) as well as the deconfinement transition
(dashed). The critical temperatures $T_{\chi}$ and $T_P$ have been normalized
to the common pseudocritical temperature $T_0=175$ MeV at $B=0$.
We note that the transition temperature $T_{\chi}$ is increasing 
more with the magnetic
field than $T_P$ is.
The shaded area corresponds to a phase where the quarks
are deconfined but where chiral symmetry is still broken.

The PNJL model has been extended by several authors
by using a non-local NJL vertex ~\cite{hell,sasa,kondo}.
In the paper by~\textcite{kondo}, he derives from QCD a non-local
vertex that depends explicitly on $T$ as well as the 
phase of the Polyakov loop. Such as a model is called the
entangled Polyakov loop model (EPNJL), which was used by~\textcite{gattopnjl}
at finite $B$ to study the chiral and deconfinement transitions.
In contrast to the PNJL model, there is basically no
splitting of the two transitions in the EPNJL model.

Another mean-field analysis was carried out by~\textcite{fragapoly}
using the PQM model focusing on the physical point.
Renormalization is carried out by subtracting the 
divergent fluctuation determinant for $B=0$.
The authors make several interesting observations.
If the fermionic vacuum fluctuations are neglected~\footnote{Note that there
are still some $B$-dependent vacuum terms that have not been removed
by the renormalization procedure.}, the
transition temperature for the deconfinement transition coincides with
that of the chiral transition, and they are both first order, except
for very small values of the magnetic field, where they are crossovers.
Moreover, the transition temperatures
are decreasing with increasing $B$.
If the vacuum fluctuations are included, the transition temperatures
are increasing with $B$ and the resulting phase diagram is qualitative
the same as in Fig.~\ref{ruggi}. The chiral transition is now a crossover.

\subsection{Two-color QCD}
So far we have been discussing QCD with three colors.
In this section, we consider two-color QCD. Two-color QCD is interesting
for a number of reasons. In contrast to three-color QCD, one can
perform lattice simulations at finite baryon chemical potential $\mu_B$.
This is a consequence of the special properties of the gauge group
$SU(2)_c$ which leads to a real-valued Dirac determinant even for $\mu_B\neq0$.
Hence, the sign problem is absent in this case and one can use 
importance sampling technques as usual.
Moreover, the order of the deconfinement transition for pure-glue QCD
is different in $SU(2)_c$ and $SU(3)_c$.
For $N_c=2$ it is second order, while for $N_c=3$ it is first order.
In two-color QCD, the critical
exponents are expected to be those of the two-state Potts model, which follows
from universality arguments.

In this section. we discuss two-color QCD in a strong magnetic
field. While there is a number of model calculations in two-color
QCD at finite temperature and baryon chemical potential, there
is only a single calculations at finite $B$~\cite{amador}.

In the Polyakov gauge, the background nonabelian gauge field
is diagonal in color space,
\bqa
A_4&=&\sigma_z\theta\;,
\eqa
where $\theta$ is real. The thermal Wilson line can then be written as
\bqa
L=\left(
\begin{array}{cc}
e^{i\phi}&0\\
0&e^{-i\phi} \\
\end{array}\right)\;,
\eqa
where $\phi=\beta\theta$. 
The Polyakov loop variable becomes
\bqa
\Phi&=&\cos(\phi)\;,
\eqa
In analogy with Eq.~(\ref{fermipol}), the Fermi-Dirac distribution 
function becomes 
\bqa
n_F(\beta E_q)&=&{1+\Phi e^{-\beta E_q}\over1+2\Phi e^{-\beta E_q}+e^{-2\beta E_q}}\;.
\label{fer2c}
\eqa
At low temperature, $\Phi\approx0$ and so Eq.~(\ref{fer2c}) 
describes excitations with
energy $2E_q$, i.e. that of a bound state.\footnote{For $N_c=2$,
two (anti)quarks can form a color singlet and belongs to the same
multiplet as the usual three quark anti-quark bound states. These states
are the "baryons'' of two-color QCD. 
} Again this is referred to as statistical confinement.
At high temperature, $\Phi\approx1$ and Eq.~(\ref{fer2c}) 
describes excitations with energy $E_q$, i.e. deconfined quarks.

The Polyakov loop potential in the gauge sector
used is~\cite{bruuner}
\bqa\nonumber
\Omega_{\rm gauge}&=&-bT\left[
24\Phi^2e^{-\beta a}+\ln(1-\Phi^2)
\right]\;,
\\ &&
\label{ppot}
\eqa
where $a$ and $b$ are constants. This form is motivated
by the lattice strong coupling expansion~\cite{polyakovpot4} 	 	
In the gauge theory without dynamical 
quarks, one can find an explicit expression for the Polyakov-loop variable
$\Phi=\sqrt{1-{1\over24}e^{-\beta a}}$ as a function of $T$
and so $a=T_c\ln24$. Moreover, $\Phi$ goes to zero in a continuous
manner and the theory exhibits a second-order transition.

A few remarks about the parameters in 2-color QCD are in order.
For $N_c=2$, there are no experimental results to guide us in the
determination of the parameters. A common way of determining them is to 
use $N_c$ scaling arguments~\cite{bruuner}. 
The pion decay constant scales as $\sqrt{N_c}$
and the pion mass scales as $N_c$. This yields the two parameters
$f_{\pi}=75.4$ MeV and $m_{\pi}=93.3$  MeV.

\begin{figure}
\begin{center}
\setlength{\unitlength}{1mm}
\includegraphics[width=7.0cm]{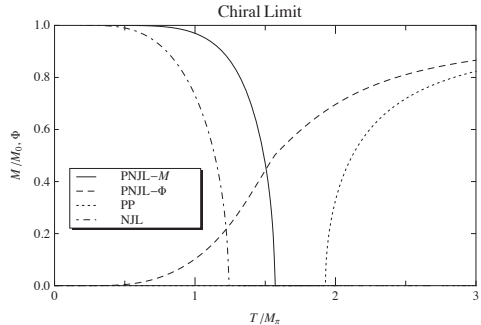}
\caption{Normalized constituent quark mass and Polyakov loop in the chiral 
limit as a function of $T/m_{\pi}$, where $m_{\pi}=140$ MeV 
is the pion mass for $N_c=3$. 
Figure taken from~\textcite{amador}.}
\label{kopling}
\end{center}
\end{figure}

\begin{figure}
\begin{center}
\setlength{\unitlength}{1mm}
\includegraphics[width=7.0cm]{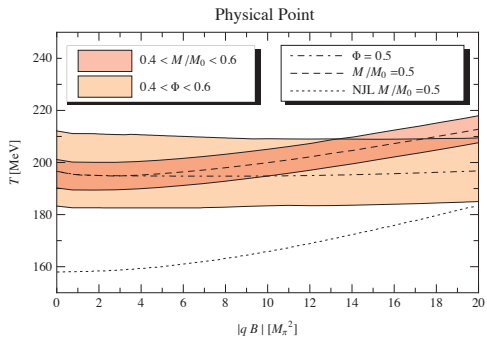}
\caption{Transition temperatures for the 
deconfinement and the chiral transition in the 
PNJL model as a function of $|qB|/m_{\pi}^2$ in 2-color QCD.
Figure taken from~\textcite{amador}.}
\label{2ctc}
\end{center}
\end{figure}

In Fig.~\ref{kopling}, we show the Polyakov loop as a function of
$T/m_{\pi}$ (with $m_{\pi}=140$ MeV)
found by minimizing the Polyakov loop potential~(\ref{ppot})
(dotted line). We also show the normalized quark condensate
obtained in the NJL model (dashed-dotted line).
The dashed and solid lines show the Polyakov loop and
normalized quark condensate as functions of $T/m_{\pi}$
in the PNJL model. 
From the figure. we see that the coupling between the two variables,
forces the 
curve for the normalized quark condensate to the right and the
curve for the Polyakov to the left so that the two transitons have a
common transition temperature.
\\
\\
\indent
In Fig.~\ref{2ctc}, we show the transition temperatures 
for the chiral and deconfinement transitions as a function of $|qB|/m_{\pi}^2$
where $m_{\pi}=140$ MeV is the physical pion mass for $N_c=3$.
The band shows the values of the order parameters 
$0.4<M/M_0<0.6$ and $0.4<\Phi<0.6$, where $M_0$ is the chiral condensate
at $T=0$.
For comparison we also show the chiral transition for NJL model.
The transitions coincide for $B=0$, cf. Fig.~\ref{kopling}
but split at finite $B$. Note the similarity with the curves
in Fig.~\ref{ruggi} and
that the deconfinement temperature is almost independent of
temperature.

\section{Functional renormalization group}
\label{funxio}
The functional renormalization group (FRG)~\cite{wetterich93} is a powerful 
nonperturbative method that has gained popularity since its formulation more
than two
decades ago. It is one way of implementing the 
renormalization group ideas of Wilson from the early 1970s.
The average effective action, which is denoted by $\Gamma_k[\phi]$
is a function of a set of fields collective denoted by $\phi$.
The subscript $k$ indicates that the effective action is a function
of a momentum scale. 

This sliding scale acts as an infrared
cutoff, such that all momenta $q$ between $k$ and the ultraviolet
cutoff of the theory, $\Lambda$, have been integrated out.
At $k=\Lambda$, no momenta have been integrated out and 
the effection action equals the bare action,
$\Gamma_{\Lambda}[\phi]=S[\phi]$. Thus the value $S[\phi]$
is the boundary
condition for the effective action.  Moreover, when $k=0$, all the
quantum and thermal fluctuations have been integrated out and
$\Gamma_0[\phi]$ is equal to the full quantum effective action.
The average effective action satisfies a
functional integro-differential equation, which reads
\begin{widetext}
\bqa
\partial_k\Gamma_k[\phi]
&=&
{1\over2}{\rm Tr}\left[
\partial_kR_k^B(p)\left[{\Gamma_k^{(2)}+R_k^B(p)}\right]^{-1}_{p,-p}
\right]
-{\rm Tr}\left[\partial_kR_k^F(p)\left[\Gamma_k^{(2)}+R_k^F(p)\right]^{-1}_{p,-p}
\right]\,,
\label{flow}
\eqa
where the superscript $n$ means the $n$th functional derivative
of $\Gamma_k[\phi]$. Moreover, the trace is over the variable $p$, 
which includes spacetime, field indices, and Dirac indices.
The socalled regulator functions 
$R_k^B(p)$ and $R_k^F(p)$ are added to implement the renormalization group
ideas mentioned above. These functions are large for $p<k$
and small for $p>k$ if $0<k<\Lambda$. The regulator functions also satisfy 
$R_{\Lambda}^B(p)=R^F_{\Lambda}(p)=\infty$. These properties guarantee
that the modes with $q<k$ are heavy and decouple and only the
modes $q$ between the sliding scale $k$ and the ultraviolet cutoff $\Lambda$
are light and integrated out. 
We will return to the choice of regulator functions below.

\begin{figure}
\vspace{2mm}
\setlength{\unitlength}{1mm}
\hspace{-2.cm}
\includegraphics[width=4.7cm]{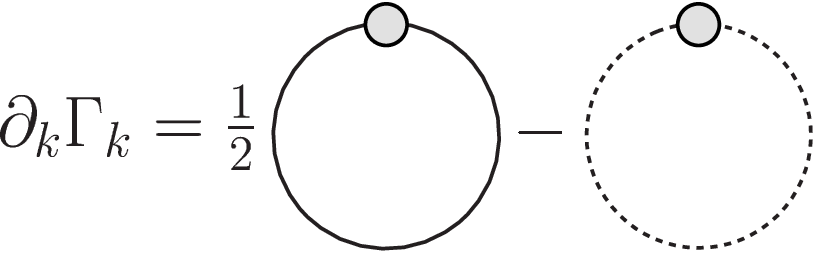}
\caption{Diagrammatic representation of the exact flow equation for 
the effective action $\Gamma_k[\phi]$. The lines denote
the exact field-dependent propagators for bosons (solid)
and fermions (dashed), while the circle denotes the insertion of the 
regulator function $R_k^B(p)$ or $R_k^F(p)$.}
\end{figure}

\subsection{Local-potential approximation}
Of course one cannot solve the flow equation exactly - then one would
have solved the theory exactly - and so one needs to make 
approximations. A framework for systematic approximations is the 
derivative expansion. 
The leading-order approximation in the derivative expansion is called
the local-potential approximation (LPA) since the full quantum effective
action is approximated by the action
\bqa
\Gamma[\phi]&=&\int_0^{\beta}d\tau\,\int d^3\left\{
{1\over2}(\partial_{\mu}\sigma)^2+
{1\over2}(\partial_{\mu}{\boldsymbol \pi})^2
+U_k(|v|,|\Delta|)
\right\}\;,
\eqa
where $U_k(|v|,|\Delta|)$ is a $k$-dependent local potential
and $v={1\over\sqrt{2}}(\sigma+i\gamma_5\pi^0)$
and $\Delta={1\over\sqrt{2}}(\pi_1+i\pi_2)$.
Thus the potential depends on two $O(2)$
invariants in accordance with the
discussion in Sec.~\ref{qmsec}.
We are not including a pion condensate and therefore the local potential
is evaluated at $|\Delta|=0$. However, the flow equation still depends
on the two partial derivatives ${\partial U_k\over\partial|v|}$
and ${\partial U_k\over\partial|\Delta|}$.
In the mean-field approximation, these partial derivatives are identical, but 
beyond they 
generally are not. In order to make the flow equation numerically
tractable,~\textcite{robb} made the approximation that they are equal.


Using the chain rule, the matrix appearing in the first
term in Eq.~(\ref{flow})
\bqa
\left[{\Gamma_k^{(2)}+R_k^B(p)}\right]^{-1}_{p,-p}
&=&
\left(\begin{array}{cccc}
\hspace{2mm}p^2+R_k^B(p)
+U_k^{\prime}+2\rho U_k^{\prime\prime}&0&0&0\\
&&&\\
0&\hspace{2mm}p^2+R_k^B(p)+U_k^{\prime}&0&0\\
0&0&
....\hspace{2mm}
&0\\
0&0&0&....\hspace{2mm}
\end{array}\right)\;,
\label{matrix2}
\eqa
where 
the .... indicates  $p^2+R_k^B(p)+U_k^{\prime}$,
$U_k^{\prime}={\partial U_k\over\partial\rho}$ and
$U_k^{\prime\prime}={\partial^2U_k\over\partial\rho^2}$
with $\rho={1\over2}\phi^2$.
We notice that the matrix Eq.~(\ref{matrix2}) is simply the
inverse tree-level propagator if we make the substitution
${\cal V}_0\rightarrow U_k+R_k^B(p)$.
The second term in Eq.~(\ref{matrix2}) has the form of an inverse
fermion tree-level propagator with a similar substitution.
We will use a modification of the regulators~\cite{litim,stok}
\bqa
\label{reg1}
R^B_k(p)&=&(k^2-{\bf p}^2)\theta(k^2-{\bf p}^2)\;,\\
R_k^F(p)&=&
\left(
\sqrt{{p_0^2+k^2\over p_0^2+{\bf p}^2}}-1\right)
/\!\!\!p
\theta(k^2-{\bf p}^2)\;.
\label{reg2}
\eqa
Knowing the spectrum in a constant magnetic field, 
we modify the regulators by making the replacements
${\bf p^2}\rightarrow p_z^2+(2n+1)|qB|$
and ${\bf p^2}\rightarrow p_z^2+(2n+1-s)|q_fB|$ above.
Note that we here and in the remainder of this section denote the
Landau levels by $n$ so there is no confusion with the sliding scale
$k$. 
The regulators Eq.~(\ref{reg1})--(\ref{reg2}) are 
very  convenient in practical
calculations since we can carry out the integral over 
the momentum $p_z$. The integro-differential flow
equation reduces to a partial differential equation that is easier
to solve numerically. 
Integrating over $p_z$ and summing over Matsubara frequencies $P_0$,
the flow equation can be written as~\cite{skokov}
\bqa\nonumber
\partial_k U_k
&=&
{k^4\over12\pi^2}
\left[
{1\over\omega_{1,k}}\left(1+2n_B(\omega_{1,k}\right)
+{1\over\omega_{2,k}}\left(1+2n_B(\omega_{2,k}\right)
\right]
\\ && \nonumber
+{|qB|\over2\pi^2}\sum_{n=0}^{\infty}
{k\over\omega_{1,k}}
\sqrt{k^2-p^2_{\perp}(q,n,0)}\,\theta\left(k^2-p^2_{\perp}(q,n,0)\right)
\left[1+2n_B(\omega_{1,k})\right]
\\ &&
-{N_c\over2\pi^2}\sum_{s,f,n=0}^{\infty}{|q_fB|k\over\omega_{q,k}}
\sqrt{k^2-p^2_{\perp}(q_f,n,s)}\,\theta\left(k^2-p^2_{\perp}(q_f,n,s)\right)
\left[1-2n_F(\omega_{q_f,k})
)\right]\;,
\label{flowb}
\eqa
where we have defined $\omega_{1,k}=\sqrt{k^2+U^{\prime}}$,
$\omega_{2,k}=\sqrt{k^2+U^{\prime}+2\rho U^{\prime}}$,
$\omega_{q,k}=\sqrt{k^2+2g^2\rho}$,
$p^2_{\perp}(q,n,s)=(2n+1-s)|qB|$.

In the limit $B\rightarrow0$, the sum over Landau levels becomes an integral
via defining the variable $p_{\perp}^2=2|qB|m$ which yields 
$p_{\perp}dp_{\perp}=|qB|dm$. Replacing the sum by an integral
one finds the flow equation first derived by~\textcite{stok}
\bqa\nonumber
\partial_k U_k
&=&
{k^4\over12\pi^2}
\left[
{1\over\omega_{1,k}}\left(1+2n_B(\omega_{1,k}\right)
+{1\over\omega_{2,k}}\left(1+2n_B(\omega_{2,k})\right)
\right]
+{k\over2\omega_{1,k}\pi^2}\int_0^{\infty}dp_{\perp}p_{\perp}
\sqrt{k^2-p^2_{\perp}}
\\ && \nonumber\times
\theta\left(k^2-p^2_{\perp}\right)
\left[1+2n_B(\omega_{1,k})\right]
-{N_cN_fk\over\omega_{q,k}\pi^2}\int_0^{\infty}dp_{\perp}p_{\perp}
\sqrt{k^2-p^2_{\perp}}
\theta\left(k^2-p^2_{\perp}\right)
\left[1-2n_F(\omega_{k})
\right]  \\ \nonumber
&=&
{k^4\over12\pi^2}
\left\{
{3\over\omega_{1,k}}\left[1+2n_B(\omega_{1,k})\right]
+{1\over\omega_{2,k}}\left[1+2n_B(\omega_{2,k})\right]
\right\}
\\ &&
-{N_cN_fk^4\over3\pi^2}
\left\{\left[
{1\over\omega_{q,k}}\left(1-2n_F(\omega_{q,k}\right)
\right]\right\}\;.
\eqa
\end{widetext}
Sometimes, one defines a so-called extended mean-field equation by
omitting the bosonic terms on the right-hand side of the flow equation.
Then the terms that depend on the derivatives of $U_k$ on the right-hand
side drop out and one can formally integrate the flow equation to 
obtain the effective potential~\cite{smekkers}. For $T=0$, this
can be done analytically even for nonzero magnetic field $B$.

The $k$-dependent minimum $f_{\pi,k}$ is found by solving
\bqa
{\partial U_k\over\partial\phi}\bigg|_{\phi=f_{\pi,k}}&=&h\;,
\label{kdep}
\eqa
i.e by minimizing the modified effective potential $\tilde{U}_k=U_k-h\phi$.
The $k$-dependent masses $m_{\pi,k}^2$ and $m_{\sigma,k}^2$
can be expressed in terms of the second derivatives of the $k$-dependent
(modified) effective potential at the $k$-dependent
minimum $f_{\pi,k}$ as follows
\bqa
\label{kpi}
m_{\pi,k}^2&=&{\partial\tilde{U}_k\over\partial\rho}\bigg|_{\phi=f_{\pi,k}}\;,\\
m_{\sigma,k}^2&=&m_{\pi}^2+
\rho{\partial^2\tilde{U}_k\over\partial\rho^2}\bigg|_{\phi=f_{\pi,k}}\;.
\eqa
Combining Eqs.~(\ref{kdep}) and~(\ref{kpi}), we find
$f_{\pi,k}m_{\pi,k}^2=h$. For $k=h=0$, this is Goldstone's theorem.

The boundary condition for the effective potential 
at $k=\Lambda$ is chosen to have the form
\bqa
U_{\Lambda}&=&{1\over2}m_{\lambda}\phi^2
+{\lambda_{\Lambda}\over24}\phi^4\;.
\eqa
The bare parameters $m_{\Lambda}^2$ and $\lambda_{\Lambda}$
are tuned such that one obtains the correct pion mass and
pion decay constant in the vacuum at $k=0$. 
In the calculation we use an ultraviolet cutoff $\Lambda=800$ MeV, although 
the results are not too sensitive to the exact value.
We ignore the running of the Yukawa coupling and set $g=g_k=3.2258$
for all values of $k$. This gives a constituent quark mass of 
$m_q=g\phi_0=gf_{\pi}=300$ MeV. 
Further details of the numerical implementation can be found 
in~\textcite{anders,robb}.

\begin{figure}
\begin{center}
\setlength{\unitlength}{1mm}
\includegraphics[width=6.0cm]{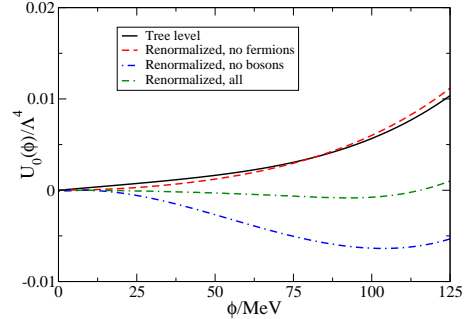}
\caption{Tree-level potential and renormalized potential 
$U_{0}(\phi)$
in different approximations.}
\label{renormeringi}
\end{center}
\end{figure}

In Fig.~\ref{renormeringi}, we show the 
effective potential $U_0(\phi)$ normalized to $\Lambda^4$ in various 
approximations. The black curve is the tree-level potential i.e.
the boundary condition $U_{\Lambda}(\phi)$ that gives the correct 
quantum effective potential given by the green curve.
In order to investigate the effects of the different terms in the
flow equation, we have solved it with the same boundary condition, but
omitted the bosonic terms (blue curve) and omitted the fermionic terms
(red curve). 
We see that the bosonic vacuum fluctuations have a tendency to decrease
symmetry breaking in the vacuum while fermions vacuum fluctuations
have a tendency to enhance symmetry breaking.
Thus we have competition between the two terms in the flow equation
and their relative importance depends on the momentum scale
$k$. 

\begin{figure}
\begin{center}
\setlength{\unitlength}{1mm}
\includegraphics[width=6.0cm]{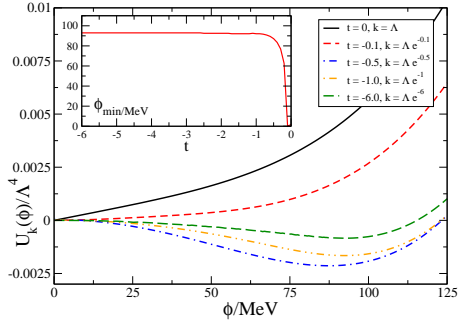}
\caption{Effective potential $U_k(\phi)$ for different values of 
$t=\ln{\Lambda\over k}$. Inset: $t$-dependent minimum of $U_k(\phi)$.
See main text for details.}
\label{renormering2}
\end{center}
\end{figure}

In Fig.~\ref{renormering2}, we show the scale-dependent effective
potential $U_k(\phi)$ for different values of $t=\ln{\Lambda\over k}$.
The black curve is the tree-level potential and the green line is the
fully renormalized potential $U_{0}(\phi)$.
It is interesting to note that the potential at intermediate stages, 
here shown as the red, blue, and orange
curves, do not evolve monotonically. This shows that the bosonic and fermionic 
terms in the flow
equation dominate in different regions of integration.
The inset shows how the $k$-dependent minimum $f_{\pi,k}$ expressed
in terms of dimensionless variable $t=\ln{k\over\Lambda}$.
We see that the minimum mainly gets renormalized 
for $t$ between $t=0$ and $t=-2$, whereafter it levels off.

\begin{figure}
\setlength{\unitlength}{1mm}
\includegraphics[width=7.0cm]{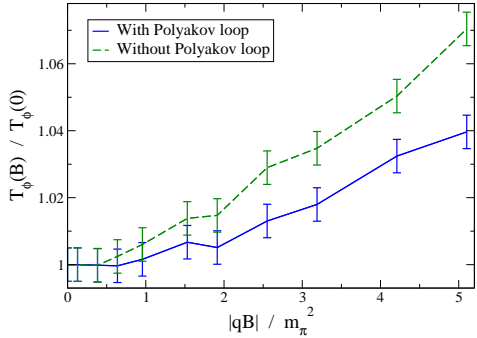}
\caption{Normalized critical temperature $T_{\phi}(B)/T_{\phi}(0)$
for the chiral transition as a function
of the magnetic field $B$ with and without the inclusion of the Polyakov loop.}
\label{tcB}
\end{figure}

In Fig.~\ref{tcB}, we show the critical temperature as a function of the
magnetic field $B$ at the physical point.
The dashed curve is the quark-meson model and solid curve is 
the Polyakov-loop extended quark-meson model.
In agreement with various mean-field results, the critical temperature
is an increasing function of $B$. It is interesting to note that the
coupling to the Polyakov-loop variable $\Phi$ 
lowers $T_c$. This is an interesting observation, in particular  
since the Polyakov loop has no influence on magnetic catalysis 
at zero temperature.
This effect can be understood by calculating the free energy density
in a given background $\phi_1$ and comparing it with 
the free energy density
in the 
deconfining background $\phi_1=0$
~\cite{endrodi3}.\footnote{Recall $\Phi={1\over3}[1+2\cos(\phi_1)]=1$ for 
$\phi_1=0$.}
The difference between these two free energy densities
is
\bqa\nonumber
\Delta{\cal F}&=&{|q_fB|\over\pi^2}
\int_0^{\infty}{ds\over s^2}
e^{-m^2s}
\coth(|q_fB|s)
\\ && \nonumber
\left[\theta_3(\phi_1+\mbox{$1\over2$}\pi,e^{-{1\over4sT^2}})
-\theta_3(\mbox{$1\over2$}\pi,e^{-{1\over4sT^2}})\right]
\;,
\\ &&
\eqa
where the elliptic theta function 
$\theta_3(u,q)$ is defined by
\bqa
\theta_3(u,q)&=&1+2\sum_{n=1}^{\infty}q^{n^2}\cos(2nu)\;.
\eqa
The function $|q_fB|\coth(|q_fB|s)$ increases with $s$
for all values of $s$. Thus a magnetic field favors deconfined Polyakov loops
and therefore tends to lower the transition temperature.
We also note that this effect decreases with larger quark masses $m$.

We are not aware of any mean-field calculations that directly
compare the chiral transition temperature with and without the Polyakov 
loop.\footnote{There are of course many mean-field calculations with 
and without the Polyakov loop, but a comparison between them requires that
physical observables in the vacuum are the same.}
However, based on the above argument as well as the renormalization 
group calculations, we expect to see the same behavior in the 
mean-field 
approximation.

\subsection{Beyond the LPA}
The results we have been discussing so far are obtained using the
local-potential approximation. In a recent paper,~\textcite{kami}
go beyond the LPA by including the 
wave-functional renormalization terms $Z_{\perp}$ and $Z_{\parallel}$.
In order to avoid the complication of having two invariants
$\rho$ and $\Delta$ on which the effective potential $U_k$ depends, they
consider the case $N_f=1$. In this case, the symmetry is
$U(1)_V\times U(1)_A$ in the chiral limit or $U(1)_V$
at the physical point. Either way, 
the quark condensate gives rise to a single (pseudo)Goldstone boson.
One expects the charged pions to decouple for sufficiently large magnetic
fields since they become heavy and therefore the $N_f=2$ model essentially
reduces to the model they considered. Moreover, their calculations
were at the physical point.

\begin{figure}
\setlength{\unitlength}{1mm}
\includegraphics[width=7.0cm]{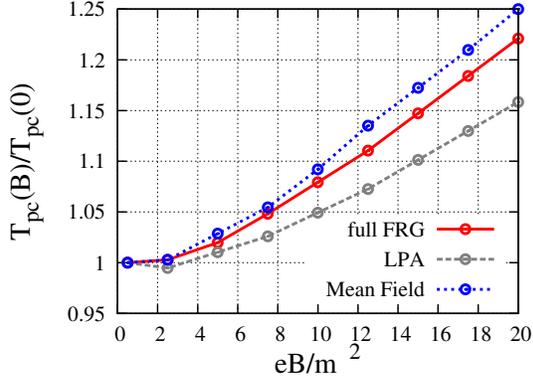}
\caption{The 
transition temperature for the chiral transition as a function
of $|qB|$ for three different approximations
normalized to the transition temperature at $B=0$.
Figure taken from~\textcite{kami}.
}
\label{tcB2}
\end{figure}

One of the interesting aspects
of this work is the systematic study of 
the various approximations. For example, they studied
the transition temperature in the mean-field approximation, 
using the LPA, and beyond the LPA. The transition temperature
is determined by the peak of $dM_q/dT$, where $M_q$
is the constituent quark mass.
The result is shown
in Fig.~\ref{tcB2}, where the blue, grey, and red lines show the
normalized transition temperature in
the three approximations. The inclusion of the
mesonic fluctuations lowers the transition temperature 
compared to the mean-field
approximation, 
while the inclusion of wavefunction renormalization
effects increases the slope somewhat.
It would be of interest to see the effects of including
the Polyakov loop as well. 

\begin{figure}
\setlength{\unitlength}{1mm}
\includegraphics[width=7.0cm]{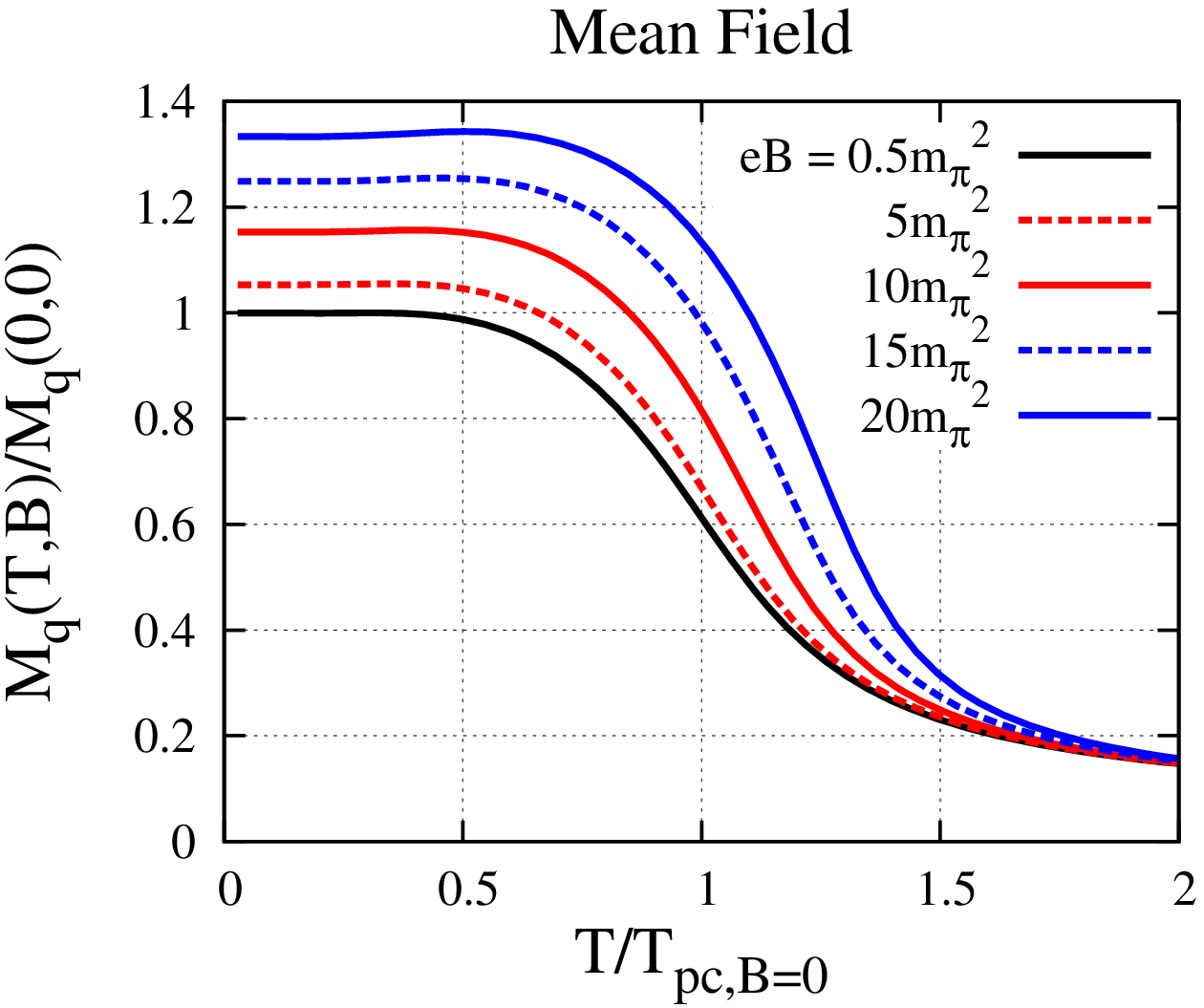}
\includegraphics[width=7.0cm]{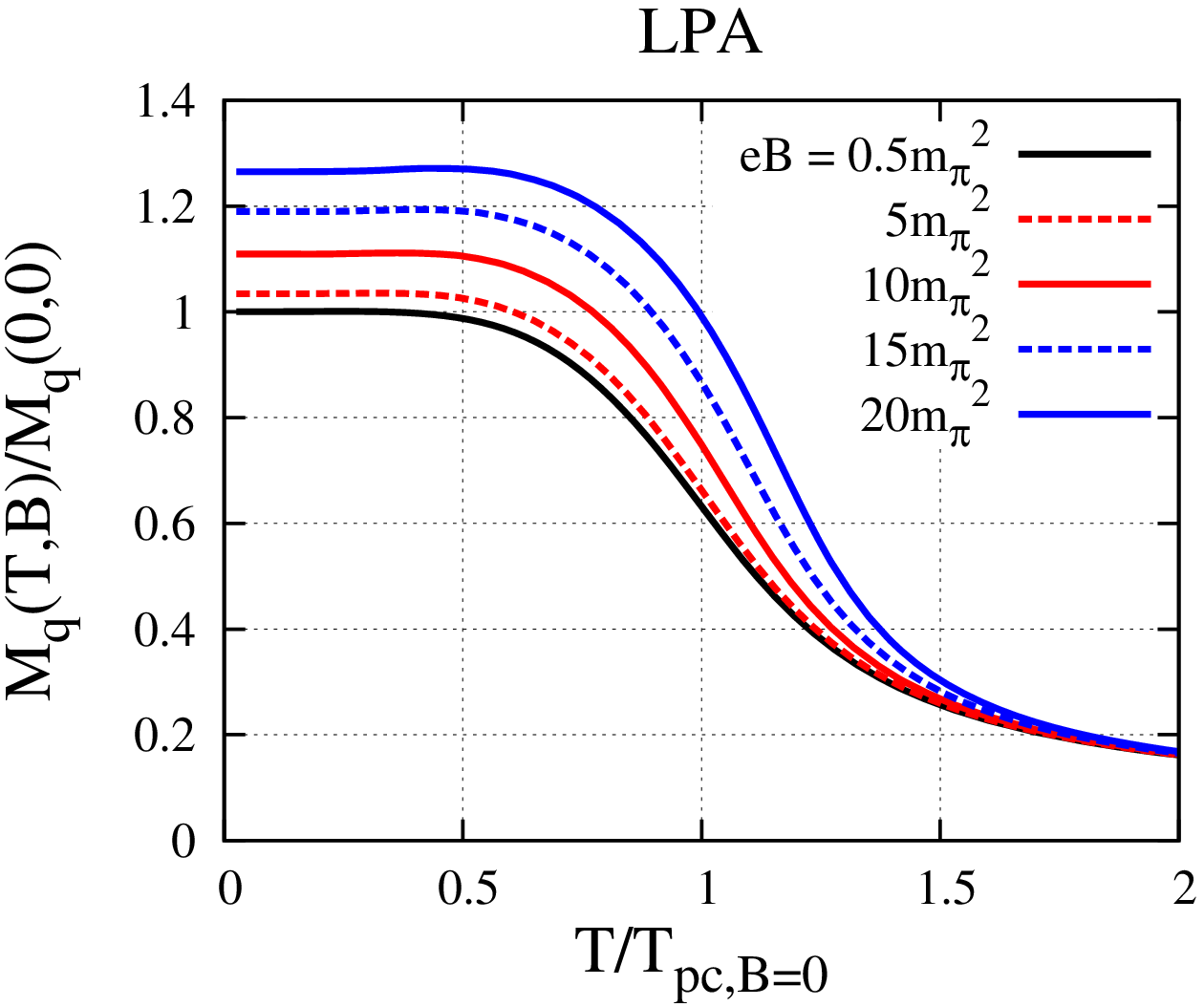}
\includegraphics[width=7.0cm]{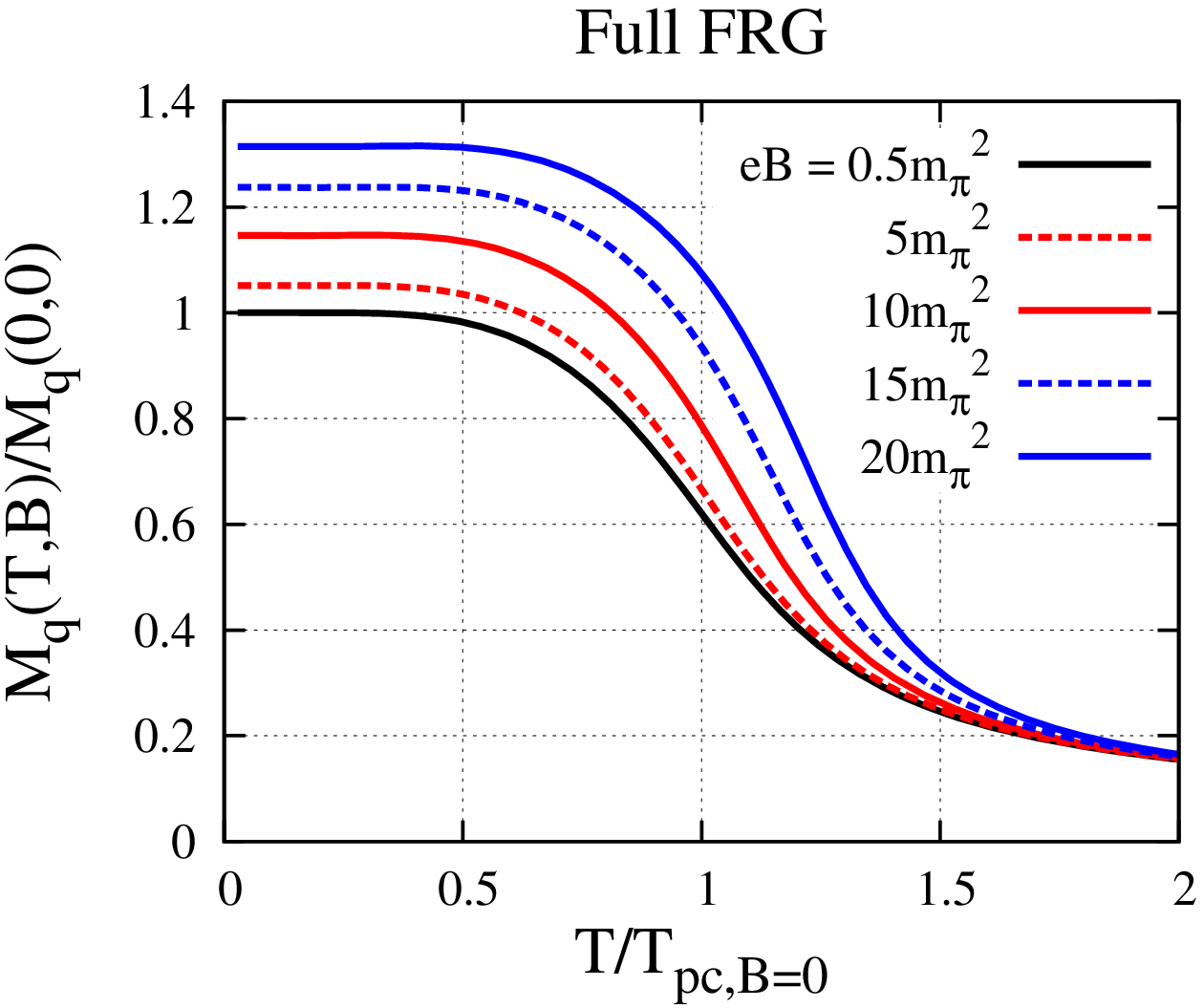}
\caption{Normalized constituent quark mass as a function of the 
normalized temperature 
for the three different approximations
and different values $|qB|$.
Figures taken from~\textcite{kami}.}
\label{mqB}
\end{figure}

The constituent quark mass normalized
to the constituent quark mass for $B=0$ as a function of the temperature
normalized to the transition temperature $T_{\rm pc,B=0}$
for $B=0$ for the three
different approximations is shown in Fig.~\ref{mqB}. 
For all temperatures, we see that
the constituent quark mass is an increasing function of the magnetic field,
thus the system shows magnetic catalysis. This is the reason for the increase 
of the transition temperature as a function of $B$ displayed in Fig.~\ref{tcB2}.
We note that magnetic catalysis is less pronounced
in the LPA as compared to the
mean-field approximation and this can probably be attributed to the
mesonic fluctuations that tend to counteract symmetry breaking
(c.f. Fig.~\ref{renormeringi}). The inclusion of the wavefunction
renormalizaton terms increases magnetic catalysis as a function of $B$
such that the transition temperature lies between the mean-field and the LPA
curves. Again, it would be of interest to see the effects of adding the
Polyakov loop variable.

We next consider the wavefunction renormalization terms $Z_k^{\parallel}$
and $Z_k^{\perp}$. 
The regulator functions chosen are the anisotropic functions
\bqa
R_k^B(p)&=&(k^2-p_z^2)Z_{\parallel}\theta(k^2-p_z^2)\\
R_k^F(p)&=&-/\!\!\!p_z\left(\mbox{$k\over |p_z|$}-1\right)\theta(k^2-p_z^2)\;,
\eqa
where $/\!\!\!p_z=\gamma^3p_z$.
The regulators clearly break rotational invariance also for $B=0$.
However, they give rise to a very simple scale-dependent fermion
propagator and so it is very useful for practical calculations.
The boundary condition for the wavefunction renormalization
terms at $k=\Lambda$ and $B=0$ is $Z_{k=\Lambda}^{\parallel}=Z_{k=\Lambda}^{\perp}=1$.
Due to the $O(4)$ symmetry at $B=0$, 
we have $Z_{k=0}^{\parallel}=Z_{k=0}^{\perp}$.
However, due to the above-mentioned breaking of rotational invariance,
the authors instead fine-tuned the
$Z_{k=\Lambda}^{\parallel}$ and $Z_{k=\Lambda}^{\perp}$, such that
$Z_{k=0}^{\parallel}=Z_{k=0}^{\perp}$ at $T=3$ MeV and $|qB|=0.5m_{\pi}^2$.
This gives the values $Z_{k=\Lambda}^{\parallel}=0.002$ and
$Z_{k=\Lambda}^{\perp}=0.236$, respectively.

\begin{figure}
\setlength{\unitlength}{1mm}
\includegraphics[width=7.0cm]{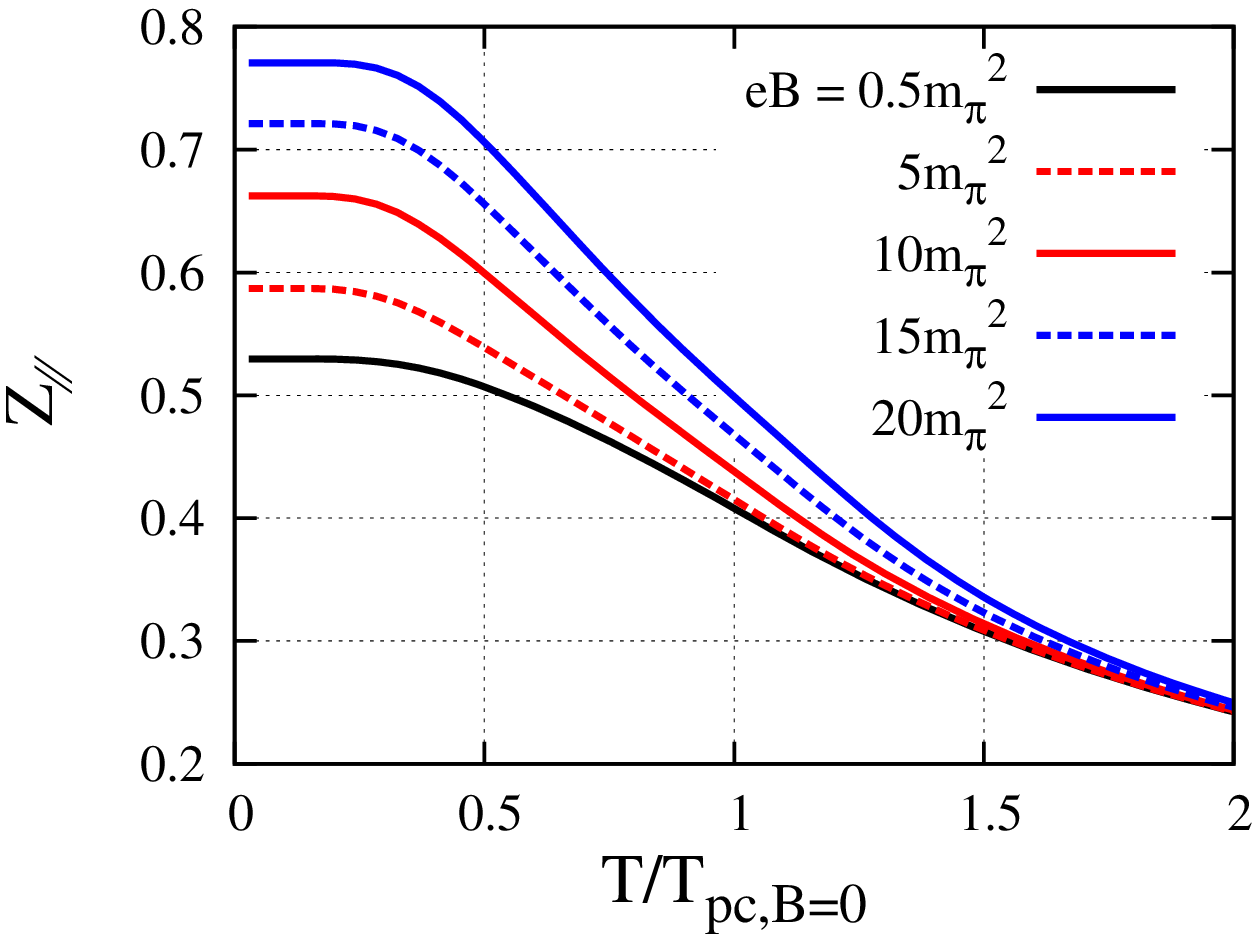}
\includegraphics[width=7.0cm]{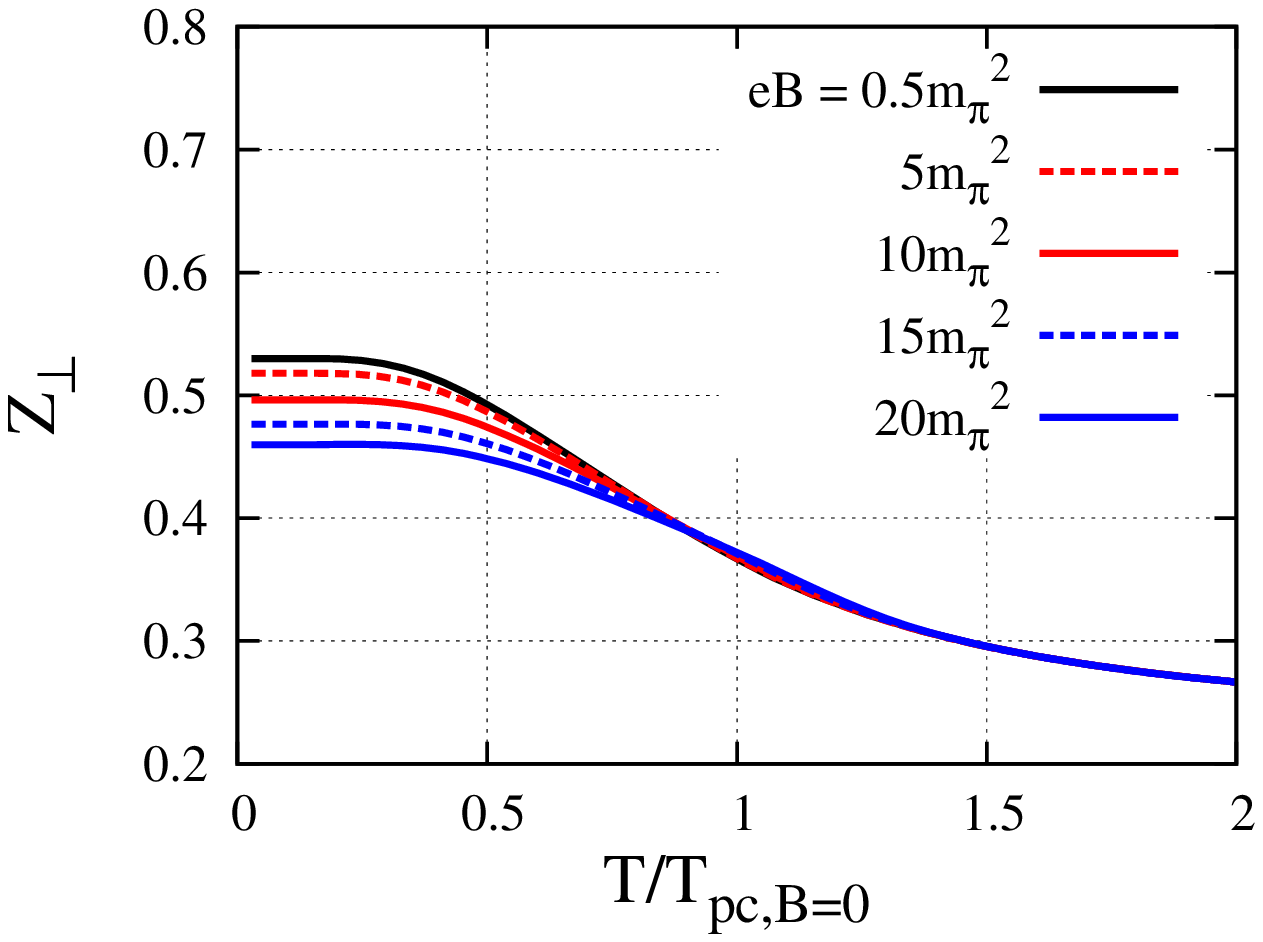}
\caption{Wavefunction renormalization terms $Z^{\parallel}_{k=0}$
and $Z^{\perp}_{k=0}$ as a function of the normalized temperature for
various magnetic fields. Figures taken from~\textcite{kami}.}
\label{zren}
\end{figure}

The wavefunction renormalization terms $Z^{\parallel}$
and $Z^{\perp}$ as functions of the normalized temperature for
various strengths of the magnetic field is shown in Fig.~\ref{zren}.
We first notice that while $Z^{\parallel}_{k=0}$ increases with the magnetic
field, $Z_{k=0}^{\perp}$ decreases.
This can probably 
be attributed to the fact that the flow equation of the former
has an explicit $B$-dependence, while the flow equation of the latter does not.
Secondly, all curves meet for sufficiently large temperatures
and that the curves for $Z_{\perp}$ have done so already before the
transition temperature. 

\section{Magnetic catalysis}
\label{magcal}
In this section, we discuss magnetic catalysis at $T=0$.
Magnetic catalysis is the effect that either
\begin{quote}
(1) The magnitude of a condensate is enhanced by the 
presence of an external magnetic field $B$ if the condensate is already present
for zero magnetic field.

(2) An external magnetic field induces symmetry breaking and the
appearence of a condensate when the symmetry is intact for $B=0$.

\end{quote}
Case (2) is also referred to as dynamical symmetry breaking 
by a magnetic field.
In context of low-energy effective theories of QCD, 
the condensate is the nonzero expectation value
of the sigma field or the quark condensate. 
The early works on magnetic catalysis date
back to the late 80s and early 90s, and focused on the
NJL model in 2+1 dimensions~\cite{konstantin1,konstantin2,konstantin3}
and in 3+1 dimensions~\cite{klevansky2,ebert20,konstantin4},
and QED~\cite{shovcat1}.
Other applications are in QCD~\cite{shovmir,ozaki} and the Walecka model in 
nuclear physics~\cite{andreas}.
Magnetic catalysis is now considered a generic feature of matter
in an external magnetic field~\cite{shovrev}.

Inspecting the dispersion relation for fermions in a magnetic 
field, $E_k=\sqrt{m_f^2+p_z^2+(2k+1-s)|q_fB|}$, we see that it
resembles the dispersion relation for a massive particle in one spatial
dimension with an effective mass $M^2_{\rm eff}=m_f^2+(2k+s-1)|q_fB|$.
Only for the lowest Landau level, is this effective mass independent
of the magnetic field. When the fermionic mass scale is much smaller
than the magnetic mass scale, $m_f^2\ll|q_fB|$, 
the higher Landau levels decouple from the long low-energy dynamics
and the long-distance behavior is determined by the lowest Landau level.
Since the particles in the lowest Landau level essentially are confined
to move along the magnetic field, i.e. the $z$-axis, 
the system becomes effectively one-dimensional and the system
exhibits dimensional reduction, $D=3+1\rightarrow1+1$.
The $1+1$-dimensional character of the lowest Landau at low momentum
can also be inferred from the form of fermion propagator given by
Eq.~(\ref{decomp}). Isolating the $k=0$ contribution, we find 
\begin{widetext}
\bqa
\tilde{S}_0(p)&=&i\exp\left(-\mbox{$p_{\perp}^2\over|q_fB|$}\right)
{\gamma^0p_0-\gamma^3p_3+m\over p_0^2-p_3^2-m_f^2}
\left[
1-is_{\perp}\gamma^1\gamma^2
\right]\;,
\eqa
where we have used $L_{-1}^a(x)=0$.
Note that dimensional reduction is not taking place for bosons as 
the ground-state energy is not vanishingly small compared to the
energy of the first excited state, $k=0$ and $k=1$ in Eq.~(\ref{ebos}).

At this point, a few remarks on dimensional reduction and 
spontaneous symmetry breaking are in order.
We have seen that a magnetic field enhances (spontaneous) symmetry
breaking as well as reduces the system to being essentially 1+1 dimensional.
However, we know from the Coleman theorem that there is 
no spontaneous symmetry breaking of a continuous symmetry in 1+1 dimension
and therefore no massless Nambu-Goldstone boson can 
exist~\cite{coleman}.\footnote{This applies to massless 
excitations that are linear in the momentum $p$ for small $p$.
Magnons are massless excitations in ferromagnets that 
are quadratic in the momentum $p$ for small $p$
and exist in 1+1 dimension. Linear Goldstone modes exist in 2+1
dimensions at $T=0$. See e.g.~\textcite{watanabe} for a detailed discussion.}
The point here is that $\langle\bar{\psi}\psi\rangle$
is neutral with respect to the magnetic field 
and that the Goldstone boson $\pi^0$ is a neutral 
excitation with respect
to the magnetic field~\cite{shovcat2}. The motion of the center of mass of
$\pi^0$ is not restricted to being along the magnetic field as it is an 
electrically neutral particle.

Let us discuss the NJL model first.
For simplicity, we consider the case $N_c=N_f=1$.
If we denote the quark condensate by $M$, the mean-field contribution
to the free energy density is given by ${M^2\over2G}$, 
cf. the first two terms in Eq.~(\ref{njleffpot1}) with $M_0=M_3=M$ for $c=0$.
Using a four-dimensional ultraviolet cutoff $\Lambda$, 
the one-loop contribution to the 
effective potential for $B=T=0$ is given by Eq.~(\ref{4dcut}).
In the limit $M\ll\Lambda$, we find
\bqa
{\cal F}&=&{M^2\over2G}+{1\over(4\pi)^2}\bigg[
{1\over2}\Lambda^4-2\Lambda^2M^2
+{1\over2}M^4+M^4\ln{\Lambda^2\over M^2}
\bigg]\;.
\eqa
The minimum is found by solving the gap equation, which reads
\bqa
M\left[{4\pi^2\over G}-\Lambda^2+M^2\ln{\Lambda^2\over M^2}
\right]&=&0\;.
\eqa
$M=0$ is always a solution. However a nontrivial solution exists
for $G>G_c={4\pi^2\over\Lambda^2}$.
Hence, for couplings larger than the critical value $G_c$, quantum fluctuations
induce symmetry breaking in the model.
The possible solutions to the gap equation in a constant magnetic field
were first considered by~\textcite{klevansky2}.
For finite magnetic field, the gap equation is
\bqa
{4\pi^2\over G}-\Lambda^2+M^2\ln\left({\Lambda^2\over M^2}\right)
-{|2q_fB|\over(4\pi)^2}\left[
\zeta^{(1,0)}(0,x_f)+x_f-{1\over2}(2x_f-1)\ln x_f
\right]&=&0\;,
\eqa
where $x_f={M^2\over2|q_fB|}$ as before.
For nonzero magnetic field $B$ and any $G$, this equation has only a nonzero 
solution for $M$. 
Consequenly, for $G<G_c$,
a nonzero magnetic field induces symmetry breaking
when the symmetry is intact for $B=0$.
This effect was first observed in the context of the NJL model 
in 2+1 dimensions by~\textcite{konstantin1,konstantin2,konstantin3}.
For $G<G_c$, one finds~\cite{shovcat2}
\bqa
M^2&=&
{|q_fB|\over\pi}\exp\left[
-{1\over|q_fB|}\left({4\pi^2\over G}-\Lambda^2\right)\right]\;.
\label{functional}
\eqa
\end{widetext}
The gap vanishes in the limit $|q_fB|\rightarrow0$ as it should.
Moreover, Eq.~(\ref{functional}) has an essential singularity at $G=0$, which
shows its nonperturbative
nature: i.e.~it is obtained by summing Feynman graphs from all orders
of perturbation theory. Any finite-order perturbative calculation yields
a vanishing gap.\footnote{Recall that the $N_c$-expansion is nonperturbative
in the sense that each order corresponds to a sum of Feynman diagrams from 
all orders of perturbation theory. The large-$N_c$ is a sum of all
daisy and superdaisy graphs.} Furthermore, it is interesting
to note the dependence on $G$ in~(\ref{functional})
is the same dependence as the solution to the gap equation in 
the BCS theory for superconductivity 
(albeit at zero magnetic field)~\cite{shovrev}.

We next turn to the QM meson model.
The value $\phi$ of the scalar field is determined by solving
${d{\cal F}_{0+1}\over d\phi}=0$.  The value $\phi$
is an increasing function of the magnetic field in the same manner
as $\langle\bar{\psi}\psi\rangle$ is in the NJL model.
From Eq.~(\ref{finalle}), we find for small values of $\phi$
\bqa\nonumber
{d{\cal F}_{0+1}\over d\phi}
&\approx&\phi\left
[m^2+{N_cg^2\over4\pi^2}\sum_f|q_fB|\ln{\pi m_q^2\over2|q_fB|}
\right]
\\ &=&0\;,
\eqa
whose to nonzero solution is approximately 
\bqa
\phi^2&\approx&{|qB|\over g^2}\exp\left[-{m^2\over N_cg^2|qB|}\right]\;.
\label{dynamic}
\eqa
If the mass parameter $m^2$ is positive, there is no symmetry breaking at
tree level. Eq.~(\ref{dynamic}) then shows that quantum fluctuations
induce symmetry breaking in the same manner as in the NJL model, 
cf. Eq.~(\ref{functional}).

\begin{figure}
\setlength{\unitlength}{1mm}
\includegraphics[width=7.0cm]{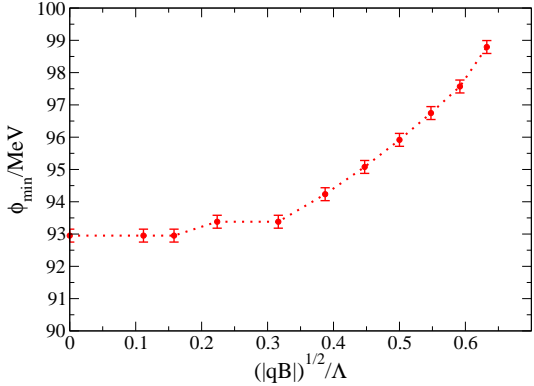}
\caption{Magnetic catalysis in the quark-meson model at the physical
point and at $T=0$. The vacuum expectation value of the field
$\phi$ as a function of the magnetic field scaled by the ultraviolet
cutoff $\Lambda$.}
\label{catal2}
\end{figure}

In Fig.~\ref{catal2}, we show the minimum of the effective potential
$U_{k=0}(\phi)$ at $T=0$
as a function of $(|qB|)^{1\over2}/\Lambda$
in the quark-meson model using the functional renormalization
group. We see that the minimum is an increasing function of the
magnetic field, so the system shows magnetic catalysis.

There have been a number of lattice calculations of the chiral condensate
at $T=0$ as a function of the magnetic field both in the 
quenched approximation~\cite{buido1,buido2,kala1} 
and with dynamical quarks~\cite{negro,endrodires}.

\begin{figure}
\setlength{\unitlength}{1mm}
\includegraphics[width=7.0cm]{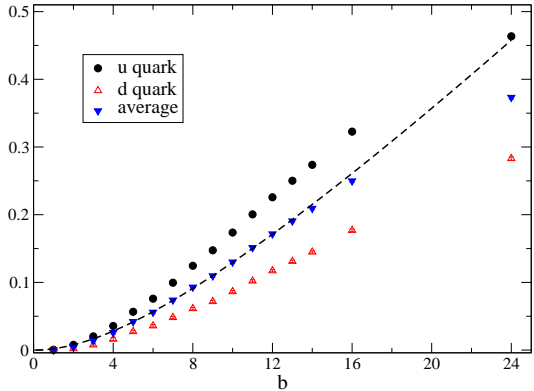}
\caption{The $\langle\bar{u}u\rangle$ (black points)
and $\langle\bar{d}d\rangle$ (red points) condensates as well as their average
(blue points)
as a function of $B$.
Figure taken from~\textcite{negro}.} 
\label{diffud}
\end{figure}

In Fig.~\ref{diffud}, the results for the condensates
$\langle\bar{u}u\rangle$ and $\langle\bar{d}d\rangle$ as well as their average
are shown as functions of the magnetic field~\cite{negro}. 
We notice that the $\bar{u}u$ condensate 
is larger than the $\bar{d}d$ in agreement with
model calculations, cf. Fig.~\ref{boomud}.

\begin{figure}
\setlength{\unitlength}{1mm}
\includegraphics[width=7.0cm]{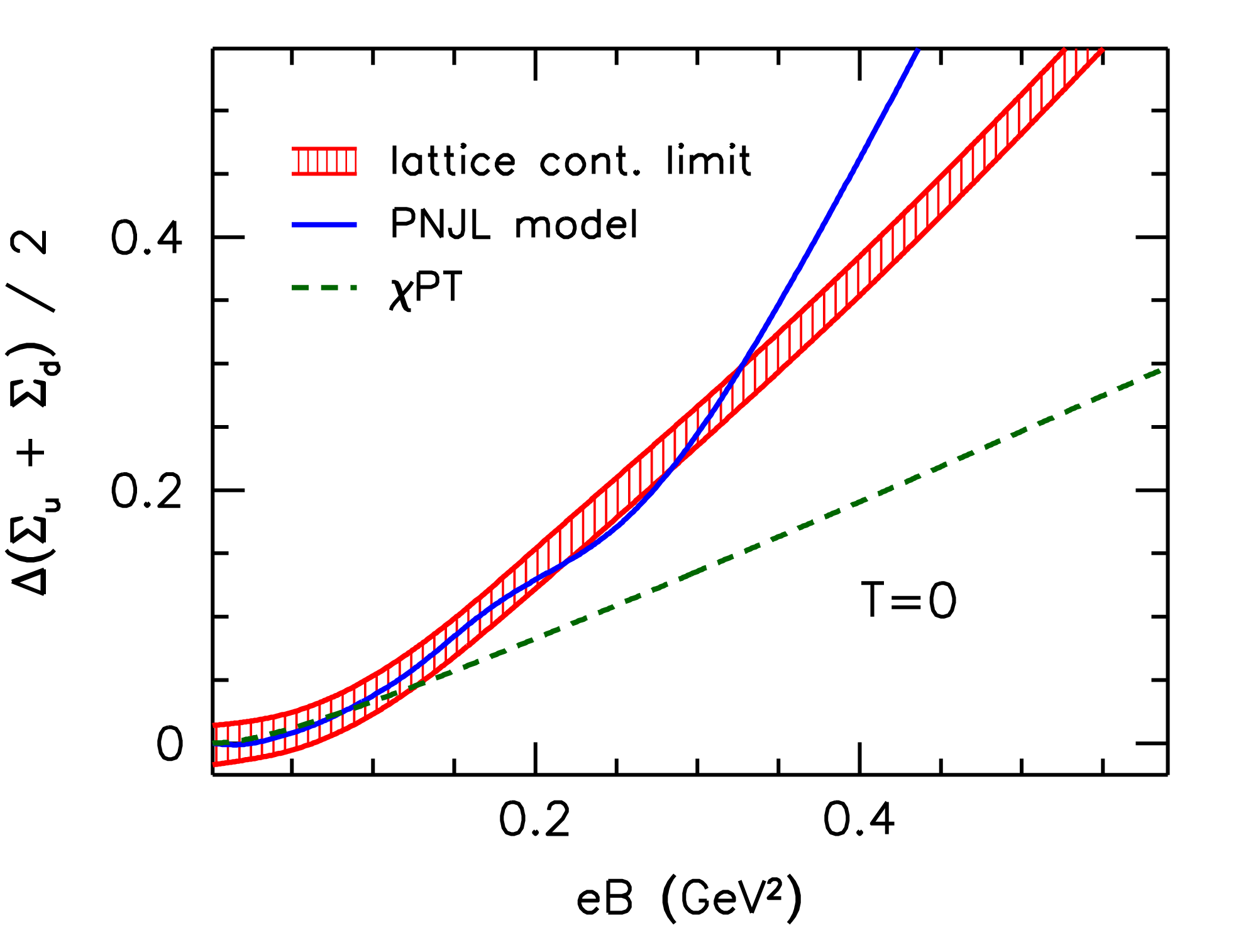}
\caption{Comparison of the continuum limit of the change of the condensate to 
with that of chiral perturbation theory~\cite{werbos,jensa,jensb} 
and the (P)NJL model~\cite{gattopnjl}.
Figure taken from~\textcite{endrodi1}.} 
\label{catal1}
\end{figure}

In Fig.~\ref{catal1}, the change of the condensate
$\mbox{$1\over2$}\Delta(\langle\Sigma_u\rangle+\langle\bar\Sigma_d\rangle)$ 
is shown as a function of $|qB|$ at $T=0$
and at the physical point~\cite{endrodires}.
The lattice results are 
continuum extrapolated. The model calculations are from one-loop
chiral perturbation theory~\cite{werbos,jensa,jensb}
as well as the Polyakov-loop extended NJL model~\cite{gattopnjl}. 
Notice that at $T=0$, the PNJL model reduces to the NJL model.
Clearly, the result of the Chpt results are in quantitative
agreement with lattice simulations for magnetic fields up to 
$|qB|\approx0.15$ GeV$^2$. 
For the (P)NJL model, the agreement with
lattice extends up to $|qB|\approx0.30$ GeV$^2$. 
The quark condensate in chiral perturbation theory is
given by Eq.~(\ref{condchpt}). Expanding around $B=0$ to
at $T=0$
using Eq.~(\ref{largexgap}), we find the shift due to the magnetic field
\bqa
\Delta\langle\bar{q}q\rangle&=&
{1\over2}(qB)^2{1\over3(4\pi)^2}
{\langle\bar{q}q\rangle\over m_{\pi}^2f_{\pi}^2}
+{\cal O}(B^4)
\;,
\eqa
where we have identified $M=m_{\pi}$ and $F=f_{\pi}$, correct
at this order. The interesting observation here, first made
by~\textcite{endrodires},  
is that the prefactor is proportional to
the one-loop $\beta$-function of scalar QED, 
$\beta={1\over3(4\pi)^2}$. A similar result is obtained for 
fermions, which can be shown using the relation
$\langle\bar{q}q\rangle\sim\mbox{$\partial{\cal F}\over\partial m_f$}$
~\cite{baliqed}.

The behavior of the quark condensate as a function of $B$
can be understood in terms of the Banks-Casher relation~\cite{cashin}.
The quark condensate $\langle\bar{\psi}\psi\rangle$
is proportional to the spectral density $\rho(\lambda)$
of the Dirac operator around zero. 
The Dirac operator
depends on the magnetic field, and therefore the 
spectral density depends on $B$. A constant magnetic field enhances
the spectral density around zero and as a result it enhances 
the quark condensate, see also the discussion in 
Sec.~\ref{latsec}.
This behavior of the spectral density is already
found in the quenched approximation~\cite{buido1,buido2,kala1} 
in which there is no back-reaction
from the quarks to the nonabelian gauge fields.
In model calculations, the quark condensate is given by the
expectation value of the operator 
${\rm Tr}(D\!\!\!\!/(B)+m)^{-1}$, which is enhanced by the magnetic field.
This enhancement is due to an increase of the spectral density, which
is a consequence of the degeneracy being proportional to the
magnetic flux, cf the discussion after Eq.~(\ref{fermspec}).

In a recent paper,~\textcite{bonnet} investigate 
dynamical quark mass generation and spin polarization
in a strong magnetic field $B$ using
the Dyson-Schwinger (DS) equations. They do this in both the quenched and
unquenched approximations at $T=\mu_B=0$.
The starting point is the Dyson-Schwinger equation for the fermion
propagator $S(x,y)$ in coordinate space
\bqa
S^{-1}(x,y)&=&S_0^{-1}(x,y)+\Sigma(x,y)\;,
\eqa
where $S_0(x,y)$ is the free fermion propagator and 
$\sum(x,y)$ is the fermion self-energy
\bqa\nonumber
\Sigma(x,y)&=&ig^2C_F\gamma^{\mu}S(x,y)\Gamma^{\nu}(y)D_{\mu\nu}(x,y)\;,
\\ &&
\eqa
where $C_F=\mbox{$N_c^2-1\over2N_c$}$,
$\Gamma^{\nu}(y)$ is the dressed fermion vertex and $D_{\mu\nu}(x,y)$
is the quenched gluon propagator.
The quenched gluon propagator in momentum space
can be written as $D_{\mu\nu}(k^2)=D(k^2)P_{\mu\nu}$, 
where the projection operator is $P_{\mu\nu}=\delta_{\mu\nu}-k_{\mu}k_{\nu}/k^2$.
The fermion propagator in the Ritus 
representation~\cite{ritus} is
\begin{widetext}
\bqa
S(x,y)&=&\sum_{k=0}^{\infty}
\int{d^2p_{\parallel}\over(2\pi)^4}\int_{-\infty}^{\infty}dp_2
E_p(x){1\over i\gamma\cdot p_{\parallel}A_{\parallel}(p)
+i\gamma\cdot p_{\perp}A_{\perp}(p)+B(p)}\bar{E}_p(y)\;,
\eqa
\end{widetext}
where $A_{\parallel}(p)$, $A_{\perp}(p)$, and $B(p)$ are the so called
dressing functions. By taking the trace in the Dyson-Schwinger equation,
one finds a set of coupled equations for the dressing functions.

The gluon propagator function $D(k^2)$ is written in terms of the
dressing function $Z(k^2)$ via $D(k^2)=Z(k^2)/k^2$.
The function $D(k^2)$
has been calculated to high precision both on the 
lattice~\cite{lein1,lein2}
and by solving the Dyson-Schwinger equations~\cite{maas,huba}. 
The quenched gluon propagator is used as input to the Dyson-Schwinger
equation together with the dressed vertex $\Gamma^{\mu}(p)$.
The latter is, however, poorly known, and~\textcite{bonnet}
made a simple ansatz for it.

\begin{figure}
\begin{center}
\setlength{\unitlength}{1mm}
\includegraphics[width=8.0cm]{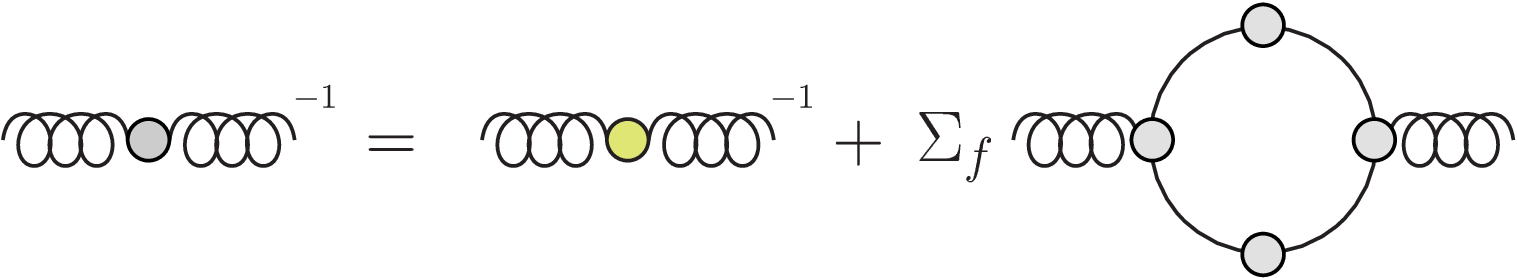}
\caption{Dyson-Schwinger equation for the inverse gluon propagator.}
\label{DS}
\end{center}
\end{figure}

In the unquenched approximation, the gluon propagator is improved
by taking into account the quark loop in the Dyson-Schwinger equation.
This is shown diagrammatically in Fig.~\ref{DS} and the 
Dyson-Schwinger equation in momentum space
can then be written as 
\bqa\nonumber
D_{\mu\nu}^{-1}(k)&=&
(D^{-1}_{\mu\nu})_0(k)+\Pi_{\mu\nu}^{\rm g}(k)
+\Pi_{\mu\nu}^{\rm q}(k)\\
&\approx&
D_{\mu\nu}^{-1,\rm eff}(k)
+\Pi_{\mu\nu}^{\rm q}(k)\;,
\eqa
where $D_{\mu\nu}^{-1}(k)$ is an effective inverse propagator corresponding
to the first diagram on the right-hand side in Fig.~\ref{DS}.
The self-energy $\Pi_{\mu\nu}^{\rm q}(k)$ corresponds to the quark loop
in Fig.~\ref{DS}.
The big blobs represent dressed propagators and dresses vertices.
Since the term $D_{\mu\nu}^{-1,\rm eff}(k)$ is isotropic, it is the quark loop
that generates the anisotropies in the dressed gluon propagator.

\begin{figure}
\begin{center}
\setlength{\unitlength}{1mm}
\includegraphics[width=7.0cm]{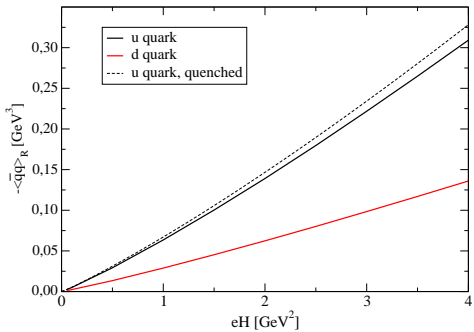}
\caption{$\langle\bar{u}u\rangle$ (black) and $\langle\bar{d}d\rangle$ 
quark condensates (red) together with the 
$\langle\bar{u}u\rangle$ condensate in the quenched approximation
(dashed). 
Figure taken from~\textcite{bonnet}.}
\label{quq}
\end{center}
\end{figure}

Fig.~\ref{quq} displays the $\langle\bar{u}u\rangle$ 
(solid black line)
and $\langle\bar{d}d\rangle$ quark condensates
(solid red line) in the unquenched approximation
as a function of the magnetic field.
For comparison, the $\langle\bar{u}u\rangle$ condensate 
(dashed black line)
in the quenched approximation has been shown as well.
One observes that the condensates are different. This simply
reflects the isospin breaking due to the different electric charges 
of the $u$ and the $d$ quark.
The most interesting result is that the quenched condensate is larger than 
the unquenched condensate. Taking the  
back-reaction of the quarks on the gluonic sector
leads to reduced magnetic catalysis.
Wheather this leads to inverse magnetic catalysis around $T_c$
is an open question, but it is certainly
of interest to investigate it.

A similar approach was used by~\textcite{reinhardt1}, in which the
Dyson-Schwinger equation was studied in the rainbow approximation.
In this approximation, the dressed quark-gluon vertex is replaced
with the bare (tree-level) vertex, while the quark propgator and its
inverse are dressed. The gluon dressing function has a phenomenological 
form that has been used to study dynamical chiral symmetry breaking.
The authors pay particular attention to the weak-field limit and so
this is complementary to the paper by~\textcite{bonnet}.
In order to connect to the case $B=0$, a nonperturbative approximation
to the quark propagator is constructed, which involves
a summation over the Landau levels. If one does sum over Landau levels,
the mass gap vanishes in the limit $B\rightarrow0$, which is incorrect
(see Fig.~\ref{quq}).
In Fig.~\ref{DS2}, the relative increment (see also 
Eq.~(\ref{relinc}) below) is shown using the Dyson-Schwinger approach
as well lattice results from by~\textcite{negro}.
The agreement is very good up to field strengths of approximately
$|qB|=0.3$ GeV$^{2}$. One must be cautious, however, as the
DS equations are solved in the chiral limit, while the lattice results
are for quark masses that correspond to $m_{\pi}\approx200$ MeV.

\begin{figure}
\begin{center}
\setlength{\unitlength}{1mm}
\includegraphics[width=7.0cm]{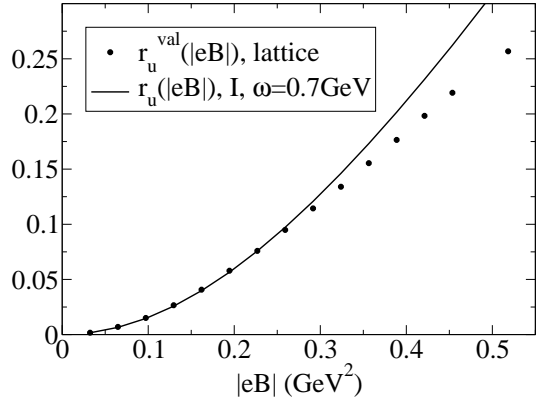}
\caption{Comparison of the up-quark relative increment with the
lattice results of~\textcite{negro}. 
$\omega$ is a parameter of the gluon dressing function.
Figure taken from \textcite{reinhardt1}.}
\label{DS2}
\end{center}
\end{figure}

\section{Lattice simulations and inverse magnetic catalysis}
\label{latsec}
As discussed in the introduction, QCD at zero baryon chemical potential
$\mu_B$ in an Abelian background field $A_{\mu}$
is free of the sign problem and so QCD can in principle
be straightforwardly simulated on the lattice using standard
Monte Carlo algorithms. This statement is independent of the
color gauge group, which opens up the possibility for
doing lattice simulations for the theories where the physics is very different.
For example, for two colors, $N_c=2$, and two massless
flavors, $N_f=2$, the symmetry
group of the Lagrangian is $SU(4)\sim SO(6)$, which is broken down
in the vacuum to the group $Sp(2)\sim SO(5)$. In the process, 
five generators are broken leading to five massless bosons
according to Goldstone's theorem.
These Goldstone particles are the pions, $\pi^{\pm}$ and $\pi_0$ as well
a diquark $\Delta$ and an antidiquark $\Delta^*$.
Due to the color group $SU(2)_c$, two quarks can form color singlets
and therefore are part of the physical spectrum. The diquarks are thus the
``fermions'' of two-color QCD. However, being bosons, they behave
differently. 
For example, they bose condense when the baryon chemical potential exceeds the 
mass of the diquarks.\footnote{Note that due to the special properties of 
the Pauli matrices, $SU(2)_c$ does not have a sign problem and so
one can perform lattice simulations at finite $\mu_B$.}

\subsection{$SU(3)_c$}
\begin{figure}
\begin{center}
\setlength{\unitlength}{1mm}
\includegraphics[width=7.0cm]{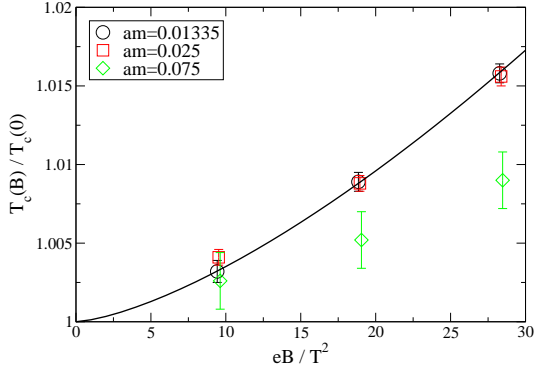}
\caption{Transition temperature normalized to $T_c$ at $B=0$
for the deconfinement and the chiral transition as a function $|qB|/T^2$. 
Figure taken from~\textcite{latticemassimo}.}
\label{slow}
\end{center}
\end{figure}

In the physical case $N_c=3$ and with $N_f=2$, 
the first lattice simulations at finite
magnetic field were carried out by~\textcite{latticemassimo}.
They used different values of the bare quark masses corresponding to
a pion mass in the 200-480 MeV range. The magnetic field strengths
were up to $|qB|\sim0.75$ GeV$^2$ and the calculations were carried out with a
lattice spacing of $0.3$ fm and the results were not continuum extrapolated. 
The authors found no evidence for a splitting between the chiral and
deconfinement transition as found in PNJL and PQM model calculations.
They also found that the critical temperature increases very slowly
with the magnetic field as can be seen in Fig.~\ref{slow}.
These results have later been confirmed by~\textcite{endrodi1,endrodi2}
and seem to be in agreement with model calculations presented
in Secs.~\ref{efttm} and~\ref{funxio}.

\begin{figure}
\begin{center}
\setlength{\unitlength}{1mm}
\includegraphics[width=7.0cm]{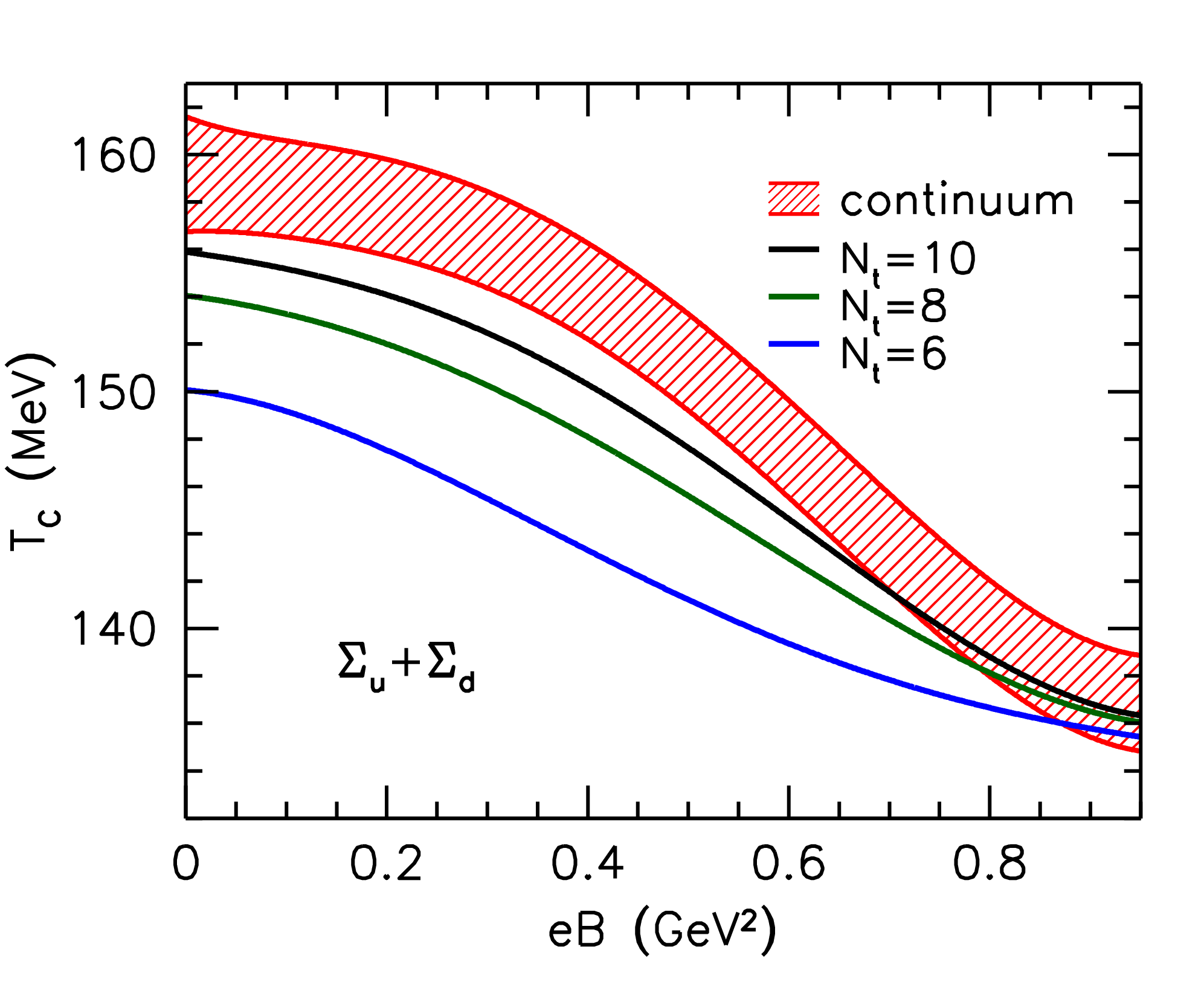}
\caption{Transition temperature
for the 
deconfinement as a function of $|qB|$ for different lattice spacings
(solid curves)
and the continuum-extrapolated result (band)
Figure taken from~\textcite{bali11}.}
\label{newtc}
\end{center}
\end{figure}

\textcite{endrodi1,endrodi2}
have also carried out lattice simulation using a
physical pion mass of $m_{\pi}=140$ MeV. 
Their results are shown in Fig.~\ref{newtc}. 
Their results have been continuum 
extrapolated and show, perhaps somewhat surprisingly, that the 
transition temperature is decreasing with the magnetic field $B$.
The fact that $T_c$ is a decreasing function of the magnetic field
suggests that the results obtained for larger pion masses will survive
the continuum extrapolation. If this is correct, the transition temperature
is a complicated function of the magnetic field and the quark masses.

In the recent papers of~\textcite{endrodi1,endrodi2}
and~\textcite{endrodi3}, the authors 
analyze in detail lattice results and thereby explain
the discrepancy for $T_c$ as a function of $B$
between the
model calculations such as (P)NJL and (P)QM models
and their results.

The chiral condensate can be written as
\begin{widetext}
\bqa
\langle\bar{\psi}\psi\rangle
&=&{1\over{\cal Z}(B)}\int d{\cal U}
e^{-S_g}\det(D\!\!\!\!/(B)+m)
{\rm Tr}(D\!\!\!\!/(B)+m)^{-1}\;,
\label{det}
\eqa
where the partition function ${\cal Z}(B)$ is
\bqa
{\cal Z}(B)&=&\int d{\cal U}
e^{-S_g}\det(D\!\!\!\!/(B)+m)\;,
\eqa
and $S_g$ is the pure-glue action. 
The magnetic field enters via the operator 
${\rm Tr}(D\!\!\!\!/(B)+m)^{-1}$ as well as the fermion functional 
determinant $\det(D\!\!\!\!/(B)+m)$.
We can think of 
${\cal P}(m,U,B)\equiv{1\over{\cal Z}(B)}e^{-S_g}\det(D\!\!\!\!/(B)+m)$,
where $U$ denotes the gauge-field configuration that corresponds to
$e^{-S_g}$,
as a measure. 
In order to study the contributions to magnetic catalysis coming 
separately from the change in the operator and in measure, one
defines the two condensates
\bqa
\label{val}
\langle\bar{\psi}\psi\rangle^{\rm val}
&=&{1\over{\cal Z}(0)}\int d{\cal U}
e^{-S_g}\det(D\!\!\!\!/(0)+m){\rm Tr}(D\!\!\!\!/(B)+m)^{-1}\;,
\\
\langle\bar{\psi}\psi\rangle^{\rm sea}
&=&{1\over{\cal Z}(B)}\int d{\cal U}
e^{-S_g}\det(D\!\!\!\!/(B)+m){\rm Tr}
(D\!\!\!\!/(0)+m)^{-1}\;.
\label{sea}
\eqa
\end{widetext}
These are the socalled valence and sea condensates.
The valence condensate is the average of the trace of the propagator
in a constant magnetic 
background, but where the sampling of the nonabelian gauge configurations
is done at $B=0$. The sea contribution is the 
average of the same operator in zero magnetic field, but where
the sampling is done at nonzero $B$.
The sea effect is absent in the quenched approximation.
More generally, a sea observable is an observable that does not depend
explicitly on the magnetic field. The Polyakov loop is another
example of a sea observable.
We note that the sea condensate equals a condensate of a neutral
quark in a two-flavor theory with one electrically charged and one neutral
quark since the magnetic field does not appear in the operator,
but in the determinant.

\begin{figure}
\begin{center}
\setlength{\unitlength}{1mm}
\includegraphics[width=7.0cm]{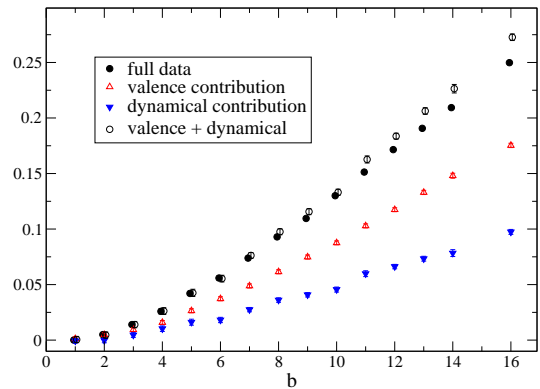}
\caption{Relative increments 
$r(B)$, $r^{\rm val}(B)$, and $r^{\rm sea}(B)$ at zero temperature
See main text for details.
Figure taken from~\textcite{negro}.}
\label{valsea}
\end{center}
\end{figure}

A useful quantity is the relative increment $r(B)$
of the quark condensate as a function of $B$, which is defined by 
\bqa
\label{relinc}
r(B)={\langle\bar{\psi}\psi\rangle(B)\over\langle\bar{\psi}\psi\rangle(0)}-1\;.
\eqa
The relative increments $r^{\rm val/sea}(B)$ are defined in a similar manner.
\textcite{negro} calculated the three quantities 
$r(B)$, $r^{\rm val}(B)$, and $r^{\rm sea}(B)$ at zero temperature.
The result is shown in Fig.~\ref{valsea}. 
The valence contribution $r^{\rm val}(B)$ (red data points)
and the sea contribution $r^{\rm sea}(B)$
(blue data points) are both positive.
The sum of the two (open circles) and $r(B)$ (full circles)
are shown as well. We notice that 
the open circles are very close to the full circles, except for
very large values of $B$, which suggests that 
that the relative increment can be written as a sum of the
valence and sea contributions.
The same behavior of $r^{\rm val}(B)$, and $r^{\rm sea}(B)$ 
are found in the simulations by~\textcite{endrodi3} for physical 
quark masses at $T=0$.

\begin{figure}
\begin{center}
\setlength{\unitlength}{1mm}
\includegraphics[width=7.0cm]{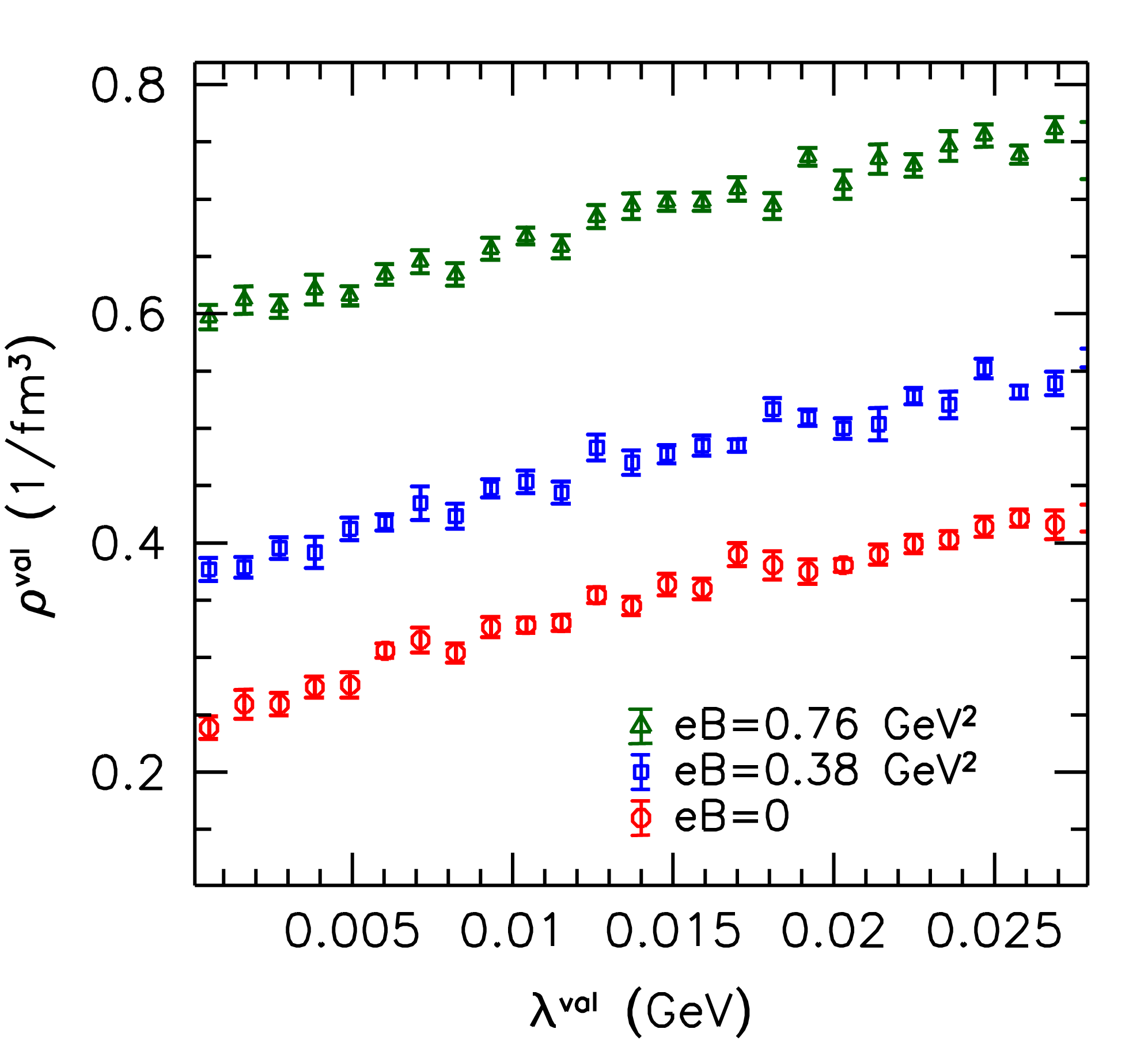}
\caption{Spectral density of the Dirac operator 
for three different values of the magnetic field $B$.
Figure taken from~\textcite{endrodi3}.}
\label{spec}
\end{center}
\end{figure}

As mentioned, earlier, 
it is possible to understand the behavior of the valence condensate
by employing the Banks-Casher relation~\cite{cashin}.
In the chiral limit, the chiral condensate is proportional to the
spectral density $\rho(\lambda)$ of the Dirac operator around zero.
In Fig.~\ref{spec}, \textcite{endrodi3} show the spectral density
for three values of the magnetic field $B$.
The ensemble of nonabelian gauge field backgrounds are generated
at zero magnetic field and at $T=142$ MeV.
It is evident from Fig.~\ref{spec} that the spectral density
and therefore the valence condensate increases with the 
strength of the magnetic field.
This behavior is independent of the temperature.

At temperatures around the transition temperature,
the valence condensate is still positive while the sea condensate
is negative. Hence there is a competition between the two, leading to a
net inverse catalysis. The sea contribution can be viewed as a back reaction
of the fermions on the gauge fields and this effect is not present in the
model calculations as there are no dynamical gauge fields.
The behavior of the sea contribution was also carefully analyzed
by~\textcite{endrodi3}. Introducing 
$\Delta S_f(B)=\log\det(D\!\!\!\!/(B)+m)-\log\det(D\!\!\!\!/(0)+m)$,
one can rewrite the full condensate as
\bqa
\langle\bar{\psi}\psi\rangle
&=&{\langle e^{-\Delta S_f(B)}{\rm Tr}(D\!\!\!\!/(B)+m)^{-1}\rangle_0\over
\langle e^{-\Delta S_f(B)}\rangle_0}\;,
\label{fully}
\eqa
where the subscript $0$ indicates that the expectation values
are at $B=0$. We note that Eq.~(\ref{fully}) reduces to the 
valence condensate if we replace the exponential factor
$e^{-\Delta S_f(B)}$ by unity.


\begin{figure}
\begin{center}
\setlength{\unitlength}{1mm}
\includegraphics[width=7.0cm]{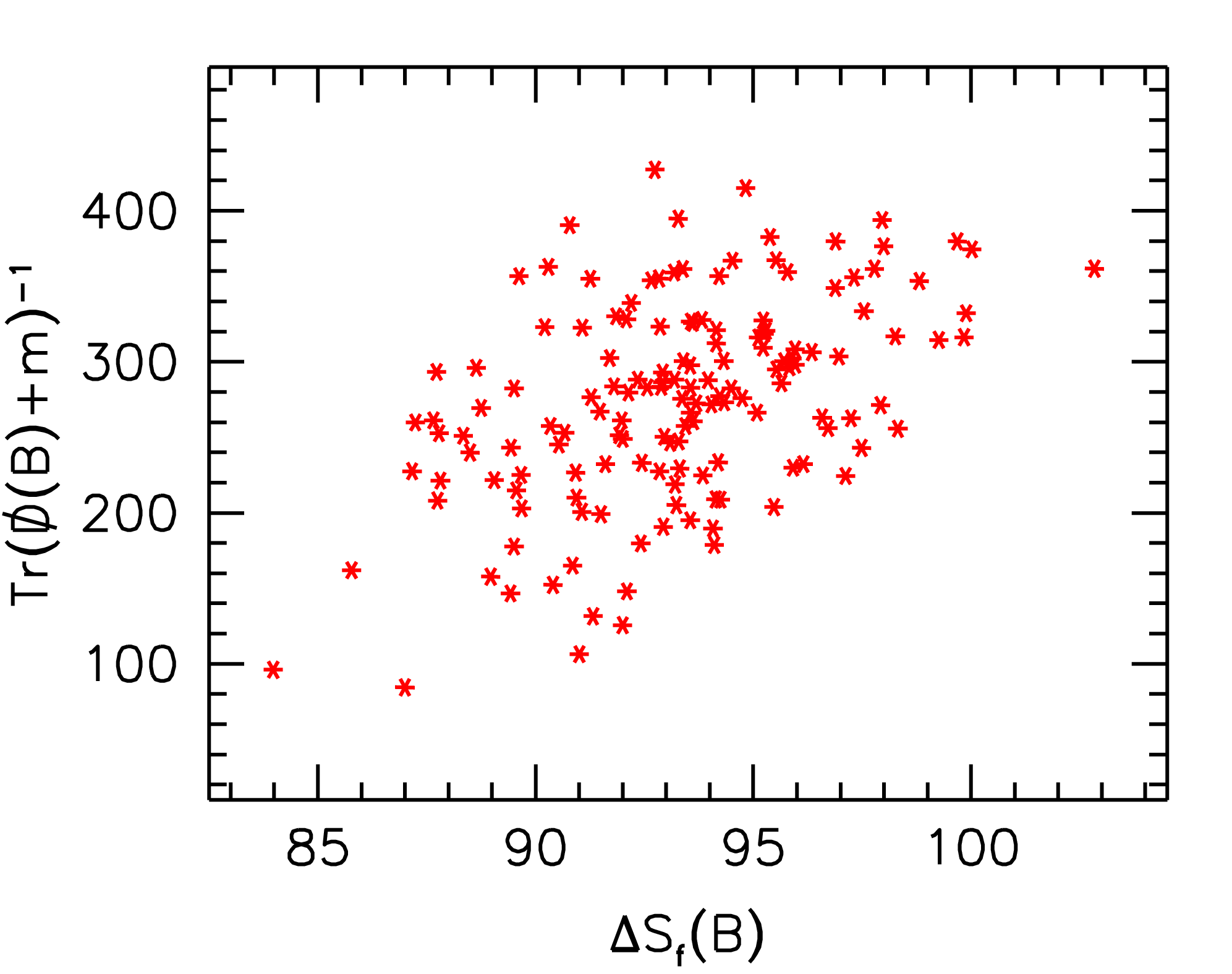}
\caption{Scatter plot of the down-quark condensate as a function
of $\Delta S_f(B)$. The magnetic field strength is $|qB|\approx0.5$ GeV$^2$
and $T\sim T_c$.
Figure taken from~\textcite{endrodi3}.}
\label{scat}
\end{center}
\end{figure}

In Fig.~\ref{scat},~\textcite{endrodi3} show a scatter plot of 
the condensate as a function of the change 
in the action $\Delta S_f(B)$ due to the magnetic
field for a magnetic field strength of $|qB|\approx0.5$ GeV$^2$
and $T$ around the transition temperature.
Each point represents a gauge configuration and they were
generated at $B=0$, and therefore a simple averaging of
${\rm Tr}(D\!\!\!\!/(B)+m)^{-1}$ without
weighting each configuration in the ensemble
with the Boltzmann factor $e^{-\Delta S_f(B)}$ gives the valence
condensate.
In order to calculate the full quark condensate, one must average
${\rm Tr}(D\!\!\!\!/(B)+m)^{-1}$ 
over the gauge configurations
including the weight factor $e^{-\Delta S_f(B)}$.
Generally, larger values of the condensate correspond to larger values
of $\Delta S_f(B)$ and as a result,
the weight of the associated gauge configuration
is suppressed. This suppression is particularly effective
around $T_c$ and in fact overwhelms the valence effect and therefore
leads to inverse magnetic catalysis in the transition region.
This suppression is not present for larger quark masses, cf.
Fig.~\ref{valsea}.
One therefore might expect the sea effect to be even more pronounced
in the chiral limit.
Let us finally add that the recent simulations of~\textcite{borna}
with large quark masses that correspond to a pion mass of approximately
$500$ MeV show a different behavior. Using chiral fermions instead
of staggered fermions, the authors find clear evidence for 
inverse magnetic catalysis also for large pion masses.
The analysis was based on the behavior of 
the chiral condensate, the expectation value of the 
Polyakov loop as well as the spectral density as 
functions of $B$ for two different values of
the temperature. The results seem to indicate that the
chiral properties are an important ingredient in inverse
magnetic catalysis.

\subsection{$SU(2)_c$}
Recently, ~\textcite{ilgen1,ilgen2} have carried out lattice
simulations with dynamical fermions. We focus on their second paper, which
is extension of the first to smaller quark masses.
They used $N_f=4$ and equal electric charge as well as quark masses
that correspond to a pion mass $m_{\pi}$
of approximately 175 MeV. 
The transition temperature for $B=0$ is in this case $T_c\approx m_{\pi}$.
\begin{widetext}
Fig.~\ref{ilgenmass} shows the mass dependence of the bare chiral condensate
for three different values of the magnetic field,
$B=0$ (grey), $|qB|=0.67$ GeV$^2$ (blue), and $|qB|=1.69$  GeV$^2$ (red) and
two different temperatures: $147$ MeV (left) and $195$ MeV (right).
Inspecting the left panel, the data points
suggest that the system is in the chirally
broken phase for all three values of the magnetic field.
In contrast, the data points in right panel indicate
that the chiral condensate is zero
(extrapolating to the chiral limit) for $B=0$ and $|qB|=0.67$ GeV$^2$,
while the chiral condensate is nonzero for $|qB|=1.69$ GeV$^2$.
This behavior suggests that the critical temperature grows with $B$
for very strong magnetic fields.

\begin{figure}
\begin{center}
\setlength{\unitlength}{1mm}
\includegraphics[width=16.0cm]{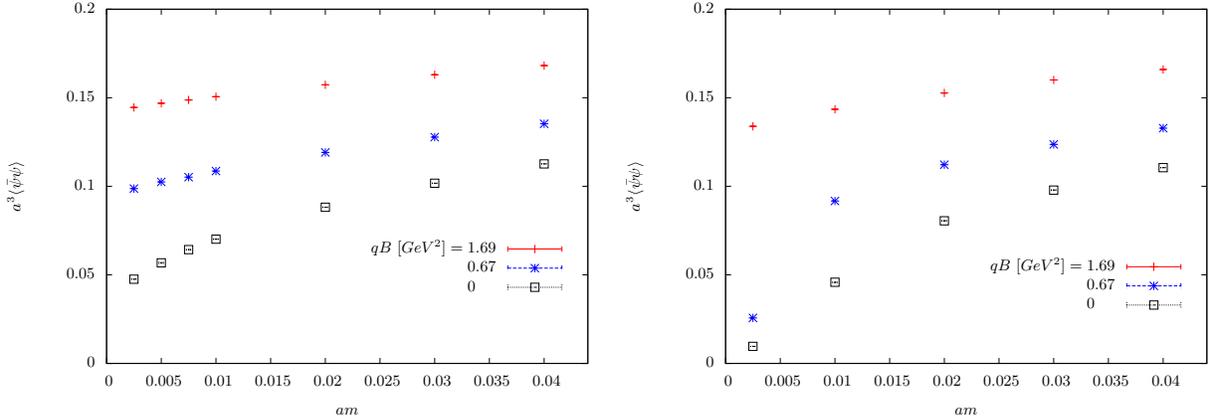}
\caption{Mass dependence of the bare chiral condensate for three
different values of the magnetic field and two different temperatures:
147 MeV (left) and 195 MeV (right).
Figure taken from~\textcite{ilgen2}.}
\label{ilgenmass}
\end{center}
\end{figure}

Further insight can be gained from 
Fig.~\ref{ilgenpol}, where
the authors show the expectation values of the Polyakov loop (left panel)
and the chiral condensate (right panel) at $T=195$ MeV
as functions of $|qB|$ up to
$|qB|=1.69$ GeV$^2$. The left panel shows a rise of the Polyakov loop
for magnetic fields up to approximately $|qB|=0.7$ GeV$^2$.
This suggests that one goes deeper into the deconfinement region and
that the system exhibits inverse magnetic catalysis at low
values of the magnetic field.
For values of the magnetic field larger than $|qB|=0.7$ GeV$^2$,
there is a significant drop of the expectation value of the Polyakov loop. 
This indicates that we are going back into the confinement region and
that the system exhibits magnetic catalysis for large values of the
magnetic field.

\begin{figure}
\begin{center}
\setlength{\unitlength}{1mm}
\includegraphics[width=16.0cm]{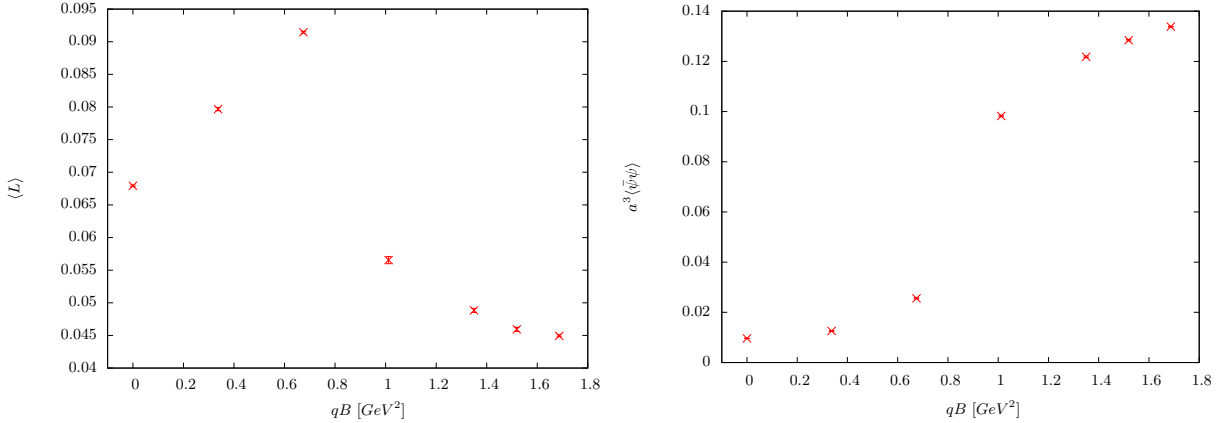}
\caption{Expectation values of the Polyakov loop (left)
and the chiral condensate (right)
vs $|qB|$ at $T=195$ MeV.
Figure taken from~\textcite{ilgen2}.}
\label{ilgenpol}
\end{center}
\end{figure}

\end{widetext}

\begin{figure}
\begin{center}
\setlength{\unitlength}{1mm}
\includegraphics[width=7.0cm]{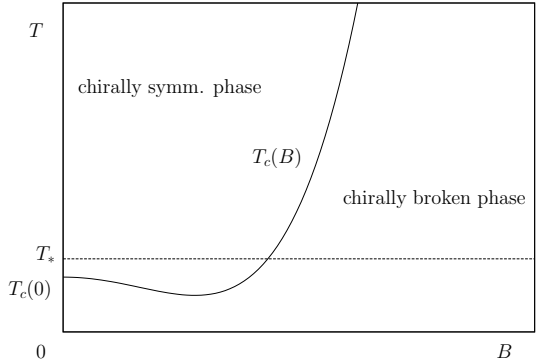}
\caption{Conjectured phase diagram in the $B$--$T$ plane.
The horizontal line is $T^*=195$ MeV. Figure taken from~\textcite{ilgen2}.}
\label{ilgentc}
\end{center}
\end{figure}

The results suggest that the critial temperature 
decreases for weak magnetic fields and increases for strong magnetic fields.
A conjectured phase diagram based on these observations is
shown in Fig.~\ref{ilgentc}.
A direct comparison with the 
only model calculation that 
exists~\cite{amador} is not straightforward since $N_f=2$ 
with $q_u={2\over3}$ and $q_d=-{1\over3}$ were used.
Nevertheless, note the similarity with 
Fig.~\ref{2ctc}, where a small minimum can be seen.
However, to firmly conclude, a thorough analysis involving
separating the valence and sea effects along the lines 
of~\textcite{endrodi3} would be of interest.

\section{Model calculations revisited}
\label{revisit}
After it was realized that most model calculations were in disagreement
with the lattice calculations, there
has been significant efforts to modify them such that they
reproduce the correct behavior of $T_c$ as a function of $B$,
or to propose a mechanism for inverse magnetic catalysis around 
$T_c$~\cite{pintoleik,costa,costainv,ferry,ayala1,ayala2,sado2,huang1,huang2,fukuinv,kojo1,palhares,tawfik,ferro2,mintz,pavlov,dors14,braun1}.
A large number of papers have been focusing on $B$-dependent
coupling constants or $B$-dependent parameters in the model and we
discuss some of them below.

\subsection{$B$-dependent transition temperature $T_0$}
The parameter $T_0$ that enters the Polyakov loop potential depends on
the number of quarks (and on the chemical potential at finite density).
At finite $B$, one expects $T_0$ to depend on the magnetic field
as well as $N_f$, which can be taken into account by using a $B$-dependent
function $b=b(N_f,B)$ in analogy with Eq.~(\ref{nfmub}).
The first attempt to incorporate a $B$-dependent transition temperature
$T_0(qB)$ was made by ~\textcite{ferry} using the (E)PNJL model.
They made the ansatz
\bqa\nonumber
T_0(qB)&=&T_0(qB=0)+\zeta(qB)^2+\xi(qB)^4\;,
\\ &&
\eqa
and fitted the parameters $\zeta$ and $\xi$
to reproduce the transition temperature extracted from 
the strange quark number susceptibility data~\cite{endrodi1}.
This approach gives a crossover
for $|qB|<0.25$ GeV$^2$ and a first-order transition 
for $|qB|>0.25$ GeV$^2$, 
when $T_0(qB)=186$ MeV, which corresponds to the critical
temperature for 2+1 massless flavors. 
The range of crossover transitions increases significantly
by using $T_o(qB=0)=270$ MeV, which corresponds to the transition 
temperature for pure-glue, i.e by omitting the backreaction from the
fermions at $B=0$.

Recently~\textcite{mintz} analyzed the possibility of inverse magnetic
catalysis by allowing the model parameter $T_0$ to be a 
a function of $B$. They calculated the transition temperature $T_c$
for the chiral transition as a function of the parameter $T_0$
in the PQM model in the mean-field approximation.

\begin{figure}
\begin{center}
\setlength{\unitlength}{1mm}
\includegraphics[width=7.0cm]{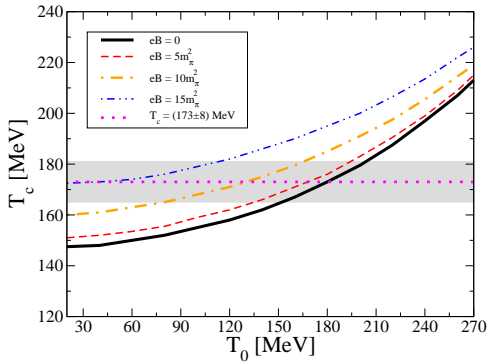}
\caption{Transition temperature $T_c$ for the chiral transition 
as a function of $T_0$ for different values of the magnetic field.
Figure taken from~\textcite{mintz}.}
\label{fragatype}
\end{center}
\end{figure}

\begin{figure}
\begin{center}
\setlength{\unitlength}{1mm}
\includegraphics[width=7.0cm]{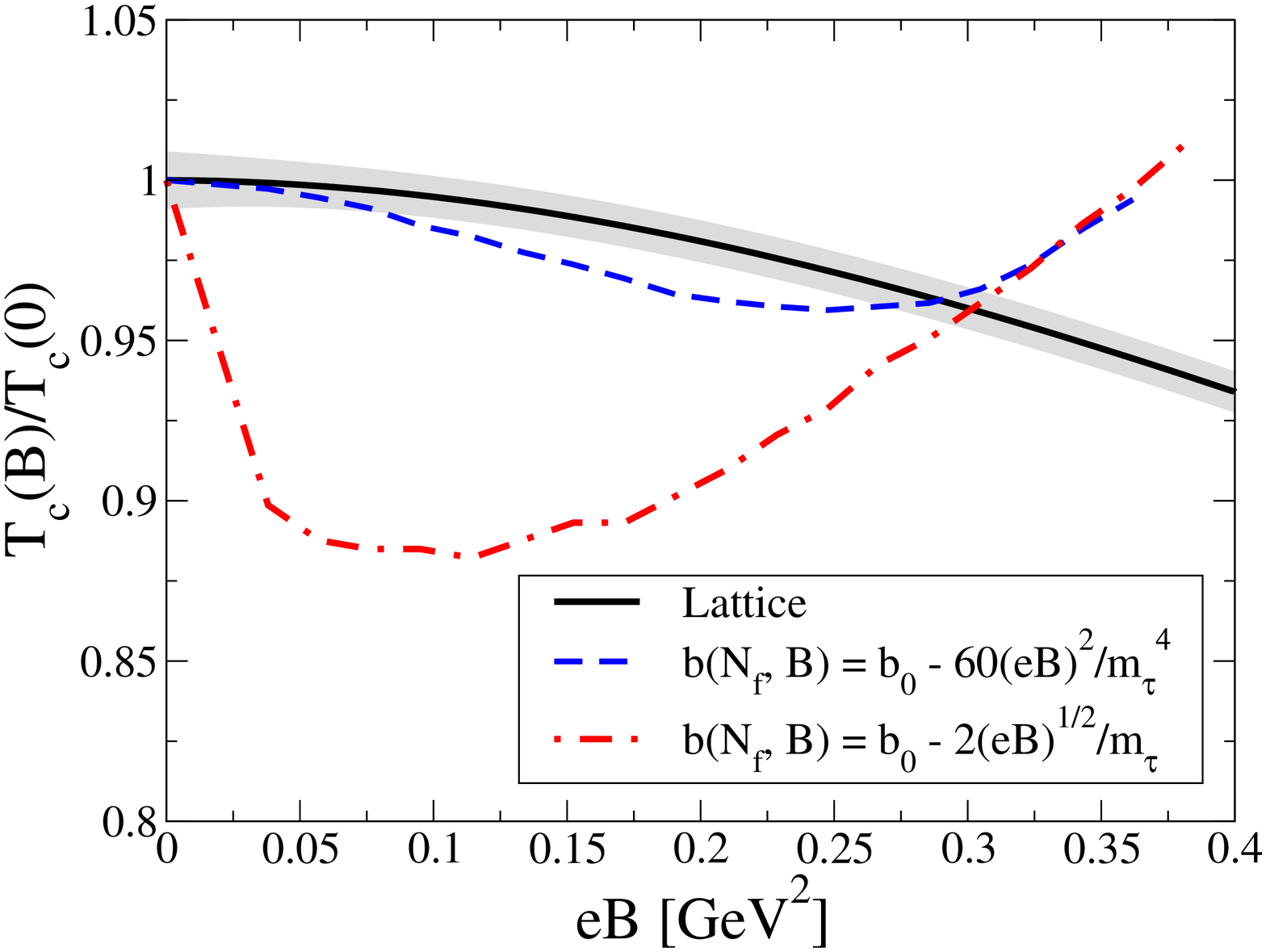}
\caption{Normalized transition temperature $T_c$ for the chiral transition 
as a function of $|qB|$. The grey band is the lattice results
from~\textcite{endrodi2}
Figure taken from~\textcite{mintz}.}
\label{fragatype2}
\end{center}
\end{figure}

Any parametrization of $T_0(B)$ gives rise to a continuous curve that
starts at some point on the black curve
corresponding to $B=0$ and crosses the other curves as $B$ is varied.
$T_c$ as a function of $B$ can be a decreasing function only if
$T_0(B)$ decreases sufficiently fast.
This can be the case for low values of the magnetic field, if 
the point $T_0(B=0)$ is sufficiently far to the right on the black curve.
However, since the curves become flatter as one moves to the
left in the figure, it is clear that this behavior cannot be sustained.
In other words, even if the critical temperature initially is decreasing
with $B$, eventually it will have a minimum and start increasing 
again for larger values of $B$.
Specific parametrizations $b(n_f,B)=b(N_f)-60(qB)^2/m_{\tau}^4$
and $b(n_f,B)=b_0-60\sqrt{|qB|}/m_{\tau}$ 
were given by~\textcite{mintz} to illustrate this point.
The result is shown in Fig.~\ref{fragatype2}.

~\textcite{robb2} have very recently 
performed the same type of calculations using the functional
renormalization group. The inclusion of mesonic fluctuations does
not change the results and conclusions, as anticipated 
by~\textcite{mintz}.

\subsection{$B$-dependent coupling constant}
~\textcite{pintoleik} investigated the 
possibility of obtaining inverse magnetic catalysis in 
the NJL model by  using an effective  coupling constant that is a
function of the magnetic field and the temperature $T$.
Motivated by the running of the QCD coupling, 
they proposed a $B$-dependent coupling 
$G(B)$ given by
\bqa
\label{rung0}
G(B)&=&{G_0\over1+\alpha\ln\left(1+\beta{|qB|\over\Lambda^2_{\rm QCD}}\right)}\;,
\eqa
where $G_0=5.022{\rm \,GeV}^{-2}$ is the value of the coupling at $B=0$.
We notice that $G(B)\rightarrow0$ as $B\rightarrow\infty$, a behavior
that is inspired by the running of $\alpha_s$ at very large magnetic fields
$|qB|\gg\Lambda_{\rm QCD}^2$:
${1\over\alpha_s}\sim\ln{|qB|\over\Lambda^2_{\rm QCD}}$~\cite{shovmir}.
Here $\alpha$ and $\beta$ are free parameters that are determined
such that one obtains a reasonable description of the average 
${1\over2}(\Sigma_u+\Sigma_d)$ calculated on the 
lattice at $T=0$, where the dimensionless quantity $\Sigma_f$
is defined by
\bqa
\Sigma_f&=&{2m_f\over m_{\pi}^2f_{\pi}^2}\left(\langle\bar{\psi}_f\psi_f\rangle_B
-\langle\bar{\psi}_f\psi_f\rangle_0\right)+1\;.
\eqa
At finite temperature the authors propose a
coupling $G(B,T)$ given by 
\bqa
G(B,T)&=&G(B)\left(1-\gamma{|qB|\over\Lambda_{\rm QCD}^2}
{T\over\Lambda_{\rm QCD}}\right)\;.
\label{rung}
\eqa
Here $\gamma$ is another parameter that is
fitted to reproduce the lattice results of~\textcite{endrodi2} 
for ${1\over2}(\Sigma_u+\Sigma_d)$ 
at the highest temperatures available.

\begin{figure}
\begin{center}
\setlength{\unitlength}{1mm}
\includegraphics[width=7.0cm]{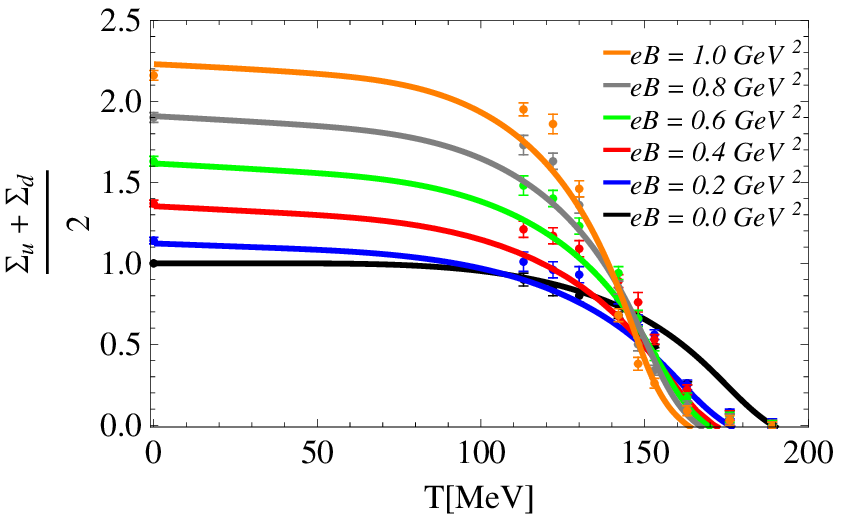}
\caption{The average ${1\over2}(\Sigma_u+\Sigma_d)$
as a function of temperature $T$ for different values of
the magnetic field. The data points are the lattice results
from~\textcite{endrodi2}.
Figure taken from~\textcite{pintoleik}.}
\label{pinto1}
\end{center}
\end{figure}

In Fig.~\ref{pinto1}, the average ${1\over2}(\Sigma_u+\Sigma_d)$
is shown as a function of temperature $T$ for different values of
the magnetic field. The data points are from the lattice simulations 
of~\textcite{endrodi2}.

The ans\"atze for the 
coupling, Eqs.~(\ref{rung0}) and~(\ref{rung})
then give a reasonable description of the lattice data.
At $T=0$, increasing magnetic field implies larger average
${1\over2}(\Sigma_u+\Sigma_d)$.
However, 
for $T\approx$ 140 MeV, the curves cross each another and the order 
of the curves is reversed beyond this temperature.
This shows inverse magnetic catalysis 
around the transition temperature.
The curves in Fig.~\ref{pinto1} become steeper around the transition
temperature as the magnetic field
increases, suggesting that transition becomes first order for
sufficiently large values of $B$.

\begin{figure}
\begin{center}
\setlength{\unitlength}{1mm}
\includegraphics[width=7.0cm]{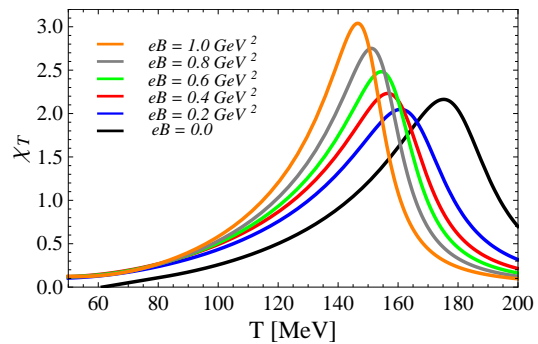}
\caption{Susceptibility $\chi$
as a function of the temperature $T$
for different values of the magnetic field.
Figure taken from~\textcite{pintoleik}.}
\label{pinto2}
\end{center}
\end{figure}

\begin{figure}
\begin{center}
\setlength{\unitlength}{1mm}
\includegraphics[width=7.0cm]{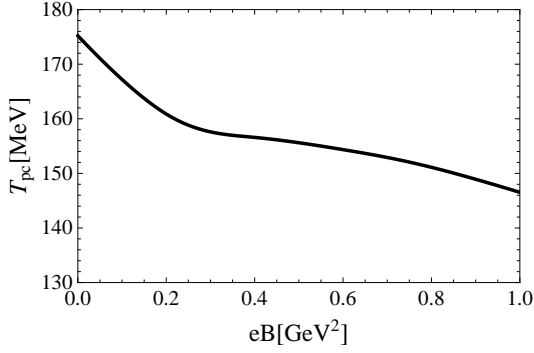}
\caption{
Pseudocritical temperature $T_{\rm pc}$
temperature as a function of the magnetic field $B$
in the NJL model. Figure taken from~\textcite{pintoleik}.}
\label{pinto4}
\end{center}
\end{figure}

The chiral susceptibility is defined by
\bqa
\chi={\partial\sigma\over\partial T}\;,
\eqa
where
\bqa
\sigma&=&
-m_{\pi}{(\langle\bar{u}u\rangle+\langle\bar{d}d)\rangle(B,T)
\over(\langle\bar{u}u\rangle+\langle\bar{d}d)\rangle(B,0)}\;,
\eqa
and is shown in Fig.~\ref{pinto2} as a function of $T$
for different values of $B$. 
The peaks move to the left as a function of
the magnetic field.
The peak of the susceptibility $\chi$ defines the 
pseudocritical temperature $T_{\rm pc}$ and in
Fig.~\ref{pinto4} the 
pseudocritical temperature 
is shown as a function of $|qB|$. 

\begin{figure}
\begin{center}
\setlength{\unitlength}{1mm}
\includegraphics[width=7.0cm]{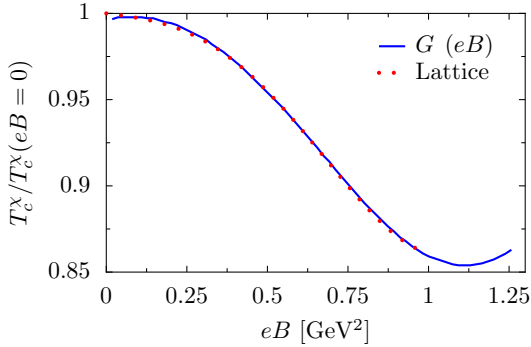}
\caption{Normalized pseudocritical temperature 
$T_c^{\chi}/T_c^{\chi}(B=0)$
as a function of $|qB|$
in the NJL model. Figure taken from~\textcite{costa}.}
\label{costa1}
\end{center}
\end{figure}

A similar approach was used by~\textcite{costa}, where an effective
coupling $G_s(|qB|/\Lambda_{\rm QCD}^2)$ was determined such that the
NJL model reproduces the normalized
transition temperature determined on the lattice.
In the fit, the lattice data points are for magnetic fields in the
range $0<|qB|<1$ GeV$^2$.
This way of determining the effective coupling leads to 
a temperature-dependent average
${1\over2}(\Sigma_u+\Sigma_d)$ that qualitative
looks like the plot in Fig.~\ref{pinto1}. 
The resulting normalized
transition temperature $T_c^{\chi}/T_c^{\chi}(B=0)$ together with lattice
data points are shown in Fig.~\ref{costa1}.

\begin{figure}
\begin{center}
\setlength{\unitlength}{1mm}
\includegraphics[width=7.0cm]{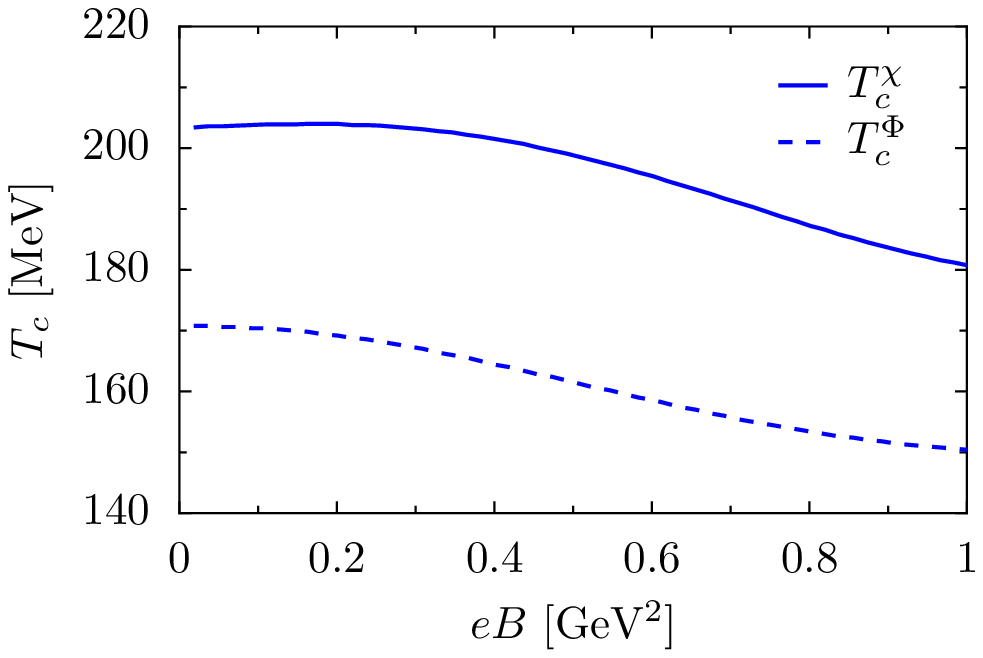}
\caption{Pseudocritical temperatures for the chiral (solid line)
and deconfinement (dashed line) transitions 
as functions of $|qB|$
in the PNJL model. Figure taken from~\textcite{costa}.}
\label{costa2}
\end{center}
\end{figure}

The $B$-dependent coupling $G_s(|qB|/\Lambda_{\rm QCD}^2)$ was subsequently
used as input to a PNJL calculation of the critical temperature for
the chiral as well as the deconfinement transition.
In the calculations, they used the value
$T_0=210$ MeV for the parameter in the Polyakov loop potential.
The result is displayed in Fig.~\ref{costa2}, where it is seen that
a gap of approximately 30 MeV between the two transitions persists for
all values of $|qB|$, with $T_c$ for the chiral transition being higher
as before. The interesting feature here is not the gap as such
since this can probably be tuned by using a different value of $T_0$;
rather it is the similar behavior of the curves.

In a recent paper,~\textcite{ferro2} studied the possibility of
inverse magnetic catalysis using the NJL model in the lowest-Landau-level
approximation. The starting point is the gap $M$
at zero temperature, which
is given by
\bqa
M&=&{2G\Lambda\over G+G^{\prime}}
\exp\left[{-2\pi^2\over(G+G^{\prime})N_c|q_fB|}\right]\;,
\eqa
where $G^{\prime}$ is given by Eq.~(\ref{int2b}) and 
the condensate $\xi$ is given by Eq.~(\ref{propcond}).
In this approximation, the phase transition is of second order and
the critical temperature is given by
\bqa\nonumber
T_c&=&1.16\sqrt{|q_fB|}
\exp\left[-{2\pi^2\over(G+G^{\prime})N_c|q_fB|}\right]\;.
\\ &&
\eqa
In the absence of a magnetic field, the coupling constant $G$ is
related to the strong coupling constant $\alpha_s$
via one-gluon exchange as $G=4\pi\alpha_s/\Lambda^2$.
In a magnetic field, the strong coupling splits into $\alpha_s^{\parallel}$
and $\alpha_s^{\perp}$, and only the latter depends on $B$.
Since $|q_fB|$ effectively acts as a cutoff in the LLL approximation,
the effective coupling becomes 
$G=4\pi\alpha_s^{\parallel}/|q_fB|$ and so the critical temperature
goes like
\bqa
T_c&=&1.16\sqrt{|q_fB|}
\exp\left[-{\pi\over2N_c\alpha_s^{\parallel}}\right]\;.
\eqa
Since $\alpha_s^{\parallel}$ is a decreasing function of the 
magnetic field~\cite{ferro2}, it is clear that $T_c$ decreases with $B$.

\begin{figure}
\begin{center}
\setlength{\unitlength}{1mm}
\includegraphics[width=7.0cm]{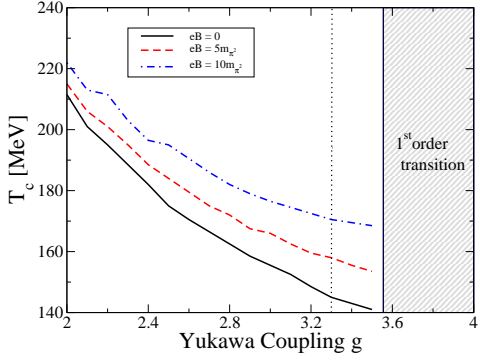}
\caption{$T_c$ as a function of the Yukawa coupling $g$
for various values of $B$. $B=0$: black solid line, 
$|qB|=5m_{\pi}^2$: red dashed line, and $|qB|=10m_{\pi}^2$: blue dash-dotted
line.Figure taken from~\textcite{mintz}.}
\label{fraga23}
\end{center}
\end{figure}

Finally, we mention that there have been attempts at obtaining
inverse
magnetic catalysis by varying the Yukawa coupling $g$ in the QM 
model~\cite{mintz}. 
This is possible if $g$ is an increasing function of $B$, see
Fig.~\ref{fraga23}. However,any curve $g(B)$ must start at $g(0)=3.22$
(indicated by the vertical dotted line) and successively cross the
red and black curves. One therefore soon enters the shaded region which
indicates a first-order transition in the QM model.
Since lattice results show that the transition is a crossover,
magnetic catalysis is ruled out.

\begin{figure}
\begin{center}
\setlength{\unitlength}{1mm}
\includegraphics[width=7.0cm]{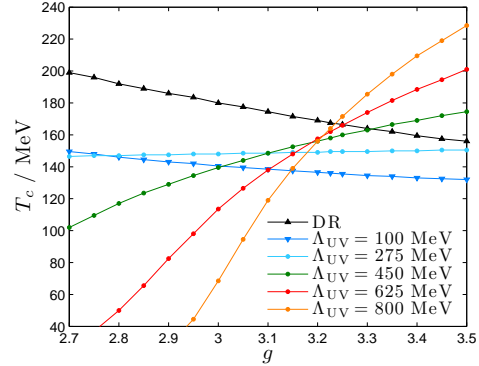}
\caption{Critical temperature as a function of the Yukawa coupling $g$
for various values of the sharp cutoff as well as a renormalization
scale of $\Lambda=100$ MeV. Figure taken from~\textcite{robb2}.}
\label{robb2}
\end{center}
\end{figure}

Motivated by the recent work on inverse magnetic catalysis at finite 
temperature,~\textcite{robb2} studied the quark-meson model using both 
dimensional regularization and a sharp cutoff $\Lambda_{\rm UV}$.
The critical temperature for the chiral transition was calculated as 
a function of the Yukawa coupling using different values of a sharp
cutoff. The results are shown in Fig.~\ref{robb2}.
The results using dimensional regularization and
a renormalization scale of $\Lambda=100$ MeV and a low value for the  
sharp cutoff are in qualitative agreement with the results of~\textcite{mintz},
namely a decreasing transition temperature as a function of 
$g$.\footnote{The result in DR is insensitive to
precise value of $\Lambda$.} At larger values of
the sharp cutoff, i.e. for more reasonable cutoffs, the transition
temperature is an increasing function of the Yukawa coupling.
This suggests that magnetic catalysis is much more delicate than
using a $B$-dependent coupling constant, cf. the discussion of the
sea and valence effects in Sec.~\ref{latsec}.

\section{Anisotropic pressure and magnetization}
\label{anis}
A constant magnetic field $B$ along the $z$-axis breaks 
Lorentz invariance. One of the consequences is that the pressure
is anisotropic: the pressure parallel to the magnetic field
${\cal P}_{\parallel}\equiv{\cal P}_z$ 
is different from the pressure perpendicular
to the magnetic field ${\cal P}_{\perp}\equiv{1\over2}({\cal P}_x+{\cal P}_y)$.
In this section, we review this topic and show how 
${\cal P}_{\parallel}$ and ${\cal P}_{\perp}$ are related to the 
different components of the energy-momentum tensor~\cite{mike+}.
We will use a Bose gas as a specific example.

The energy density ${\cal E}$ and the components of the
pressure ${\cal P}_i$ ($i=x,y,z$)
are given by the total
derivatives of the partition function
\bqa
{\cal E}=-{1\over V}{d\ln{\cal Z}\over d\beta}\;,\\
{\cal P}_i=
{L_i\over\beta V}{d\ln{\cal Z}\over dL_i}\;,
\label{pressuredef}
\eqa
where the quantization volume is $V=L_xL_yL_z$.
Taking the derivatives in Eq.~(\ref{pressuredef})
with respect to $L_x$ or $L_y$, it 
essential to distinguish between two cases; in the first case,
one keeps the magnetic field $B$ fixed and in the second case,
one keeps the magnetic flux $\Phi$ fixed~\cite{balianis}.
In the first case, a total derivative with respect to
$L_x$ or $L_y$ can be replaced by a partial derivative, while in
the second case, we must take into account the implicit dependence
of $L_x$ or $L_y$ on the magnetic field $B$ into account.
These two cases are referred to as the $B$-scheme and the $\Phi$-scheme,
respectively.
With obvious notation, we can write the pressures as 
\bqa
\label{pressurede2}
{\cal P}_i^B&=&{L_i\over\beta V}{\partial\ln{\cal Z}\over\partial L_i}\;,
\\ 
{\cal P}_i^{\Phi}&=&
{L_i\over\beta V}{\partial\ln{\cal Z}\over\partial L_i}
+{L_i\over\beta V}{\partial\ln{\cal Z}\over\partial B}
{\partial B\over\partial L_i}\;.
\label{pressurede3}
\eqa
Using the definition of the magnetization, 
$q{\cal M}={1\over\beta V}{\partial\ln{\cal Z}\over\partial B}$
and $BL_xL_y=\Phi={\rm const}$, we can write
\bqa
{\cal P}_{x,y}^{P}&=&{\cal P}_z^{B}\;,\\
{\cal P}_z^{\Phi}&=&{\cal P}_z^{B}\;,\\
{\cal P}_{x,y}^{\Phi}&=&{\cal P}_{x,y}^{B}
-qBM\;.
\label{pperp}
\eqa
Finally, we note that the energy density is the same, 
${\cal E}^B={\cal E}^{\Phi}$.

We next relate the pressure defined above to the 
expectation value of various components
of the energy-momentum tensor.
The conventional energy-momentum tensor ${\cal T}_{\mu\nu}$ in 
a constant magnetic background is given by
\bqa
{\cal T}_{\mu\nu}&=&
{\partial{\cal L}\over\partial(\partial^{\mu}\phi)}\partial_{\nu}\phi
-\eta_{\mu\nu}{\cal L}\;.
\eqa
For a massive complex bosonic field coupled to 
an external Abelian gauge field $A_{\mu}$ with 
Lagrangian ${\cal L}=(D_{\mu}\Phi)^{\dagger}(D^{\mu}\Phi)-m^2\Phi^{\dagger}\Phi$,
the energy-momentum tensor reads
\bqa\nonumber
{\cal T}_{\mu\nu}&=&
(\partial_{\mu}\Phi)^{\dagger}(D_{\nu}\Phi)
+(D_{\mu}\Phi)^{\dagger}(\partial_{\nu}\Phi)
-\eta_{\mu\nu}{\cal L}\;.
\\ &&
\eqa
However, this definition is neither gauge invariant nor 
symmetric~\cite{portillo,mike+}. 
There is another definition of ${\cal T}_{\mu\nu}$ 
that guarantees it is gauge invariant and 
symmetric. 
The method is that of metric perturbations, which
is based on the fact that matter fields couple to gravity and that
the energy-momentum tensor acts as a source for it.
The energy-momentum tensor is found by calculating the relation
between the variation of the metric and the variation of the action
according to 
\bqa\nonumber
\label{deltas}
\delta S&=&{1\over2}\int d^4x\sqrt{-g}\,\,{\cal T}^{\mu\nu}\delta g_{\mu\nu}
\;,
\eqa
where the action is
\bqa
S=\int d^4x\sqrt{-g}\,\left[{\cal L}_{\rm scalar}+{\cal L}_{\rm EM}\right]\;.
\eqa
Here $g$ is minus the determinant of the metric 
$g_{\mu\nu}$ and 
the Lagrangian densities are~\cite{birr}
\bqa
\label{cont1}
{\cal L}_{\rm EM}&=&-{1\over4}F^{\alpha\beta}F^{\gamma\delta}
g_{\alpha\gamma}g_{\beta\delta}\;,
\\
{\cal L}_{\rm scalar}&=&(D^{\alpha}\Phi)^{\dagger}(D^{\beta}\Phi)g_{\alpha\beta}
-m^2\Phi^{\dagger}\Phi\;.
\label{cont2}
\eqa
In order to proceed, we need the variation of $\sqrt{-g}$, which 
is given by 
$\delta\sqrt{-g}=-{1\over2}\sqrt{-g}g^{\mu\nu}\delta g_{\mu\nu}$.
Using this and calculating the variation of the terms in Eqs.~(\ref{cont1})
and~(\ref{cont2}), one can calculate the variation $\delta S$
and read off the energy-momentum tensor using Eq.~(\ref{deltas}).
This yields 
${\cal T}_{\mu\nu}={\cal T}_{\mu\nu}^{EM}+{\cal T}_{\mu\nu}^{\rm scalar}$,
where
\bqa
\label{tmunu1}
{\cal T}^{\rm EM}_{\mu\nu}
&=&-{\cal F}_{\mu\alpha}F_{\nu}^{\,\,\alpha}-\eta_{\mu\nu}{\cal L}_{\rm EM}\;,\\
{\cal T}^{\rm scalar}_{\mu\nu}
&=&2(D_{\mu}\Phi)^{\dagger}(D_{\nu}\Phi)
-\eta_{\mu\nu}{\cal L}_{\rm scalar}\;,
\label{tmunu2}
\eqa
where we have made the replacement $g_{\mu\nu}\rightarrow\eta_{\mu\nu}$
at the end.
Clearly, the expressions Eqs.~(\ref{tmunu1}) and~(\ref{tmunu2})
are symmetric and gauge invariant.
Note that ${\cal T}^{\rm EM}_{\mu\nu}$ is traceless, while 
${\cal T}_{\mu\nu}^{\rm scalar}$ is traceless only in 
the massless case in 1+1 dimensions.

\begin{figure}
\begin{center}
\setlength{\unitlength}{1mm}
\includegraphics[width=7.0cm]{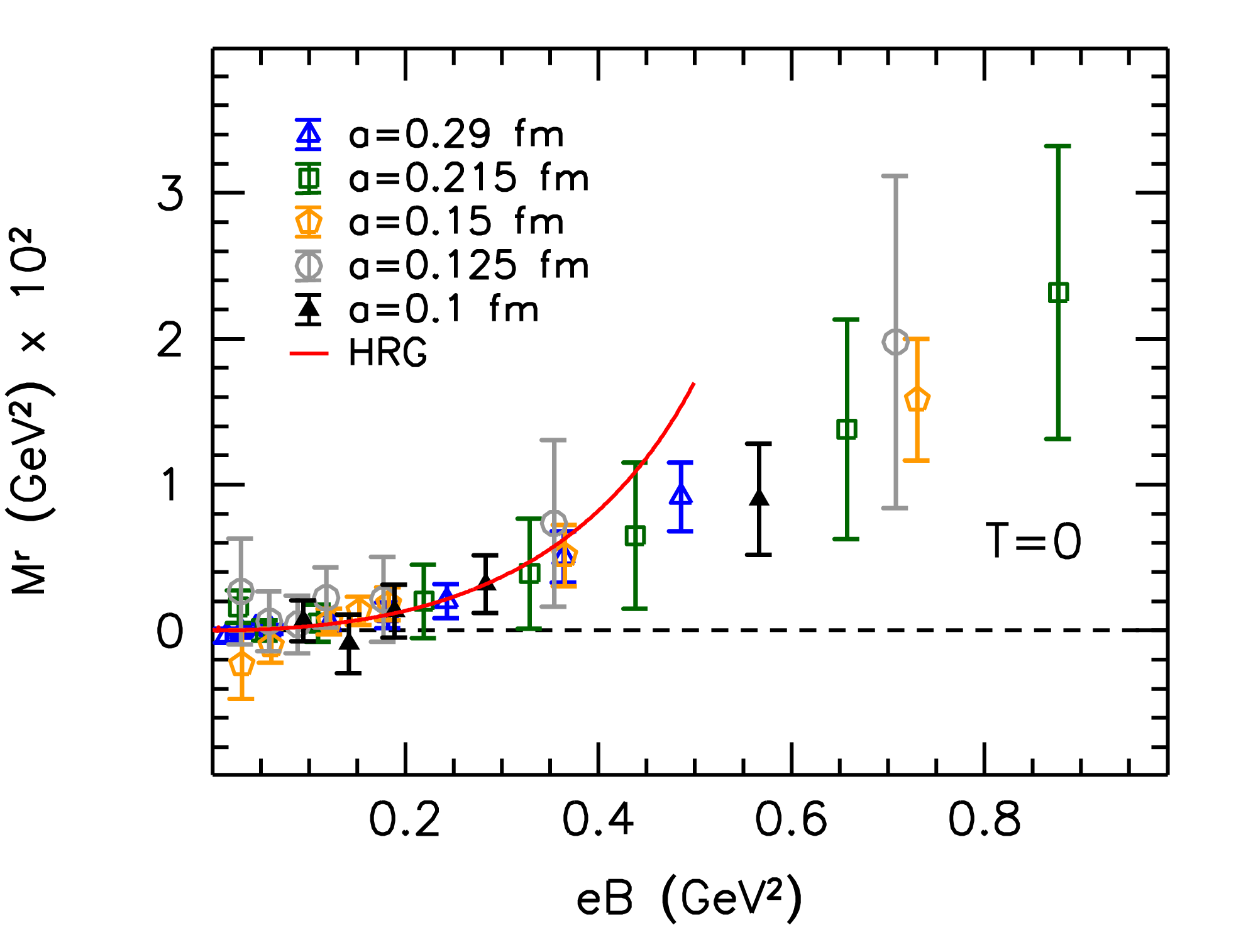}
\caption{Magnetization at zero temperature as a function of the magnetic
field calculated on the lattice for different lattice
spacings. Red curve is the result from the HRG calculation
by~\textcite{endrodires}.
Figure taken from~\textcite{balianis}.}
\label{magnetization}
\end{center}
\end{figure}

Specializing to a constant magnetic field $B$, one finds
${\cal T}_{\mu\nu}^{\rm EM}={1\over2}{\rm diag}(B^2,B^2,B^2,-B^2)$
and 
\begin{widetext}
\bqa
{\cal T}_{00}^{\rm scalar}
&=&
(\partial_0\Phi)^{\dagger}(\partial_0\Phi)
+(\partial_z\Phi)^{\dagger}(\partial_z\Phi)
+(D_{\perp}\Phi)^{\dagger}(D_{\perp}\Phi)
+m^2\Phi^{\dagger}\Phi
\;,
\\ 
{\cal T}_{zz}^{\rm scalar}
&=&
(\partial_0\Phi)^{\dagger}(\partial_0\Phi)
+(\partial_z)\Phi^{\dagger}(\partial_z\Phi)
-(D_{\perp}\Phi)^{\dagger}(D_{\perp}\Phi)
-m^2\Phi^{\dagger}\Phi
\;,\\ 
{1\over2}({\cal T}_{xx}^{\rm scalar}+{\cal T}_{yy}^{\rm scalar})
&=&
(\partial_0\Phi)^{\dagger}(\partial_0\Phi)
-(\partial_z\Phi)^{\dagger}(\partial_z\Phi)
-m^2\Phi^{\dagger}\Phi
\;,
\eqa
where we have defined 
$(D_{\perp}\Phi)^{\dagger}(D_{\perp}\Phi)
=(D_x\Phi)^{\dagger}(D_x\Phi)+(D_y\Phi)^{\dagger}(D_y\Phi)$.
The next step is to calculate expectation values of
the different components of ${\cal T}_{\mu\nu}$. 
By scaling the coordinate $z$ by a factor $\xi$, 
the partial derivative transformas as 
$\partial_z\rightarrow{1\over\xi}{\partial}_z$.
One can show that
the expectation value of ${\cal T}_{zz}$ can be expressed as
a derivative of the partition function~\cite{portillo},
\bqa
\langle{\cal T}_{zz}\rangle
&=&{L_z\over\beta V}{d\ln{\cal Z}\over dL_z}\;,
\eqa
and equals the pressure ${\cal P}_{zz}$ in both schemes.
If one scales the $x$ and $y$ coordinates in the same manner, 
the covariant derivative transforms in the same way as a partial derivative
only if the strength of the magnetic field transforms as 
$B\rightarrow\mbox{$B\over\xi^2$}$. This is the $\Phi$-schemed
as defined above. One then finds
\bqa
{1\over2}\langle{\cal T}_{xx}^{\rm scalar}+{\cal T}_{yy}^{\rm scalar}\rangle
&=&{1\over2}{1\over\beta V}
\left[{L_x}{d\ln{\cal Z}\over dL_x}+L_y{d\ln{\cal Z}\over dL_y}\right]\;.
\eqa
\end{widetext}
Since this corresponds to the $\Phi$-scheme, we conclude that 
\bqa
\langle{\cal T}_{xx}+{\cal T}_{yy}\rangle&=&
{\cal P}^{\Phi}_{xx}+{\cal P}^{\Phi}_{yy}\;.
\eqa
Let us close this section by discussing the magnetization, which is
defined by 
\bqa
{\mathcal M}&=&{1\over\beta V}{1\over q}{\partial\ln{\cal Z}\over\partial B}\;.
\eqa
One is not interested in the ${\cal O}(B^2)$ contribution 
to the pressure coming from the 
external magnetic field. Using the renormalization 
given by Eq.~(\ref{endren}), this term is given by $B_r^2$
in Eq.~(\ref{endren1}) and is subtracted~\cite{endrodires,balianis}.
The magnetization as a function of the magnetic field at $T=0$
has been calculated on the lattice by~\textcite{balianis}
using various lattice spacings and the results are shown in 
Fig.~\ref{magnetization}. For comparison, the result from the HRG
calculation of \textcite{endrodires} is also shown (red line).
Note that as a result of the subtraction mentioned above, the magnetization
is positive, suggesting that the QCD vacuum is paramagnetic.

\section{Conclusions and outlook}

In this review, we have discussed a number of low-energy models and 
theories that are used to describe QCD in a magnetic background
at zero and finite temperature. 

One aspect we think is missing in the literature is systematic
studies of various approximations. As we have discussed, 
parameter fixing is important and nontrivial.
There are many papers in which
the authors employ a certain model and a specific set of parameters.
However, it would be very useful to compare various 
approximations and levels of sophistication using the 
same values for physical quantities.
For example, it would be useful to calculate
the critical temperature or the magnetization 
in the NJL model and its Polyakov-loop extended version and compare the two.
One of the few papers where a systematic study is carried out is 
by~\textcite{kami}, where the critical temperature is 
calculated in the mean-field approximation, in the LPA, and beyond the LPA.

It is instructive to compare the results of the (P)NJL model and the 
(P)QM model at the mean-field level. In this case, only fermionic fluctuations
are taken into account. If one takes into account the vacuum fluctuations
of the (P)QM model, the results are similar. For example, the 
nature of the transition is the same and the phase diagrams
closely resemble each other. This is not surprising as one
is essentially evaluating the same fermionic functional determinant.
However, we have seen that if the vacuum contribution is
omitted in the (P)QM model, the crossover at finite pion mass turns into
a first-order transition. In the same manner, the second-order
transition in the chiral limit becomes first order.
In the (P)NJL model it makes no sense to
subtract the vacuum fluctuations as they are
responsible for spontaneous symmetry breaking in the vacuum.
In the (P)QM model, spontaneous symmetry breaking is present already at
tree level provided by the quartic Higgs potential.

Regarding the calculations using the functional renormalization group,
there are several directions for further improvement.
For the physically most interesting cases $N_f=2$ and 
$N_f=3$ (see~\textcite{kana2} for a recent application with $N_f=2+1$),
one should solve the flow equation for the effective potential
as a function of the two invariants $\rho$ and $\Delta$,
including wavefunction renormalization terms in the calculations.
One might consider another regulator function that ensures
$Z_{k=0}^{\parallel}=Z_{k=0}^{\perp}$ in the vacuum.
Finally, including new condensates that are invariant under rotations
about the $z$-axis is of interest. However, this requires
the inclusion of new interaction 
terms in the Lagrangian of the quark-meson model
and the problem is that one does not know the value of their couplings.

The most important issue that we have discussed,
is the disagreement between model and lattice calculations regarding
the behavior of the transition temperature as a function of the magnetic field.
On the lattice, two contributions to the quark condensate have been identified,
namely the valence and the sea contribution. While the former 
increases as a function of $B$ for all temperatures, the behavior of the
latter is more complicated.
At zero temperature the sea contribution 
is also increasing with the magnetic field and
together with the valence contribution, they give rise to magnetic catalysis.
Around the transition temperature, however, it decreases as a function of
$B$ for physical quark masses.
The sea contribution overwhelms the valence contribution such that 
there is a net suppression of the condensate, which leads to inverse
magnetic catalysis and a decrease
of the transition temperature as a function of $B$.
The mechanism behind this effect is that the magnetic field in the 
quark determinant changes the relative weight of the gauge configurations
and that gauge configurations with 
larger values of the condensate are suppressed by the
quark determinant~\cite{endrodi3}.
Moreover, using staggered fermions, the back-reaction of the
quark determinant on the glue sector is very delicate; small quark masses
lead to inverse catalysis while large values lead to magnetic catalysis.
This dependence on the quark mass may very well be different if one
uses chiral fermions.
Calculations employing the (P)NJL model or the (P)QM show 
a different behavior; the transition temperature increases with
the strength of the magnetic field.
In hindsight, this disagreement should perhaps not be surprising as
there is no sea effect in the (P)NJL and (P)QM models.
However, it is interesting to notice that the coupling to the 
Polyakov loop in the QM model, 
gives less magnetic catalysis around the transition 
temperature than without, cf. Fig.~\ref{tcB}.
Regarding the attempts to modify models to accommodate inverse magnetic
catalysis, most of them do not couple the Polyakov loop variable
to the quark determinant and therefore does not include
the underlying mechanism. The idea of using a $B$-dependent 
parameter $T_0$ in the Polyakov loop potential 
was implemented by~\textcite{mintz}; however, it 
was shown not to lead to magnetic catalysis.
\\
\indent
In conclusion, although the level of complexity and sophistication in model 
calculations of the QCD transition  in a magnetic background
are steadily improving, it remains a 
challenge to properly incorporate the phenomenon of inverse magnetic catalysis. 

\section*{Acknowledgments}
The authors would like to thank G. Bali, D. Boer,
T. Brauner, F. Bruckmann, M. Chernodub, P. Costa,
G. Endrodi, E.~S. Fraga, V. de la Incera, T. Kalaydzhyan
K. Kamikado, T. Kanazawa, S. Ozaki, M.~B. Pinto, A. Schmitt,
L. von Smekal, R. Stiele, M. Strickland, and P. Watson.
for useful discussions.
We also thank the authors who have granted us permission to use their figures
in this review.

\appendix
\section{Notation and conventions}
We use the Minkowski metric $g_{\mu\nu}={\rm diag}(1,-1,-1,-1)$ and
natural units such that $k_B=\hbar=c=1$.

We will be using the Dirac and
chiral representations of the $\gamma$-matrices.
In Minkowski space they are given respectively by
\bqa
\gamma^0=\left(
\begin{array}{cc}
1&0\\
0&-1
\end{array}\right)\;,
\hspace{0.3cm}
{\boldsymbol\gamma}=\left(
\begin{array}{cc}
0&{\boldsymbol \sigma}\\
-{\boldsymbol \sigma}&0
\end{array}\right)\;,
\\
\gamma^0=\left(
\begin{array}{cc}
0&1\\
1&0
\end{array}\right)\;,
\hspace{0.5cm}
{\boldsymbol\gamma}=\left(
\begin{array}{cc}
0&{\boldsymbol \sigma}\\
-{\boldsymbol \sigma}&0
\end{array}\right)\;,
\eqa
where ${\boldsymbol \sigma}=(\sigma_x,\sigma_y,\sigma_z)$
are the Pauli spin matrices.
The Euclidean $\gamma$-matrices are related to the 
$\gamma^{\mu}$-matrices in Minkowski space by $\gamma_j=i\gamma^j$
and $\gamma_4=\gamma^0$. They satisfy
\bqa
\left\{\gamma_{\mu},\gamma_{\nu}\right\}
=2\delta_{\mu\nu}{\bf 1}\;,
\hspace{0.6cm}
{\rm Tr}[\gamma_{\mu}\gamma_{\nu}]=
4\delta_{\mu\nu}\;,
\eqa
where ${\bf 1}$ is the unit matrix in four dimensions.

\section{Sum-integrals}
\label{sumint}
The bosonic and fermionic sum-integrals are defined by
\bqa
\label{defsumint1}
\sumint_P&=&T\sum_{P_0=2\pi nT}\int_p\;,
\\
\sumint_{\{P\}}&=&T\sum_{P_0=(2\pi +1)nT}\int_p
\label{defsumint2}
\;,
\eqa
where $P_0=2\pi nT$ for bosons and $P_0=(2n+1)\pi T$ for fermions. 
where the integral over $p$ is 
\bqa
\int_p&=&\left({e^{\gamma_E}\Lambda^2\over4\pi}\right)^{\epsilon}
\int{d^dp\over(2\pi)^d}\;,
\eqa
and where $d=3-2\epsilon$.
The prefactor $\left({e^{\gamma_E}\Lambda^2\over4\pi}\right)^{\epsilon}$
is chosen such that $\Lambda$
is associated with renormalization scale in the modified
minimal subtraction scheme $\overline{\rm MS}$.
Here $\gamma_E$ is the Euler-Mascheroni constant.
In the case of particles with electric charge $q$ moving in a constant
magnetic field, the sum-integral is a sum over Matsubara frequencies
$P_0$, a sum of Landau levels $k$, and an integral over momenta 
in $d=1-2\epsilon$ dimensions. For fermions, we also sum over spin $s$.
We define for bosons and fermions,
respectively
\bqa
\sumint_P^B&=&{|qB|\over2\pi}T\sum_{k=0}^{\infty}
\sum_{P_0=2\pi nT}\int_{p_z}\;,
\label{defsumint3}
\\
\sumint_{\{P\}}^B&=&{|q_fB|\over2\pi}T
\sum_{s=\pm1}
\sum_{k=0}^{\infty}
\sum_{P_0=(2n+1)\pi T}\int_{p_z}\;,
\label{defsumint4}
\eqa
where the integral is
\bqa
\int_{p_z}&=&
\left({e^{\gamma_E}\Lambda^2\over4\pi}\right)^{\epsilon}
\int{d^{d-2}p\over(2\pi)^{d-2}}\;.
\eqa
\begin{widetext}
Eqs.~(\ref{defsumint3}) and~(\ref{defsumint4})
reduce to Eqs.~~(\ref{defsumint1}) and~(\ref{defsumint2})
in the limit $B\rightarrow0$.

The specific sum-integrals we need are
\bqa
\sumint_P\ln\big[P_0^2+p^2+m^2\Big]&=&
-{1\over2(4\pi)^2}\left({\Lambda^2\over m^2}\right)^{\epsilon}
\left[\left({1\over\epsilon}+{3\over2}\right)m^4+
2J_0(\beta m)T^4+{\cal O}(\epsilon)
\right]\;,
\label{sumb}
\\ \nonumber
\sumint_P^B\ln\Big[P_0^2+p_z^2+M_B^2\Big]&=&
{1\over2(4\pi)^2}\left({\Lambda^2\over|2qB|}\right)^{\epsilon}
\left[\left({(qB)^2\over3}-m^4\right)\left({1\over\epsilon}+1\right)
\right. \\ &&\left.
+8(qB)^2\zeta^{(1,0)}(-1,\mbox{${1\over2}$}+x)
-2J_0^B(\beta m)|qB|T^2+{\cal O}(\epsilon)
\right]\;,
\label{sumbose}
\\ 
\sumint_{\{P\}}\ln\Big[P_0^2+p^2+m_f^2\Big]&=&
-{1\over2(4\pi)^2}\left({\Lambda^2\over m_f^2}\right)^{\epsilon}
\left[\left({1\over\epsilon}+{3\over2}\right)m_f^4-
2K_0(\beta m_f)T^4+{\cal O}(\epsilon)
\right]\;,
\label{sumf}
\\ \nonumber
\sumint_{\{P\}}^B\ln\Big[P_0^2+p_z^2+M_B^2\Big]&=&
-{1\over(4\pi)^2}\left({\Lambda^2\over|2qB|}\right)^{\epsilon}
\left[\left({2(qB)^2\over3}+m_f^4\right)\left({1\over\epsilon}+1\right)
-8(qB)^2\zeta^{(1,0)}(-1,x)
\right.\\ &&\left. 
-2|q_fB|m_f^2\ln x_f-2K_0^B(\beta m_f)|q_fB|T^2+{\cal O}(\epsilon)
\right]\;,
\label{sumfermi}
\\
\sumint_P{1\over(P_0^2+p^2+m^2)}&=&
-{1\over(4\pi)^2}\left({\Lambda^2\over m^2}\right)^{\epsilon}
\left[\left({1\over\epsilon}+1\right)m^2-
J_1(\beta m)T^2+{\cal O}(\epsilon)
\right]\;,
\\ \nonumber
\sumint_P^B{1\over(P_0^2+p_z^2+M_B^2)}&=&
-{1\over(4\pi)^2}\left({\Lambda^2\over|2qB|}\right)^{\epsilon}
\bigg[
{1\over\epsilon}m^2
-\zeta^{(1,0)}(\mbox{0,${1\over2}$}+x_f)|qB|
\\ &&
-J_1^B(\beta m)|qB|
+{\cal O}(\epsilon)
\bigg]\;,
\\ 
\sumint_{\{P\}}{1\over(P_0^2+p_z^2+m_f^2)}&=&
-{1\over(4\pi)^2}\left({\Lambda^2\over m_f^2}\right)^{\epsilon}
\left[
\left({1\over\epsilon}+1\right)m_f^2
+K_1(\beta m_f)T^2+{\cal O}(\epsilon)
\right]
\\ \nonumber
\sumint_{\{P\}}^B{1\over(P_0^2+p_z^2+M_B^2)}&=&
-{1\over(4\pi)^2}\left({\Lambda^2\over|2qB|}\right)^{\epsilon}
\left[
{1\over\epsilon}m_f^2
-2\zeta^{(1,0)}(0,x)|q_fB|
+K_1^B(\beta m_f)|q_fB|
+{\cal O}(\epsilon)
\right]\;,
\\ &&
\eqa
where $x={m^2\over2|qB|}$, $x_f={m_f^2\over2|q_fB|}$. 
The bosonic and fermionic masses are $M_B=\sqrt{p_z^2+m^2+|qB|(2k+1)}$
and $M_B=\sqrt{p_z^2+m_f^2+|q_fB|(2k+1-s)}$, respectively.
The generalized zeta-function is defined by
$\zeta(s,q)=\sum_{k=0}^{\infty}{1\over(q+k)^s}$.
The thermal functions $J_n(\beta M)$ , $J_n^B(\beta M)$, 
$K_n(\beta m_f)$, and $K_n^B(\beta m_f)$ 
are defined
\bqa
J_n(\beta m)&=&
{4e^{\gamma_E\epsilon}\Gamma(\mbox{${1\over2}$})\over
\Gamma(\mbox{${5\over2}$}-n-\epsilon)}
\beta^{4-2n}m^{2\epsilon}
\int_0^{\infty}
{p^{4-2n-2\epsilon}dp\over\sqrt{p^2+m^2}}{1\over e^{\beta\sqrt{p^2+m^2}}-1}\;,
\\ 
J_n^B(\beta m)&=&
{8e^{\gamma_E\epsilon}\Gamma(\mbox{${1\over2}$})\over
\Gamma(\mbox{${3\over2}$}-n-\epsilon)}
\beta^{2-2n}(|2qB|)^{\epsilon}
\sum_{k=0}^{\infty}\int_0^{\infty}
{p_z^{2-2n-2\epsilon}dp_z\over\sqrt{p_z^2+M_B^2}}
{1\over e^{\beta\sqrt{p_z^2+M_B^2}}-1}\;,
\\
\label{kn}
K_n(\beta m_f)&=&
{4e^{\gamma_E\epsilon}\Gamma(\mbox{${1\over2}$})\over
\Gamma(\mbox{${5\over2}$}-n-\epsilon)}
\beta^{4-2n}m_f^{2\epsilon}
\int_0^{\infty}
{p^{4-2n-2\epsilon}dp\over\sqrt{p^2+m_f^2}}{1\over e^{\beta\sqrt{p^2+m_f^2}}+1}\;,
\\ 
\label{knb}
K_n^B(\beta m_f)&=&
{4e^{\gamma_E\epsilon}\Gamma(\mbox{${1\over2}$})\over
\Gamma(\mbox{${3\over2}$}-n-\epsilon)}
\beta^{2-2n}(|2qB|)^{\epsilon}
\sum_{s=\pm1}
\sum_{k=0}^{\infty}
\int_0^{\infty}
{p_z^{2-2n-2\epsilon}dp_z\over\sqrt{p_z^2+M_B^2}}
{1\over e^{\beta\sqrt{p_z^2+M_B^2}}+1}\;.
\eqa
The sum over Matsubara frequencies is
\bqa
{1\over2}T\sum_{P_0}\ln\left[
P_0^2+\omega^2
\right]&=&{1\over2}\omega+T\ln\left[1\pm e^{-\beta\omega}\right]\;,
\label{matsusum}
\eqa
where the upper sign is for fermions and the lower signs is for bosons.

\section{Small and large-$B$ expansions}\label{smallarge}
In this appendix we list a number small and large-$B$ expansions
of various $\zeta$-functions.
The small-$x$ expansions of the various derivatives
of the Hurwitz zeta-functions are
\bqa
\zeta^{(1,0)}(-1,x)&=&
\zeta^{\prime}(-1)+
{1\over2}x-{1\over2}\ln(2\pi)x
-x\ln x
+{\cal O}\left(x^2\right)\;,
\\
\zeta^{(1,0)}(-1,\mbox{${1\over2}$}+x)&=&
-{1\over2}\zeta^{\prime}(-1)
-{1\over24}\ln2
-{1\over2}x\ln2+{\cal O}\left(x^2\right)\;,
\\
\zeta^{(1,0)}(0,x)&=&
-{1\over2}\ln(2\pi)-\ln x 
-\gamma_Ex
+{\cal O}\left(x^2\right)\;,
\\
\zeta^{(1,0)}(0,\mbox{${1\over2}$}+x)&=&
-{1\over2}\ln2
-2\ln2x-x\gamma_E
+{\cal O}\left(x^2\right)\;,
\label{small4}
\eqa
where $\zeta^{\prime}(-1)={1\over12}-\ln(A)\approx-0.165421$ and $A$ is the
Glaisher-Kinkelin constant.
The large-$x$ expansion of the the various derivatives of the
Hurwitz zeta-functions are
\bqa
\label{largefermi}
\zeta^{(1,0)}(-1,x)&=&
-{1\over4}x^2
+{1\over2}x^2\ln x-{1\over2}x\ln x
+{1\over12}\ln x
+{1\over12}
+{\cal O}\left({1\over x^2}\right)\;,
\\
\zeta^{(1,0)}(-1,\mbox{${1\over2}$}+x)&=&
-{1\over4}x^2
+{1\over2}x^2\ln x-{1\over24}\ln x
-{1\over24}
+{\cal O}\left({1\over x^2}\right)\;,
\\
\zeta^{(1,0)}(0,x)&=&
x\ln x-x-{1\over2}\ln x+{1\over12x}+{\cal O}\left({1\over x^3}\right)\;,
\label{largexgap}
\\
\zeta^{(1,0)}(0,\mbox{${1\over2}$}+x)&=&
x\ln x-x-{1\over24x}+{\cal O}\left({1\over x^3}\right)\;.
\eqa

\section{Propagators in a magnetic background}
\label{propapp}
In this Appendix, we briefly discuss the boson and propagator
in a constant magnetic background. Denoting the 
bosonic propagator in coordinate space by $\Delta({x,x^{\prime}})$, it
satisfies the equation
\bqa
\left[
\partial_{x^0}^2-\partial_{x^3}^2-\partial_{x^1}^2
-\left(\partial_{x^2}-iqA_2\right)^2
+m^2
\right]\Delta(x,x^{\prime})&=&\delta^4(x-x^{\prime})\;,
\label{coord}
\eqa
where we have chosen the Landau gauge, $A^{\mu}=(0,0,Bx,0)$.
We next introduce the propagator 
$\Delta({\bf p}_{\parallel},{\bf x}_{\perp},{\bf x}^{\prime}_{\perp})$
via the Fourier transform  
\bqa
\Delta(x,x^{\prime})&=&\int{d^2{p}_{\parallel}\over(2\pi)^2}
e^{-i{\bf p}_{\parallel}
({\bf x}_{\parallel}-{\bf x}_{\parallel}^{\prime})}
\Delta({\bf p}_{\parallel},{\bf x}_{\perp},{\bf x}^{\prime}_{\perp})
\;,
\label{part}
\eqa
where ${\bf p}_{\parallel}=(p_0,p_3)$, 
${\bf x}_{\parallel}=(x^0,x^3)$, and ${\bf x}_{\perp}=(x^1,x^2)$.
Inserting Eq.~(\ref{part}) into Eq.~(\ref{coord}), we obtain
\bqa
\left[-p_0^2+p_3^2+m^2
-\partial^2_{x^1}
-\left({\partial\over\partial x^2}-iqBx\right)^2
\right]\Delta({\bf p}_{\parallel},{\bf x}_{\perp},{\bf x}^{\prime}_{\perp})
&=&\delta^2({\bf x}_{\perp}-{\bf x}^{\prime}_{\perp})\;.
\eqa
We next need a complete set of eigenfunctions of the operator
$\partial^2_{x^1}
+\left({\partial\over\partial x^2}+iqBx\right)^2$, which are the
well-known solutions involving the Hermite polynomials $H_k(x)$.
The normalized wavefunctions are
\bqa
\psi_{k,p_2}({\bf x}_{\perp})&=&
{1\over{\sqrt{2\pi l}}}
{1\over\sqrt{2^kk!\sqrt{\pi}}}
H_k\left(\mbox{$x^1\over l$}+p_2l\right)e^{-{1\over2l^2}(x^1+p_2l^2)^2}
e^{isx^2p_2}\;,
\eqa
where $s_{\perp}={\rm sign}(qB)$ and
$l^2=1/|qB|$. These functions satisfy the usual orthonormality
and completeness relations:
\bqa
\int d^2{x}_{\perp}\psi^*_{k,p_2}({\bf x}_{\perp})
\psi_{k^{\prime},p_2^{\prime}}({\bf x}_{\perp})&=&
\delta_{kk^{\prime}}\delta(p_2-p_2^{\prime})\;, \\
\int_{-\infty}^{\infty}dp_2\sum_{k=0}^{\infty}
\psi_{k,p_2}({\bf x}_{\perp})\psi_{k,p_2}^*({\bf x}_{\perp}^{\prime})
&=&\delta^2({\bf x}_{\perp}-{\bf x}_{\perp}^{\prime})\;.
\label{complete}
\eqa
Using the completeness relation~(\ref{complete}), the propagator can be written
as
\bqa
\Delta({\bf p}_{\parallel},{\bf x}_{\perp},{\bf x}^{\prime}_{\perp})
&=&
\int_{-\infty}^{\infty}dp_2\sum_{k=0}^{\infty}\left[
-p_0^2+p_3^2+m^2+|qB|(2k+1)
\right]^{-1}
\psi_{k,p_2}({\bf x}_{\perp})\psi_{k,p_2}^*({\bf x}^{\prime}_{\perp})\;,
\eqa
which after some algebra can be written as 
\bqa\nonumber
\Delta({\bf p}_{\parallel},{\bf x}_{\perp},{\bf x}^{\prime}_{\perp})
&=&{1\over2\pi l}\int_{-\infty}^{\infty}dp_2\sum_{k=0}^{\infty}\left[
-p_0^2+p_3^2+m^2+|qB|(2k+1)
\right]^{-1}
{1\over2^kk!\sqrt{\pi}}
\\ &&\times \nonumber
H_k(\mbox{${x^1\over l}$}+p_2l)H_k(\mbox{${x^{\prime1}\over l}$}+p_2l)
e^{-[p_2l+(x^1+x^{\prime 1})/2l-is(x^2-x^{\prime 2})/2l]^2}
\\ &&\times
e^{-\mbox{${1\over4l^2}$}(x^2-x^{\prime 2})^2}e^{\mbox{${1\over4l^2}$}(x^1-x^{\prime 1})^2}
e^{-{is\over2l^2}(x^1+x^{\prime 1})(x^2-x^{\prime 2})}\;.
\label{longex}
\eqa
We next need the following integral 
\bqa
\int_{-\infty}^{\infty}dx\,
e^{-x^2}H_k(x+z)H_k(x+w)&=&2^kk!\sqrt{\pi}L_k(-2zw)\;,
\eqa
where $L_k(x)$ is a Laguerre polynomial of order $k$.
In Eq.~(\ref{longex}),
we make the substitution 
$p_2^{\prime}=p_2+(x^1+x^{\prime 1})/2l^2-i(x^2-x^{\prime 2})/2l^2$
and so we identify $z=(x^1-x^{\prime 1})/2l-i(x^2-x^{\prime 2})/2l$
and $w=-(x^1-x^{\prime 1})/2l-i(x^2-x^{\prime 2})/2l$. This implies that
$-2zw=({\bf x}_{\perp}-{\bf x}^{\prime}_{\perp})^2/2l^2$
and we can write
\bqa\nonumber
\Delta({\bf p}_{\parallel},{\bf x}_{\perp},{\bf x}^{\prime}_{\perp})
&=&{1\over2\pi l^2}\sum_{k=0}^{\infty}\left[-p_0^2+m^2+p_3^2+|qB|(2k+1)\right]^{-1}
e^{-{1\over4l^2}({\bf x}_{\perp}-{\bf x}^{\prime}_{\perp})^2}
L_k\left(\mbox{${({\bf x}_{\perp}-{\bf x}^{\prime}_{\perp})^2\over2l^2}$}\right)
\\ &&\times
e^{-is\Phi({\bf x}_{\perp},{\bf x}_{\perp}^{\prime})}
\;,
\label{trpsch}
\eqa
where the so-called Schwinger phase is
\bqa
\Phi({\bf x}_{\perp},{\bf x}_{\perp}^{\prime})
&&={(x^1+x^{\prime1})(x^2-x^{\prime2})\over 2l^2}\;.
\label{svingerfase}
\eqa
The propagator in Eq.~(\ref{trpsch})
is now a product of a
translationally invariant part and the Schwinger phase.
The Fourier transform $\Delta(p_{\perp},p_{\parallel})$
of the translationally invariant part is
\bqa
\Delta({\bf p}_{\perp},{\bf p}_{\parallel})
&=&
-2e^{-p_{\perp}^2l^2}
\sum_{k=0}^{\infty}{(-1)^n\over p_0^2-p_z^2-m^2-|qB|(2k+1)}
L_k(2p_{\perp}^2l^2)\;.
\label{useful}
\eqa
The term $\left[p_0^2-p_z^2-m^2-|qB|(2k+1)\right]$
is rewritten using Schwinger's trick
\bqa
{i\over p_0^2-p_z^2-m^2-|qB|(2k+1)}
&=&\int_0^{\infty}ds\,e^{is[p_0^2-p_z^2-m^2-|qB|(2k+1)]}\;.
\eqa
Using the summation formula for the generalized Laguerre polynomials
$L_n^{\alpha}(x)$
\bqa
\sum_{k=0}^{\infty}L_k^{\alpha}(x)z^k
&=&(1-z)^{-(\alpha+1)}e^{{xz\over z-1}}
\;,
\eqa
the translationally invariant propagator can be written as 
\bqa
\Delta({\bf p}_{\perp},{\bf p}_{\parallel})
&=&
i\int{ds\over\cos(|qB|s)}\exp\left\{is\left[
{\bf p}_{\parallel}^2-m^2\right]-i{\bf p}_{\perp}^2{\tan(|qB|s)\over|qB|}
\right\}\;,
\eqa
Finally, the propagator $\Delta(x,x^{\prime})$ takes the form
\bqa
\Delta(x,x^{\prime})&=&e^{i\Phi({\bf x}_{\perp},{\bf x}_{\perp}^{\prime})}
\int{d^4p\over(2\pi)^4}e^{-ip(x-x^{\prime})}
\Delta({\bf p}_{\perp},{\bf p}_{\parallel})\;.
\eqa
\end{widetext}

\bibliography{refs}
\bibliographystyle{apsrmp4-1}

\end{document}